\documentclass[journal]{IEEEtran}
\usepackage{graphicx}
\usepackage{svg}
\usepackage{comment}
\usepackage{bbding}
\usepackage{xcolor}
\usepackage{multirow}
\usepackage{longtable}
\usepackage{array}
\usepackage{cite}
\usepackage{hyperref}
\usepackage{indentfirst}
\usepackage{verbatim}
\usepackage{soul}
\usepackage{tabularx}
\usepackage{lettrine}
\usepackage{colortbl} 
\newcommand*{\rom}[1]{\MakeUppercase{\romannumeral #1}}
\usepackage[shortcuts,acronym, nonumberlist]{glossaries}

\newcommand{\Rev}[1]{{\color{black}{#1}}}

\setstcolor{red}
\ifCLASSINFOpdf
\else
\fi
\makeglossaries 
\newacronym{uwb}{UWB}{Ultra-Wideband}
\newacronym{twr}{TWR}{Two-Way Ranging}
\newacronym{cw}{CW}{Continuous Wave}
\newacronym{fmcw}{FMCW}{Frequency-Modulated Continuous Wave}
\newacronym{rfid}{RFID}{Radio Frequency Identification}
\newacronym{fcc}{FCC}{Federal Communications Commission}
\newacronym{iot}{IoT}{Internet of Things}
\newacronym{hci}{HCI}{Human-Computer Interaction}
\newacronym{rf}{RF}{Radio Frequency}
\newacronym{ai}{AI}{Artificial Intelligence}
\newacronym{llnl}{LLNL}{Lawrence Livermore National Laboratory}
\newacronym{mir}{MIR}{Micropower Impulse Radar}
\newacronym{tdoa}{TDoA}{Time Difference of Arrival}
\newacronym{pdoa}{PDoA}{Phase Difference of Arrival}
\newacronym{tof}{ToF}{Time of Flight}
\newacronym{rssi}{RSSI}{Received Signal Strength Indication}
\newacronym{bpm}{BPM}{Bi-Phase Modulation}
\newacronym{pam}{PAM}{Pulse Amplitude Modulation}
\newacronym{ook}{OOK}{On-Off Keying}
\newacronym{ppm}{PPM}{Pulse Position Modulation}
\newacronym{ir-uwb}{IR-UWB}{Impulse Radio Ultra Wideband}
\newacronym{ds-uwb}{DS-UWB}{Direct Sequence Ultra Wideband}
\newacronym{pn}{PN}{Pseudo-Random Noise}
\newacronym{cir}{CIR}{Channel Impulse Response}
\newacronym{mmwave}{mmWave}{Millimeter Wave}
\newacronym{tx}{TX}{Transmitter}
\newacronym{rx}{RX}{Receiver}
\newacronym{etsi}{ETSI}{European Telecommunications Standards Institute}
\newacronym{api}{API}{Application Programming Interface}
\newacronym{tinyml}{TinyML}{Tiny Machine Learning}
\newacronym{lte}{LTE}{Long Term Evolution}
\newacronym{ccc}{CCC}{Car Connectivity Consortium}
\newacronym{dnn}{DNN}{Deep Neural Network}
\newacronym{pca}{PCA}{Principal Component Analysis}
\newacronym{amf}{AMF}{Adaptive Median Filter}
\newacronym{kld}{KLD}{Kullback-Leibler Divergence}
\newacronym{wpt}{WPT}{Wavelet Packet Transform}
\newacronym{spc}{SPC}{Statistical Process Control}
\newacronym{kpca}{KPCA}{Kernel Principal Component Analysis}
\newacronym{svm}{SVM}{Support Vector Machines}
\newacronym{cnn}{CNN}{Convolutional Neural Networks}
\newacronym{rnn}{RNN}{Recurrent Neural Networks}
\newacronym{gps}{GPS}{Global Positioning System}
\newacronym{los}{LOS}{Line of Sight}
\newacronym{dfl}{DFL}{Device-Free Localization}
\newacronym{aoa}{AoA}{Angle-of-Arrival}
\newacronym{svd}{SVD}{Singular Value Decomposition}
\newacronym{rmse}{RMSE}{Root Mean Square Error}
\newacronym{acir}{ACIR}{Accumulated Channel Impulse Response}
\newacronym{rpca}{RPCA}{Robust Principal Component Analysis}
\newacronym{cut}{CUT}{Cell Under Cut}
\newacronym{ekf}{EKF}{Extended Kalman Filter}
\newacronym{ukf}{UKF}{Unscented Kalman Filter}
\newacronym{ut}{UT}{Un-Scented Transformation}
\newacronym{pdf}{PDF}{Probability Density Function}
\newacronym{gm-phd}{GM-PHD}{Gaussian Mixture Probability Hypothesis Density}
\newacronym{phd}{PHD}{Probability Hypothesis Density}
\newacronym{dop}{DOP}{Dilution-of-Precision}
\newacronym{spt}{SPT}{Sleep Postural Transition}
\newacronym{roi}{ROI}{Region of Interest}
\newacronym{adl}{ADL}{Activities of Daily Living}
\newacronym{lstm}{LSTM}{Long Short-Term Memory}
\newacronym{knn}{KNN}{K-Nearest Neighbors}
\newacronym{cae}{CAE}{Convolutional Autoencoder}
\newacronym{tdi}{TDI}{Time-Domain Imaging}
\newacronym{hog}{HOG}{Histogram of Oriented Gradients}
\newacronym{stft}{STFT}{Short-Time Fourier Transform}
\newacronym{2d-vmd}{2D-VMD}{Two-Dimensional Variational Mode Decomposition}
\newacronym{scdae}{SCDAE}{Semi-Supervised Stacked Convolutional Denoising Autoencoder}
\newacronym{se}{SE}{Squeeze-and-Excitation}
\newacronym{scgrnn}{SCGRNN}{Segmented Convolutional Gated Recurrent Neural Networks}
\newacronym{convlstm}{ConvLSTM}{Convolutional Long Short-Term Memory}
\newacronym{hr}{HR}{Heart Rate}
\newacronym{rr}{RR}{Respiration Rate}
\newacronym{ecg}{ECG}{Electrocardiogram}
\newacronym{bmi}{BMI}{Body Mass Index}
\newacronym{fft}{FFT}{Fast Fourier Transform}
\newacronym{hht}{HHT}{Hilbert-Huang Transform}
\newacronym{mae}{MAE}{Mean Absolute Error}
\newacronym{sids}{SIDS}{Sudden Infant Death Syndrome}
\newacronym{rher}{RHER}{Respiratory and Heartbeat Energy Ratio}
\newacronym{wtd}{WTD}{Wavelength Threshold Denoising}
\newacronym{eemd}{EEMD}{Ensemble Empirical Mode Decomposition}
\newacronym{cwt}{CWT}{Continuous-Wavelet Transform}
\newacronym{vmd}{VMD}{Variational Mode Decomposition}
\newacronym{tdca}{TDCA}{Time Domain Coherent Accumulation}
\newacronym{emd}{EMD}{Empirical Mode Decomposition}
\newacronym{nlos}{NLOS}{Non Line of Sight}
\newacronym{nic}{NIC}{Network Interface Card}
\newacronym{pwr}{PWR}{Passive WiFi Radar}
\newacronym{har}{HAR}{Human Activity Recognition}
\newacronym{rtls}{RTLS}{Real-Time Location Systems}
\newacronym{wsn}{WSN}{Wireless Sensor Network}
\newacronym{tw-tof}{TW-TOF}{Two-Way Time of Flight}
\newacronym{ads}{ADS}{Advanced Design System}
\newacronym{mom}{MoM}{Method-of-Momen}
\newacronym{ml}{ML}{Machine Learning}
\newacronym{afsiw}{AFSIW}{Air-Filled Substrate-Integrated-Waveguide}
\newacronym{dra}{DRA}{Dielectric Resonator Antenna}
\newacronym{siso}{SISO}{Single Input Single Output}
\newacronym{mimo}{MIMO}{Multiple Input Multiple Output}
\newacronym{snr}{SNR}{Signal-to-Noise Ratio}
\newacronym{simo}{SIMO}{Single In Multiple Out}
\newacronym{vats}{VATS}{Variance-Based Temporary-Spatial}
\newacronym{cfar}{CFAR}{Constant False Alarm Rate}
\newacronym{isac}{ISAC}{Integrated Sensing and Communication}
\newacronym{ar}{AR}{Augmented reality}
\newacronym{vr}{VR}{Virtual Reality}
\newacronym{fim}{FIM}{Fisher Information Matrix}
\newacronym{bpsk}{BPSK}{Binary Phase Shift Keying}
\newacronym{efim}{EFIM}{Equivalent Fisher Information Matrix}
\newacronym{crb}{CRB}{Cramer-Rao Boun}
\newacronym{cr}{CR}{Cognitive Radio}
\newacronym{toa}{TOA}{Time of Arrival}
\newacronym{awgn}{AWGN}{Additive white Gaussian noise}

\begin{document}

\title{A Comprehensive Overview on UWB Radar: Applications, Standards, Signal Processing Techniques, Datasets, Radio Chips, Trends and Future Research Directions}
\author{Mohammad~Cheraghinia,~
        Adnan~Shahid,~\IEEEmembership{Senior~Member,~IEEE,}
        Stijn~Luchie, 
        Gert-Jan~Gordebeke,~\IEEEmembership{Member,~IEEE,}
        Olivier~Caytan,~\IEEEmembership{Member,~IEEE,}
        Jaron~Fontaine,
        Ben~Van~Herbruggen,
        Sam~Lemey,~\IEEEmembership{Member,~IEEE,}
        and~Eli~De Poorter%
\thanks{
The authors, M. Cheraghinia, A. Shahid, S. Luchie, G. Gordebeke, O. Caytan, J. Fontaine, B. Herbruggen, S. Lemey, and E. De Poorter, are affiliated with IDLab, Department of Information Technology, Ghent University - imec, 9052 Ghent, Belgium. Emails: (Mohammad.Cheraghinia, Adnan.Shahid, Stijn.Luchie, GertJan.Gordebeke, Olivier.Caytan, Jaron.Fontaine, Ben.VanHerbruggen, Sam.Lemey, Eli.DePoorter)@UGent.be.}}

\markboth{}%
{Shell \MakeLowercase{\textit{et al.}}: Bare Demo of IEEEtran.cls for IEEE Journals}
\maketitle

\begin{abstract}

Due to their large bandwidth, relatively low cost, and robust performance, \ac{uwb} radio chips can be used for a wide variety of applications, including localization, communication, and radar. This article offers an exhaustive survey of recent progress in \ac{uwb} radar technology. The goal of this survey is to provide a comprehensive view of the technical fundamentals and emerging trends in \ac{uwb} radar. Our analysis is categorized into multiple parts. Firstly, we explore the fundamental concepts of \ac{uwb} radar technology from a technology and standardization point of view. Secondly, we examine the most relevant UWB applications and use cases, such as device-free localization, activity recognition, presence detection, and vital sign monitoring, discussing each time the bandwidth requirements, processing techniques, algorithms, latest developments, relevant example papers, and trends. Next, we steer readers toward relevant datasets and available radio chipsets. Finally, we discuss ongoing challenges and potential future research avenues. As such, this overview paper is designed to be a cornerstone reference for researchers charting the course of \ac{uwb} radar technology over the last decade.

\end{abstract}


\IEEEpeerreviewmaketitle


\section{Introduction}

\lettrine[findent=3pt]{\textbf{R}}{ }adar is a detection system that uses radio waves to determine objects' range, angle, or velocity. A transmitter emits a radar signal, which is reflected off the target and returned to a receiver. The time it takes for the echo to return is used to calculate the distance to the object. Both radar and non-radar UWB systems utilize the unique properties of \Rev{\ac{uwb}} signals. However, they are tailored to distinct applications based on their operational needs and the nature of the information they are designed to collect or transmit.

This survey paper presents a comprehensive review of \ac{uwb} technology, focusing on radar applications. \ac{uwb} is defined as a radio frequency technology that transmits data across a wide bandwidth (500 MHz or more) or at least 20\% of the arithmetic center frequency. The use of a large bandwidth allows the transmission of very short, narrow pulses in the time domain, according to Formula $BW \times T \geq 4/\pi$, which expresses the connection between bandwidth (BW) and pulse duration (T). For traditional communication technologies such as WiFi, bandwidths are limited to e.g. 20 MHz, resulting in a pulse width larger than 4 nanoseconds. In contrast, \ac{uwb} systems of 500 MHz have time pulses of only 0.16 ns wide. This timing resolution is sufficiently accurate to distinguish between the incoming signal and the reflections of the signal, making the technology highly suitable for radar applications.

Due to this large bandwidth, \ac{uwb} was originally proposed for security radar applications and communication purposes. A larger bandwidth allows better range resolution, allowing improved radar performance. Simultaneously, the large bandwidth made the technology robust against narrowband interference, allowing highly robust communications. In 2002, the \ac{fcc} opened up the technology towards non-military applications by allowing the unlicensed use of UWB systems in radar, public safety, and data communication applications, albeit with strict rules on the allowed frequencies and the maximum transmitted power. This decision resulted in the growing popularity of the technology. Especially in the radar and wireless localization domains, where the timing resolution allows us to calculate the time-of-arrival of signals with great accuracy.

\subsection{UWB radar versus mmwave, FMCW, and WiFi radar}

Multiple technologies can be used for radar applications. Noteworthy technologies include \ac{uwb}, \ac{mmwave}, \ac{fmcw}, and WiFi. The most important factors impacting the choice of optimal radar technology are the bandwidth, the center frequency, the modulation type, and the cost and availability.

\begin{figure}[t]
\includegraphics[width=0.4\textwidth]{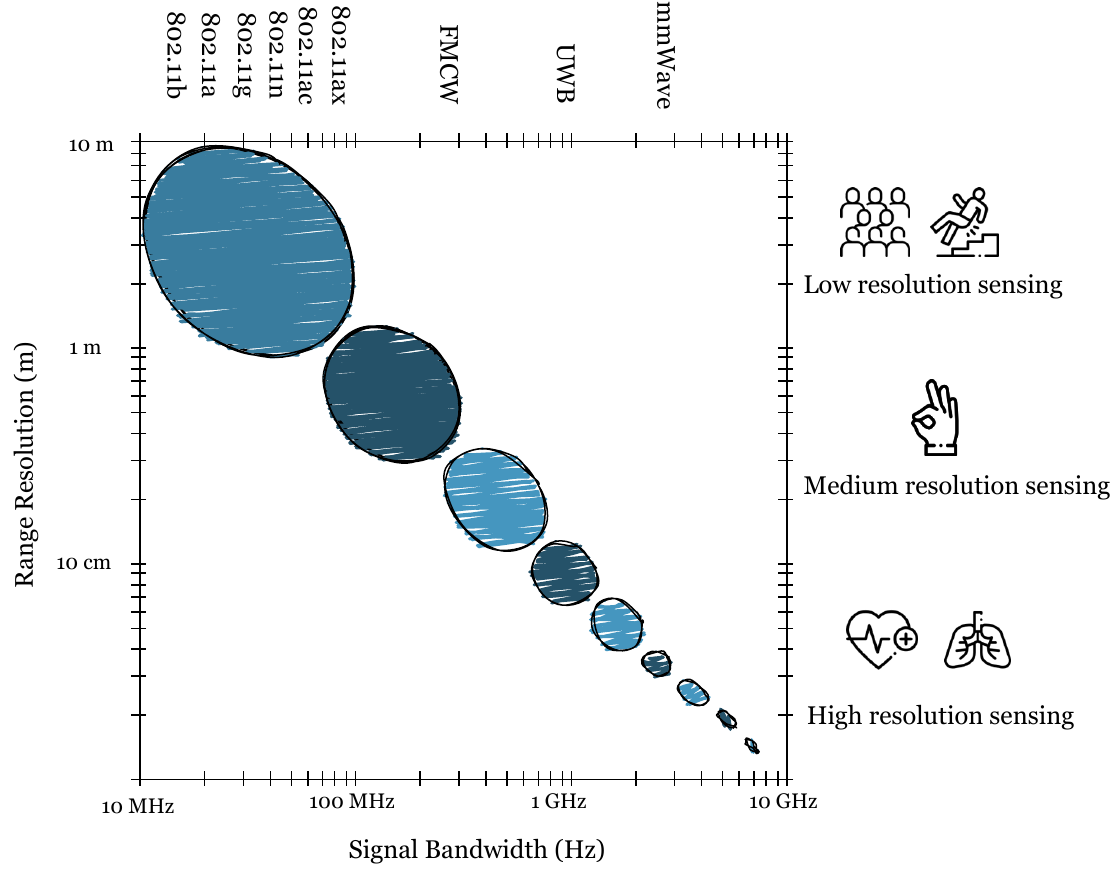}
\centering
\caption{The bandwidth of a radar determines the range resolution, with smaller range resolutions allowing to distinguish more details. This figure compares the typical bandwidth and corresponding range resolution of various popular radar technologies, as well as their potential application domains.  }
\label{BWRangeResApplication}
\end{figure}

(i) \textbf{Impact of bandwidth}. The closer the two targets are, the more difficult they are to distinguish. The radar system will no longer be able to separate them at a certain point. Since a large bandwidth allows a better range resolution, the bandwidth directly determines the size of the objects that can be distinguished from each other. It is possible to calculate the range resolution of different radar technologies using the formula $D_{\text{res}}=c/(2 \times BW)$ where $D_{\text{res}}$ represents the range resolution, and $c$ represents the speed of light. Figure \ref{BWRangeResApplication} shows the range resolutions for different types of radar technology. Besides \ac{mmwave}, \ac{uwb} has the best potential for distinguishing small obstacles. However, increasing the bandwidth also has negative consequences: the maximum radar range ($R_{\text{max}}$) is negatively impacted by increasing bandwidths, according to the formula $R_{\text{max}} = (F_s \times c) / (2 \times BW)$ for a given sampling rate $F_s$. As such, increasing the bandwidth for a given sampling rate will reduce the detection range. Hence, it is important to optimally select the bandwidth according to the application domain: high-resolution sensing applications require large bandwidths, but long-range radars can benefit from smaller bandwidths. The bandwidth of \ac{uwb} sits at a sweet spot, allowing granular range resolutions while still benefiting from relatively large detection ranges. Finally, it is worth noting that the radar resolution can be further improved by increasing the observation time, i.e., by capturing more radar samples. This resolution upscaling technique through capturing more samples is often used for \ac{uwb} radar, as will discussed in later sections.

(ii) \textbf{Impact of center frequency}. The center frequency impacts the absorption rate of wireless signals by obstacles and air molecules. Lower center frequency signals typically propagate further, increasing the range of the signals. Wi-Fi operates on the 2.4 GHz and 5 GHz bands, making it ideal for consumer electronics. It is convenient to use and can support demanding activities like streaming and file sharing. In contrast, \ac{uwb} operates between 3.1 GHz and 10.6 GHz. It offers data transmission speeds of up to 110 Mbps within 10 meters but can support ranges of up to 100 meters at lower data rates and is particularly resistant to interference. Lastly, \ac{mmwave} operates between 30 GHz and 300 GHz. These high frequencies facilitate high-speed data transmission but severely limit the range. Physical barriers can especially obstruct the signals. Again, \ac{uwb} sits at a sweet spot, offering detection ranges up to tens of meters for low-resolution radar whilst still supporting high-resolution radar in close regions. Moreover, the relatively low center frequency of \ac{uwb} radar allows \ac{uwb} signals to penetrate through several materials, making it better suitable for ground-penetrating radar, medical imaging, and through-the-wall observations than technologies such as mmWave.

(iii) \textbf{Impact of modulation type}. Finally, the modulation type impacts the radar's complexity, energy consumption, and robustness. Technologies such as WiFi and \ac{mmwave} use modulation types such as OFDM, QPSK, QAM, etc. which were originally designed for spectral efficiency and high-throughput communications. Although highly efficient for communication, such systems can be complex, requiring advanced modulation and demodulation processes in the transceiver. Basic modulation, such as QAM signals, may also be sensitive to noise and fading, impacting performance in certain conditions. In contrast, more complex modulations, such as OFDM, may face challenges in obtaining unambiguous Doppler measurements due to the multicarrier nature of the signal. On the other hand, \ac{fmcw} radars use a continuous waveform characterized by a continuous variation in carrier frequency over time, which can provide range and velocity information simultaneously. Moreover, such a continuous waveform allows advanced signal processing techniques such as Fourier transform and correlation. This continuity aids in distinguishing between the radar signal and interference. However, \ac{fmcw} does not allow dual use of the radio for both radar and communication purposes. Finally, most \ac{uwb} radar uses pulse-based modulation schemes that facilitate precise time-of-flight measurements. This temporal precision is critical for both localization and radar, whilst also supporting high data throughput. Additionally, UWB's inherent resistance to interference enhances its suitability for applications requiring robust performance in crowded electromagnetic environments.

To summarize, due to its large bandwidth, relatively low center frequency, and pulse-based modulation, \ac{uwb} offers good range resolution whilst still being able to penetrate through obstacles and/or realize long-range radar detection. It is also suitable for (simultaneous) radar and communication, making it a very versatile technology for many radar use cases. A more in-depth discussion of \ac{uwb}-specific technology aspects for radar applications will be given in Section~\ref{section:basics}.

\begin{table*}[t!]
\caption{List of Technical Abbreviations in Alphabetical Order}
\fontsize{8pt}{8.2pt}\selectfont
\label{table:acronyms}
\renewcommand{\arraystretch}{1.1}
\centering
\begin{tabularx}{\textwidth}{X X}

\begin{tabular}[t]{m{1.4cm}|m{6.4cm}} 
 \hline
Abbreviation & Definition\\ [0.5ex] 
 \hline\hline
ACIR & Accumulated Channel Impulse Response\\
 ADL & Activities of Daily Living \\
 AFSIW & Air-Filled Substrate-Integrated-Waveguide \\
 AI & Artificial Intelligence \\
 AoA & Angle-of-Arrival \\
 API & Application Programming Interface \\
 AR & Augmented Reality \\
 AWGN & Additive White Gaussian Noise \\
 BMI & Body Mass Index \\
 BPM & Bi-Phase Modulation \\
 BPSK & Binary Phase Shift Keying \\
 CAE & Convolutional Autoencoder \\
 CCC & Car Connectivity Consortium \\
 CFAR & Constant False Alarm Rate\\
 CIR & Channel Impulse Response \\
 CNN & Convolutional Neural Networks \\
 ConvLSTM & Convolutional Long Short-Term Memory \\
 CR & Cognitive Radio \\
 CRB & Cramer-Rao Bound \\
 CUT & Cell Under Test \\
 CW & Continuous Wave \\
 CWT & Continuous-Wavelet Transform \\
 DFL & Device-Free Localization \\
 DNN & Deep Neural Network \\
 DOP & Dilution-of-Precision \\
 DRA & Dielectric Resonator Antenna\\
 DS-UWB & Direct Sequence Ultra Wideband \\
 EEMD & Ensemble Empirical Mode Decomposition \\
 ECG & Electrocardiogram \\
 EKF & Extended Kalman Filter \\
 EMD & Empirical Mode Decomposition \\
 EFIM & Equivalent Fisher Information Matrix \\
 ETSI & European Telecommunications Standards Institute \\
 FCC & Federal Communications Commission \\
 FFT & Fast Fourier Transform \\
 FMCW & Frequency-Modulated Continuous Wave \\
 FIM & Fisher Information Matrix \\
 GM-PHD & Gaussian Mixture Probability Hypothesis Density \\
 GPS & Global Positioning System \\
 HAR & Human Activity Recognition \\
 HCI & Human-Computer Interaction \\
 HHT & Hilbert-Huang Transform \\
 HOG & Histogram of Oriented Gradients \\
 HR & Heart Rate \\
 IR-UWB & Impulse Radio Ultra Wideband \\
 IoT & Internet of Things \\
 ISAC & Integrated Sensing and Communication \\
 KLD & Kullback-Leibler Divergence \\

\end{tabular}

&

\begin{tabular}[t]{m{1.4cm}|m{6.4cm}}
 \hline
Abbreviation & Definition\\ [0.5ex] 
 \hline\hline
  KNN & K-Nearest Neighbors \\
 KPCA & Kernel Principal Component Analysis \\
 LOS & Line of Sight \\
 LSTM & Long Short-Term Memory \\
 MAE & Mean Absolute Error \\
 ML & Machine Learning \\
 MIMO & Multiple Input Multiple Output \\
 MIR & Micropower Impulse Radar \\
mmWave & Millimeter Wave \\
 NLOS & Non Line of Sight \\
 NIC & Network Interface Card \\
 OOK & On-Off Keying \\
 PAM & Pulse Amplitude Modulation \\
 PCA & Principal Component Analysis \\
 PDF & Probability Density Function \\
 PDoA & Phase Difference of Arrival \\
 PHD & Probability Hypothesis Density \\
 PN & Pseudo-Random Noise \\
 PPM & Pulse Position Modulation \\
 PWR & Passive WiFi Radar \\
 RF & Radio Frequency \\

 RMSE & Root Mean Square Error \\
 ROI & Region of Interest \\
 RPCA & Robust Principal Component Analysis \\
 RR & Respiration Rate \\
 RSSI & Received Signal Strength Indication \\
 SCDAE & Semi-Supervised Stacked Convolutional Denoising Autoencoder \\
 SCGRNN & Segmented Convolutional Gated Recurrent Neural Networks \\
 SIDS & Sudden Infant Death Syndrome \\
 SPC & Statistical Process Control \\
 SPT & Sleep Postural Transition \\
 SVM & Support Vector Machines \\
 STFT & Short-Time Fourier Transform \\
 TDCA & Time Domain Coherent Accumulation \\
 TDI & Time-Domain Imaging \\
 ToF & Time of Flight \\
 TinyML & Tiny Machine Learning \\
 TX & Transmitter \\
 UKF & Unscented Kalman Filter \\
 UT & Un-Scented Transformation \\
 UWB & Ultra Wideband \\
 VATS & Variance-Based Temporary-Spatial\\
 VMD & Variational Mode Decomposition \\
 VR & Virtual Reality \\
 WPT & Wavelet Packet Transform \\
 WTD & Wavelength Threshold Denoising \\
\end{tabular}

\end{tabularx}
\end{table*}

\begin{figure*}[t]
\includegraphics[width=0.7\textwidth]{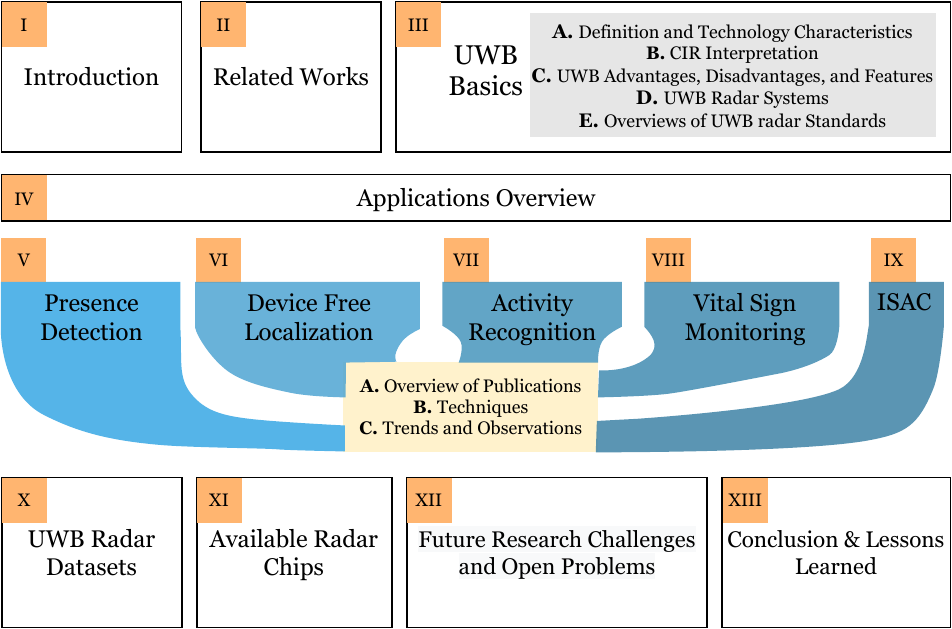}
\centering
\caption{The outline of this survey paper, indicating the structure of the paper and the section numbers for each topic.}
\label{Paper:structure:mindmap}
\end{figure*}

\subsection{Contributions}
This survey provides an up-to-date overview of \ac{uwb} radar applications, processing techniques, hardware aspects, trends, and evolutions. In this survey, we make the following contributions to the field of UWB radar:

\begin{itemize}
\item This survey categorically defines the fundamentals of UWB radar technology, offering readers a clear understanding of its main characteristics and standards.

\item The survey categorizes and contrasts UWB radar applications, providing an understanding across domains. Dedicated sections offer in-depth overviews of specific UWB radar applications, from presence detection to vital sign monitoring, each providing comprehensive insights.

\item The survey offers a technical overview of UWB radar, encompassing datasets for development, available hardwares, and details required for best hardware selection.

\item Future research areas and ongoing challenges in UWB radar are outlined, mapping out the trajectory for future advancements in the field.
\end{itemize}

\subsection{Organization of the Survey}

Our paper is structured as depicted in Fig.~\ref{Paper:structure:mindmap}. Section~\ref{section:RW} discusses related works, including UWB-specific and non-UWB-specific radar overviews. Following this, Section~\ref{section:basics} discusses the essential fundamentals and characteristics of UWB technology and UWB radar, then delves into different UWB radar systems and available standards. Section~\ref{section:AppOver} categorizes and contrasts the applications of UWB radar, thereby preparing for the exploration of emerging trends, challenges, and techniques within these applications in Section~\ref{section:presence} to Section~\ref{section:vitalsign}, each providing a comprehensive overview of a specific application: Section~\ref{section:presence} is dedicated to presence detection; Section~\ref{section:DFL} delves into device-free localization; Section~\ref{section:activity} explores activity and gesture recognition; Section~\ref{section:vitalsign} provides an in-depth analysis of vital sign monitoring; and Section~\ref{section:jsac} explores upcoming innovations related to \ac{isac}.

After exploring UWB radar applications, the following sections focus on the practical aspects of this technology. Section~\ref{section:dataset} catalogs available open-source datasets essential for developing and evaluating UWB algorithms. Complementing this, Section~\ref{section:chips} provides an overview of the available UWB chips and devices. Researchers can quickly identify appropriate datasets and devices with these essential resources for UWB radar research. Concluding the survey, Section~\ref{section:Future} outlines future research areas and ongoing challenges in UWB radar, while Section~\ref{section:Conclusion-lessons-learned} encapsulates the essence of the survey, summarizing the lessons learned and conclusions from various sections. For easy reference, this survey provides a comprehensive list of abbreviations and their corresponding definitions in Table~\ref{table:acronyms}.

\section{Related work } \label{section:RW}
This section provides a comprehensive overview of prior \ac{uwb} radar survey papers, as well as their scope and limitations. Based on this overview, we will identify the scope, research gaps, and limitations of existing work and discuss how our work will fill the gap. 

\begin{table*}[t!]
\fontsize{8pt}{8.2pt}\selectfont
\caption{Overview of prior \ac{uwb} radar surveys and how these works differ from our overview paper.} 
\label{SyrveyComp:table}
\renewcommand{\arraystretch}{1.1}
\centering
\begin{tabular}{m{0.65cm}|m{8.1cm} | m{2cm}| m{1cm} m{0.9cm} m{1.2cm} m{0.9cm} m{0.9cm}} 
 \hline
 Paper / year & Scope & Applications & \ac{uwb} specific  & UWB radar Standards & Chipsets &  Datasets\\ [0.5ex] 
 \hline\hline

\cite{WearUWB:Review} /2018 & This paper discusses and compares UWB wearable antennas for off-body and on-body communications. Furthermore, it discusses UWB propagation models for off-body and on-body channels and then provides examples of UWB wireless body area networks using indoor localization and breathing monitoring. & Wireless body area networks & \Checkmark   & - & - & - \\

 \hline

\cite{MedSenseImage:UWB} /2005 & This paper discusses UWB radar methods and techniques for medical sensing and imaging. It covers the basics of UWB technology, its advantages, and the basics of sensing with UWB. The paper also explores the medical applications of UWB radars. & Vital Signs and Imaging  & \Checkmark   & - & - & - \\

\hline

\cite{ShortRangeVehcular:UWBradar} /2004  & This paper discusses the use of \ac{uwb} radar sensors in short-range vehicular applications, including FCC compliance, system architecture, waveform design, and suitable technology for sensor integration. It analyzes potential benefits and challenges and reviews older research in this field. & Vehicular Applications & \Checkmark & \Checkmark & \Checkmark& - \\

\hline

\cite{UWBIOT:SmartCity} /2018 & This paper reviews UWB technology in IoT applications, considering its limitations and benefits, and then studies the available UWB standards in this area. The authors provide a wide range of examples of IoT and smart city applications. & IoT, Smart city & \Checkmark   & \Checkmark & - & - \\

\hline

\cite{FeatureCapabilities:UWBradar} /2009 & This paper reviews UWB radars, including signal shape, radiation patterns, detection of UWB signals, and their applications in areas such as medical and localization. & Medical and localization & \Checkmark  & - & - & - \\

\hline

\cite{ShortDistance:UWBradars} /2005 & This paper discusses short-distance UWB radars. First, it covers measuring methods and their advantages, and it describes a non-contact measuring method. The paper also presents the block diagram of a UWB radar and discusses a radar prototype. Finally, it explores the various applications of UWB radars. & Vital signs and activity recognition & \Checkmark& - & \Checkmark & - \\

\hline

\cite{PracticalApplication:UWBradar} /2006 & This paper discusses the practical application of UWB radars. It explains UWB radars and their measuring methods and presents a block scheme for a UWB radar. The paper also introduces prototypes and applications of UWB radars. & Vital Signs and Localization & \Checkmark  & - & \Checkmark & - \\

\hline \hline


\cite{radar:medical:survay}  /2016 & This paper surveys the detection of pathological conditions and vital sign monitoring for three radars, including UWB, CW, and FMCW. The authors present a comprehensive description of the three radar technologies and provide a detailed comparison; then, they go through the different radar antennas and discuss the regulation and safety issues of the radars in medical applications. & Medical applications & -   & - & - & - \\ 

\hline

\cite{IndLocDev:Survey} /2017 & This paper provides a survey on wireless indoor localization, and it covers device-based and device-free techniques, including those that rely on UWB, infrared, ultrasonic, RFID, sensors, and Wi-Fi. It also discusses open research issues in the field.& Indoor Localization & -   & - & - & - \\

\hline

\cite{UbiLoc:Survey} /2019 & This paper covers device-free localization techniques for smart environments. These techniques do not require the use of devices carried by users but rely on sensors and signal processing methods to determine location. The paper classifies device-free localization techniques into radio vision and radar-based approaches. It also discusses applications, open issues, and future directions for research in this field. & Device-free localization  & -  & - & - & - \\

\hline

\cite{HandGestureHCInteraction:Review} /2020 & This paper reviews hand gesture recognition using radar sensors for human-computer interaction. It covers the use of gestures for human-computer interaction, the process of acquiring hand-gesture signals through radar, and the representation of hand-gesture radar signals. The paper also discusses hand-gesture recognition algorithms. & Gesture Recognition  & - & -&\Checkmark& - \\

\hline

\cite{DeepHAR:Survey} /2019 & This paper surveys deep learning techniques applied to human activity recognition on radar. It covers various types of deep learning models and radar systems used for this purpose and reviews deep learning approaches for human activity recognition in radar. Finally, the paper provides an overview of this field's current state of the art and identifies key challenges and opportunities for future research. & Human activity recognition & - & - & - & - \\

\hline

\cite{DL_TL_HAR_survey} /2022 & This paper discusses the use of deep learning and transfer learning in device-free human activity recognition and covers sensor modalities, deep learning, and transfer learning techniques, as well as the current state of research in this field. & Device-free human activity recognition & -  & - & - & \Checkmark \\

\hline

\cite{UWBChip:Survay} /2022 & The paper covers various aspects of radar technology, including sensor architectures, RF transceiver systems, integrated radar sensors, and AI-assisted RF transceiver sensing and communications. It provides an overview of the current state of the art and discusses the challenges and opportunities in this field. & Chip-Scale RF Transceivers for Radar Sensing & -   & - & \Checkmark & - \\ 

\hline \hline
This paper & Applications, Standards, Signal Processing Techniques, Datasets, Radio Chips, Trends and Future Research Directions & Presence detection, Device-free localization, Activity recognition, Vital sign monitoring& \Checkmark &  \Checkmark & \Checkmark & \Checkmark \\
\hline
\end{tabular}
\end{table*}

The majority of surveys and overviews on \ac{uwb} radars can be divided into two categories: UWB-specific and non-UWB-specific. UWB-specific studies are solely concerned with the \ac{uwb} radar technology itself and cover subjects like signal processing methods, antenna design, and system architecture. Unfortunately, most \ac{uwb} technology-focused survey papers are not very recent and, as such, do not include current trends, technologies, and innovations. On the other hand, non-UWB-specific articles discussing the overall impact and relevance of UWB radars in specific use cases, such as vital signs monitoring, gesture recognition, and activity recognition, generally underscore the pros and cons of UWB radars in these applications, comparing them to other types of radars without delving into the technical details of UWB radar technology. Additionally, there are research articles that concentrate on specific radar algorithm techniques, including \ac{ml} and deep learning approaches. While these overviews emphasize \ac{ml} techniques, they provide a broader perspective on applying these algorithms in radar systems rather than an in-depth exploration of UWB-specific methodologies.

Table~\ref{SyrveyComp:table} provides a comprehensive comparison of other surveys and tutorials in the field of \ac{uwb} radar. It includes the publication year of each survey or tutorial and a brief explanation of their scope and whether they are \ac{uwb}-centric. The table also lists the \ac{uwb} radar applications included in each survey or tutorial. In addition to these elements, the table provides information about whether datasets, radar standards, and hardware are discussed in the survey or tutorial. Our survey is unique in that it concentrates solely on UWB radar, including a comprehensive overview of all available applications, solutions, algorithms, and essential components such as radar standards, available hardware platforms, datasets, and future research directions. As such, our work is a helpful tool for researchers in the field of \ac{uwb} radar to find relevant \ac{uwb} radar information and identify potential areas for further research.

\subsection{UWB-Specific Overviews}

The use of UWB radar systems has the potential to significantly impact various industries. This literature review examines studies that solely overview different aspects of UWB radar.

UWB radar technology has found an application in the medical field, where non-invasive and accurate monitoring would improve patient care. The difficulties of medical applications using UWB radars are comprehensively reviewed in \cite{MedSenseImage:UWB}, which discusses methods and techniques for medical sensing and imaging. Similarly, \cite{WearUWB:Review} examines the increasing usage of UWB technology in wireless body area networks, detailing wearable UWB antennas and their applications in on-body and off-body communication, with examples including indoor localization and breathing monitoring.

The automotive industry has benefited from the integration of UWB technology, particularly in short-range applications. \cite{ShortRangeVehcular:UWBradar} explores the potential of UWB radar sensors in vehicular contexts, considering aspects such as \ac{fcc} compliance, system architecture, and waveform design, as well as technological integration challenges. On the other hand, in the growing domain of \ac{iot} and smart cities, UWB technology stands as a cornerstone for various applications. The work by \cite{UWBIOT:SmartCity} delves into the application of UWB in \ac{iot}, addressing the technology's strengths, limitations, and the standards that have been established, alongside practical case studies demonstrating its application in smart city infrastructures.

Although the review of UWB radar technologies includes in-depth discussions on signal processing and hardware design, it is essential to note that some of the large overview papers, such as \cite{FeatureCapabilities:UWBradar}, \cite{ShortDistance:UWBradars}, and \cite{PracticalApplication:UWBradar}, may not fully represent the latest advancements in the field. To gain a complete understanding of the current state of the art in UWB radar technology, it is necessary to complement these historical perspectives with recent research that encompasses the latest trends and technological breakthroughs in UWB signal shape, radiation patterns, detection capabilities, and practical applications.

Consequently, this survey is centered on UWB radar, covering an extensive range from foundational principles to cutting-edge applications and innovative solutions. This comprehensive approach sets our work apart from existing surveys, offering a fresh and novel perspective in UWB radar studies.

\subsection{Non-UWB-Specific Overviews}

There is a wide range of radar technologies available beyond UWB technology that have found applications in various fields. This section explores the literature that offers a broader perspective on radar applications without solely focusing on UWB. Most of these overviews concentrate on specific applications or solutions like \ac{ml} and provide a comparison of different types of radars, including UWB radar.

The medical radar technology landscape is diverse, with different types of radar being employed for various applications. An extensive survey \cite{radar:medical:survay} compares the use of \ac{uwb}, \ac{cw}, and \ac{fmcw} radars for detecting pathological conditions and vital sign monitoring, emphasizing the unique attributes and challenges associated with each technology.

The field of wireless indoor localization has seen the use of various radar technologies, including UWB radar, as discussed in a detailed survey presented in \cite{IndLocDev:Survey}. The paper examines the advantages and limitations of each technology and identifies research questions that remain unresolved. Additionally, human-computer interaction is another area where radar technology has been effectively applied. The survey presented in \cite{HandGestureHCInteraction:Review} focuses on using radar sensors for recognizing hand gestures, outlining the signal acquisition process and the associated recognition algorithms.

Human activity recognition is an active research area where \ac{ml} and transfer learning techniques are applied to various sensor modalities, including radar. The papers \cite{DeepHAR:Survey} and \cite{DL_TL_HAR_survey} explore these techniques, with the former focusing on deep learning for radar-based activity recognition and the latter providing a broader view of the application of both deep learning and transfer learning in the field.

Finally, \cite{UWBChip:Survay} discusses into the latest advancements in radar chip technology, covering sensor architecture, integrated sensors, and the combination of \ac{ai} with \ac{rf} sensing and communication. It offers a valuable perspective on the current technological standards and future trends in radar chip development.

In our study, we extend the coverage of existing literature and provide a comprehensive overview of UWB radar technology, incorporating an analysis of the existing challenges and future work in this field. Unlike other overviews that touch upon UWB technology in a broader sense, our work focuses specifically on UWB radar, providing a detailed analysis. Our survey focuses on the various applications of UWB radar, such as presence detection, device-free localization, activity recognition, and vital sign monitoring. We have also studied the specific methodologies of UWB radar, including standards, hardware components, and datasets, making our survey an essential resource for researchers and practitioners looking for a comprehensive understanding of UWB radar technology. By concentrating solely on UWB radar technology, we have differentiated our survey from others and made it an essential resource for people seeking an in-depth understanding of this technology.

\section{UWB basics} \label{section:basics}

In the 1990s, while working on a new high-speed, low-cost sampler for pulse laser research, McEwan developed a \ac{mir} system, which used \ac{uwb} signals for radar applications. This discovery sparked interest in the potential uses of \ac{uwb} technology, and research on \ac{uwb} has continued to advance in the subsequent decades \cite{MedicineUWBradar:old}.

\subsection{Definition and technology characteristics}
 \ac{uwb} is a wireless technology transmitting sub-nanosecond pulses, resulting in large bandwidths, that utilizes very low-power radio waves to communicate over a short distance. Shannon's capacity formula declares that a wide bandwidth equates to a significant capacity, enabling high processing improvements and accommodating the opportunity to access a massive number of users. To further explore this idea, \ac{uwb} signal has two definition methods: first, to have a bandwidth greater than 0.5 GHz or, second, to have a fractional bandwidth greater than 0.2 \cite{SUmmaryFCC:UWBcommunication}. Fractional bandwidth ($B_F$), in this case, is the proportion of the signal's bandwidth ($BW$) to its central frequency ($f_C$), demarcated by the -10dB emission points \cite{DefClassUWB:signals_devices}. Any signal meeting one of the above conditions is defined as an \ac{uwb} signal. The conditions are defined in brief below:

	\begin{eqnarray}
	 &BW > 500 MHz \\
	 &B_F \geq 0.2
	\end{eqnarray}
And the $B_{F}$ can be defined based on the upper ($f_H$), lower ($f_L$), and $f_C$ frequencies as below:

    \begin{equation}
        B_F = \frac{BW}{f_C} = \frac{(f_H - f_L)}{(f_H + f_L)/2}
    \end{equation}

This technology exploits the spectrum between 3.1 GHz and 10.6 GHz \cite{UWBFCC15revision}, enabling data transmission speed up to reach 110 Mbps over a range of up to 10 meters \cite{OverviewUWBcommunicationSystem}. However, leading companies in the field are striving to improve the data rates, reaching up to 1.66 Gbps for short-range applications \cite{imec-ultra-wideband}. Generally, \ac{uwb} pulses are distinguished by their short pulse duration in the order of sub-nanoseconds or picoseconds \cite{UWBpicosecondRADAR:hardware}. Also, the pulse amplitude must be normalized to meet the \ac{fcc} mask requirements, which we will discuss in more detail in the Overview of UWB standards at the end of this section.

Several limitations and challenges associated with the application of UWB pulses need to be addressed \cite{carrier_free_physical_constraints}. One major challenge is the complex radio channel, which can lead to significant signal degradation due to multipath and fading effects. Another challenge is reconciling UWB and high-efficiency operation in a compact, low-frequency footprint. To overcome these challenges, researchers have explored using higher frequencies and more advanced antenna designs that can operate efficiently at UWB frequencies. Moreover, the propagation mechanisms for UWB signals in mobile and indoor settings are intricate, requiring careful consideration of the environment and signal characteristics to achieve optimal performance. Additionally, the characteristics of the UWB communication channel cause the pulse shape to undergo complete changes, including not just phase and amplitude alteration but also frequency dispersion, which can result in significant signal distortion. As such, it is crucial to develop signal-processing techniques to mitigate these effects and improve the overall performance of UWB systems.

\ac{uwb} channels can be analyzed in the time and frequency domains, each of which has specific parameters to investigate. For instance, in the time domain, temporal properties of the channel, such as amplitude fading, time of arrival, and power delay profile, are taken into consideration. On the other hand, parameters like the geometry, proximity to the unit circle of the poles, and the channel's frequency response are studied in the frequency domain. Additionally, information about the frequency-selective nature of the channel, such as the bandwidth and the frequency-dependent attenuation, is extracted \cite{UWbchannelSounding}.

The \ac{cir} is a fundamental concept in UWB communication systems that characterizes the behavior of the channel. In our study, we adopt the \ac{cir} model presented in \cite{HighAcc-MP-UWB}:

\begin{equation}\label{eq:CIR}
h(t) = \sum_{i=1}^{N} A_i \delta(t - \tau_i) + \nu(t) 
\end{equation}

This equation consists of two main components: deterministic multipath components and diffuse multipath components modeled as \ac{awgn}. The first term represents the deterministic multipath components, which account for the propagation of the UWB signal through multiple paths with different delays. An amplitude $A_i$ and a corresponding delay $\tau_i$ characterize each component. The second term represents the diffuse multipath components, which capture random variations and disturbances in the channel. These components are modeled as \ac{awgn} denoted by $\nu(t)$, enabling the realistic representation of the channel's behavior. The equation describes the autocorrelation of this noise as below:

\begin{equation}
E\left[\nu(t)\left[\nu(\tau)\right]^*\right] = S(t) \delta(t - \tau)
\end{equation}

\noindent where the autocorrelation function $S(t)$ signifies uncorrelated noise at different time instances, except for a correlation spike at $\tau=t$.

The frequency response of the CIR, denoted as \(H(f)\), provides a comprehensive view of how the channel affects different frequencies of the transmitted signal. This frequency response is the Fourier transform of the time-domain channel impulse response \(h(t)\), and it can be expressed as follows:

\begin{equation}
H(f) = \int_{-\infty}^{\infty} h(t) e^{-j2\pi ft} dt
\end{equation}

Understanding \(H(f)\) is crucial for analyzing the channel's effect on the signal spectrum and designing effective UWB systems that mitigate frequency-dependent fading and distortion.

\ac{uwb} radar is classified into \ac{ds-uwb} radar and \ac{ir-uwb} radar. The fundamental distinction between them is how they transmit and receive signals. To transmit and receive signals, \ac{ds-uwb} radar utilizes a direct-sequence spread spectrum. This method involves modulating a narrowband carrier signal with a wideband \ac{pn} code, which spreads the signal's energy across a wide frequency band. The modulation and signal process can be described as follows:

\begin{equation}
s(t) = \sqrt{\frac{2E}{T}} \cos(2\pi f_c t + \theta) \sum_{n=-\infty}^{\infty} c_n p(t - nT_c)
\end{equation}

Where \(E\) is the energy of the transmitted pulse, \(T\) is the pulse duration, \(f_c\) is the carrier frequency, \(\theta\) is the phase, \(c_n\) represents the code sequence, \(p(t)\) is the pulse shape, and \(T_c\) is the chip duration. This equation shows how the transmitted signal is constructed using the spread spectrum technique to ensure the energy is distributed over a wide bandwidth, enhancing security and resistance to interference.

The receiver can then demodulate the signal using the same \ac{pn} code applied to de-spread it. In \ac{ir-uwb}, on the other hand, the radar transmits and receives signals using an impulse-based technique. Instead of a continuous carrier signal, this approach uses short, higher-power pulses broadcast at random intervals. The receiver may then pick up these pulses and calculate the target's distance by measuring the \ac{tof} \cite{ModulationOptionsUWB}, which can be estimated as:

\begin{equation}
\label{tof:eq}
\text{ToF} = \frac{c \cdot \Delta t}{2}
\end{equation}

where \(c\) is the speed of light, and \(\Delta t\) is the measured time delay.

\ac{ir-uwb} modulations employ techniques such as \ac{pam}, \ac{ook}, \ac{bpm}, and \ac{ppm} to transmit data over a wide range of frequencies. \ac{pam} and \ac{ook} are not widely used in \ac{uwb} systems due to their susceptibility to noise interference. \ac{bpm} has the advantage of being less susceptible to distortion but only supports binary communication. \ac{ppm} is efficient in minimizing spectral peaks and less sensitive to channel noise but requires precise timing synchronization.

The mathematical expressions for these modulations for the pulse shape of $p(t)$ and pulse duration of $T$ is:

\noindent - \ac{pam}: Varies the amplitude of the pulse to encode data \begin{equation} s(t) = \sum_{n=-\infty}^{\infty} A_n p(t-nT) \end{equation}
- \ac{ook}: A special case of \ac{pam} \begin{equation} s(t) = \sum_{n=-\infty}^{\infty} A_n p(t-nT), A_n \in \{0, A\} \end{equation}
- \ac{bpm}: Involves shifting the phase of the pulse \begin{equation} s(t) = \sum_{n=-\infty}^{\infty} p(t-nT) e^{j\phi_n}, \phi_n \in \{0, \pi\} \end{equation}
- \ac{ppm}: Encodes data by varying the position of each pulse in time slot \begin{equation} s(t) = \sum_{n=-\infty}^{\infty} p(t-nT - d_n), d_n \in \{0, \Delta\} \end{equation}

These methods vary in complexity and robustness to noise, with \ac{ppm} offering significant advantages in UWB systems for its spectral efficiency and noise resilience, albeit at the cost of requiring precise timing.\cite{UWB:Trtanscievers:Overview}.

\Rev{Compared to \ac{ds-uwb}, \ac{ir-uwb} is simpler to implement, uses less power, and is more robust against multipath interference. As a result, \ac{ir-uwb} is most commonly used for radar applications. For this reason, the next Section discusses in more detail how to interpret \ac{cir} signals of \ac{ir-uwb} modulations.}

\begin{figure}[ht]
\centering
\includegraphics[width=0.4\textwidth]{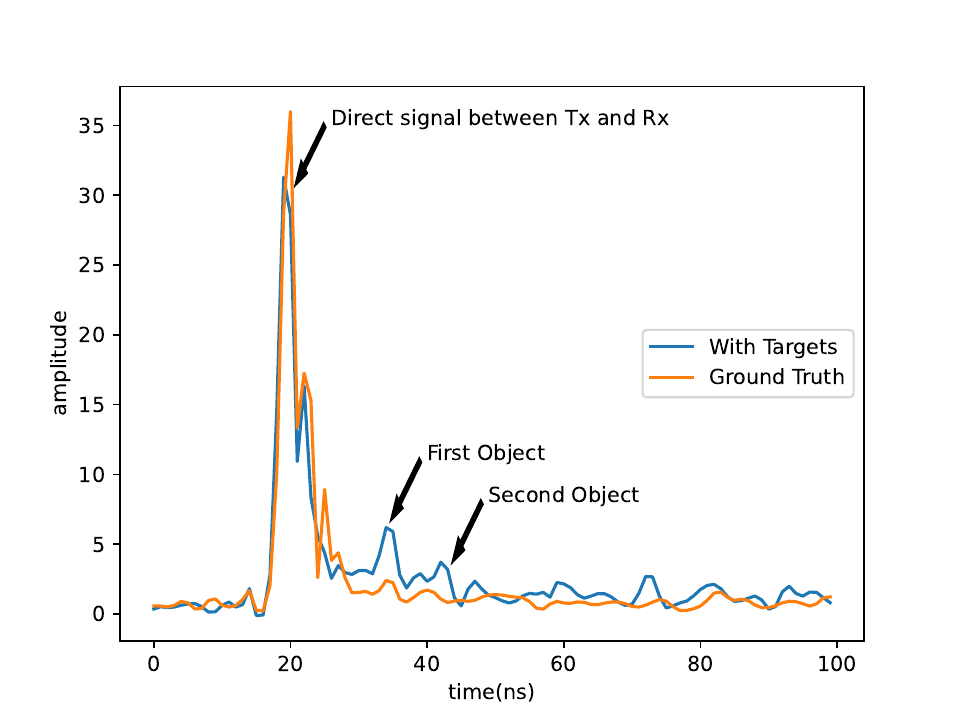}
\caption{\Rev{Illustration of a \ac{cir} sample of UWB radar in two scenarios. Scenario 1 (orange): no target in front of the radar. Scenario 2 (blue): two targets in front of the radar}}
\label{CIR:twotarget}
\end{figure}

\Rev{\subsection{\ac{cir} Interpretation}
In an \ac{uwb} radar, the \ac{cir} represents how the signal propagates from the \ac{tx} to the \ac{rx} as explained in equation~\ref{eq:CIR}. The most important segment of the \ac{cir} is the direct signal between \ac{tx} and \ac{rx}, which is typically the first and usually the strongest component in the CIR. It represents the signal traveling directly from the \ac{tx} to the \ac{rx}, assuming there is no obstacle between them.

The next crucial piece of information is the reflection of the target of interest. This is the signal reflected by the target that the radar intends to detect. This reflection appears after the direct signal in the CIR, as it typically travels a longer path (to the target and back in mono-static arrangement). This reflection's strength and time delay are critical parameters for detecting the target. The time difference with the direct signal can be used to calculate the distance using the \ac{tof} in equation~\ref{tof:eq}, and the phase information in the time-stamp of reflection can be used for \ac{aoa} estimation. However, not all reflections are beneficial. The \ac{cir} also includes unwanted reflections known as multipath components. These signals are reflected off other objects or surfaces in the environment, such as walls, furniture, or even the ground. These reflections complicate the detection of the target due to confusion in multipath components.

We provide an illustration of a typical received \ac{cir} signal for UWB radar. The \ac{cir} signal of Figure~\ref{CIR:twotarget} represents the time pulses received at the receiver side from a single transmission. The arrival time of each pulse is annotated with a timestamp with a resolution of a nanosecond or even sub-nanosecond \cite{10332499}. As such, the CIR provides an instantaneous representation of the environment. The transmission from the TX to the RX is illustrated in Fig.~\ref{CIR:twotarget} as the `direct signal.' It represents the line of sight \ac{cir}, where no targets are in front of the radar. It also indicates on the CIR two objects of interest that have been recognized, as denoted by the arrows. The CIR shows the direct signal, the reflection components, and the multipath components.  
}

\begin{figure}[ht]
\centering
\includegraphics[width=0.4\textwidth]{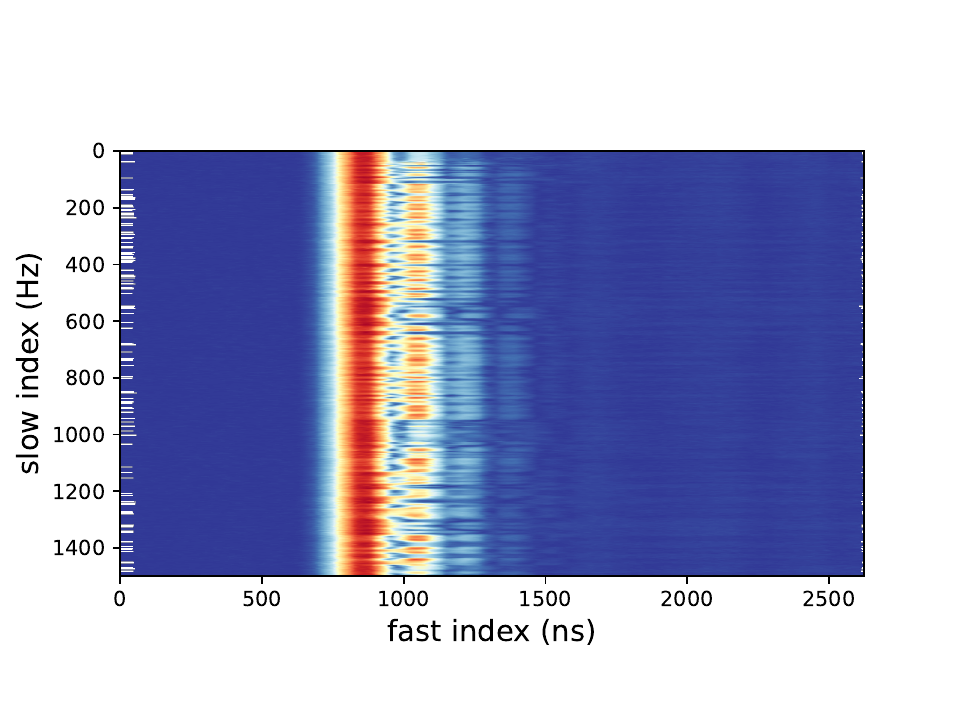}
\caption{\Rev{Illustration of the slow-fast index representation of respiration signals captured using UWB radar. The heatmap shows variations in signal intensity corresponding to respiratory chest movements.}}
\label{CIR:slow_fast_index}
\end{figure}

 \Rev{
 For radar applications, signals are typically transmitted multiple times to obtain a view of how the environment is changing over time. In this case, a slow-fast index representation is used. Fig.~\ref{CIR:slow_fast_index} illustrates the use of a slow-fast index for UWB radar for respiration detection. The slow-fast index representation plots radar data in two dimensions. (i) The x-axis represents the fast index, which shows the intensity of a single received \ac{cir} (expressed in ns or µs). (ii) The y-axis represents the slow index, corresponding to consecutive \ac{cir} measurements over time with a resolution equal to the packet transmission frequency (expressed in Hz). The figure is captured from a mono-static radar in front of a person and clearly shows periodic patterns corresponding to chest movements due to respiration. For ease of interpretation, Fig.~\ref{CIR:slow_fast_index} shows only the amplitude of each received \ac{cir}, but a similar plot can be made for the phase changes over time. As with the \ac{cir}, multipath components also affect this representation, resulting in misinterpretations of the signal. Effective signal processing techniques, such as filtering or signal separation, are often employed to mitigate these effects.}

\subsection{UWB advantages, disadvantages, and features}

The capacity of \ac{uwb} radar to deliver precise positioning is one of its main advantages. The large bandwidth resulting in short signals because of the inverse relation with bandwidth increases the time resolution of \ac{uwb} signals \cite{WLS:IR-UWB:devicefreeLoc}. It lets a \ac{uwb} radar measures the distance between the radar and the target using the \ac{tof} method. In addition to accurate positioning, \ac{uwb} systems can also offer real-time location measurement under its positioning accuracy limitations, making it ideal for applications requiring fast and stable data transmission, like fast-moving objects in diverse scenarios such as drones and automation systems\cite{UWNpositioningALG:diffINDscenes, UWBimuHRate:cm-levelACC}.

\ac{uwb} is far more accurate than the \ac{rssi} approach used by WiFi and Bluetooth, which can be impacted by multi-path fading and other variables. The resilience of \ac{uwb} radar against multi-path fading, which happens when a signal travels along numerous routes before interfering with one another, is another benefit \cite{DS-CDMA-UWBmodulation-lowfading}. Unfortunately, fading can result in substantial positioning inaccuracies. However, \ac{uwb} radar can get around this issue because of its high time resolution, short pulses, and extremely precise \ac{tof} measurement technique \cite{CCI-UWB-fading-partially}. \ac{uwb} radar is also highly resistant to narrowband interference, which can disrupt the operation of other technologies. The wide bandwidth of \ac{uwb} signals allows them to occupy a large portion of the spectrum and thus reduces the likelihood of interference from other narrowband signals. This makes it a reliable and robust choice for applications that require uninterrupted operation \cite{Trappedvictims:UWBradar}.

Additionally, \ac{uwb} technology has been shown to have the ability to coexist with other wireless technologies without causing significant interference. This is because \ac{uwb} signals have a low power density and can be highly directional, which reduces the likelihood of interference with other systems operating in the same frequency band \cite{UWB-coex-cogn-rad}.

\ac{uwb} technology can be low-power, meaning it can operate for long periods without requiring frequent battery replacements. Furthermore, \ac{uwb} radar is capable of quick and consistent data transmission over a large spectrum of several GHz ranging from 3.1 to 10.6 GHz, despite its low power consumption. Due to its wide spectrum and ability to function over short distances, it can be used for security, asset tracking, and indoor navigation applications \cite{Low-power:UWBcirbuitsArch, low-power:low-rate:UWBtransciver}. On the other hand, for the same reasons, this technology has lower costs than the other methods to provide higher-quality communication.

\begin{table*}[t]
\caption{Overview of UWB radar standards, their standardization organization, and their scope.}
\label{TBL:UWB:standards}
\fontsize{8pt}{8.2pt}\selectfont
\centering
\renewcommand{\arraystretch}{1.3}
\begin{tabular}{m{0.6cm} m{1.6cm}  m{5.7cm} m{8.4cm} } 
\hline
Institute & Number  & Application  & Scope\\ [0.5ex] 
\hline\hline

ETSI & EN 302 065-1 & Communications devices, Presence detection, Non-contact vital signs  & Applies to UWB technologies for short-range applications across a range of equipment. Restrictions for outdoor fixed locations and aviation use. \\
\hline

ETSI & EN 302 065-2 & Location tracking & Concerns UWB technologies for location tracking, detailing three types of systems (LT1, LT2, LAES). Excludes UWB transmitters used in aviation and public vehicles. \\
\hline

ETSI & EN 302 065-3 & Keyless entry  & Focuses on UWB technologies for short-range applications in road and rail vehicles. Excludes fixed road infrastructure installations and aviation use. \\
\hline

ETSI & EN 302 065-4 &  Building material analysis, Security Scanning, Ground humidity and condition, Non-Contact based external material sensing, Parking lot sensor &  Specifies requirements for UWB in material sensing applications, including non-fixed and fixed material sensors. Excludes UWB communication devices and various radar devices.\\
\hline

IEEE & 1672-2006 & UWB radar &  Organizes the terms and definitions used in the field of UWB radar\\
\hline

IEEE & 802.15.4ab & High volume consumer platforms, Public health sector, Industrial sector, Transportation sector & Focuses on enhancing UWB PHYs, MAC, and ranging techniques, improving accuracy, power efficiency, device density support, and hybrid operation with narrowband signaling with priority to low duty-cycle ranging.\\
\hline

Google & Ripple & vital sign monitoring, occupancy detection, activity and gesture recognition & A standardized open radar API designed to promote compatibility between hardware and software to boost the development of user-friendly radar applications for a wide range of consumer purposes.\\

\hline\hline
\end{tabular}
\end{table*}

Due to the wide bandwidth employed by \ac{uwb} radar systems, the peak power density of the signal is reduced, and the possibility of injuring live tissues is decreased. The signal's absorption by live tissue is diminished at high frequencies, reducing the possibility of heating \cite{HumanTissuePropag}. In contrast, traditional radar systems employ powerful, narrowband waves. Because \ac{uwb} radar can pass through non-metallic objects, it is a fantastic choice for non-invasive testing, medical imaging, and other applications needing less intrusion \cite{UWBtumorDetection}.

This characteristic of \ac{uwb} radar technology makes it an essential instrument for non-destructive testing and medical imaging since it offers an easy and convenient way to measure internal body structures and find abnormalities.

\begin{figure}[ht]
\centering
\includegraphics[width=0.4\textwidth]{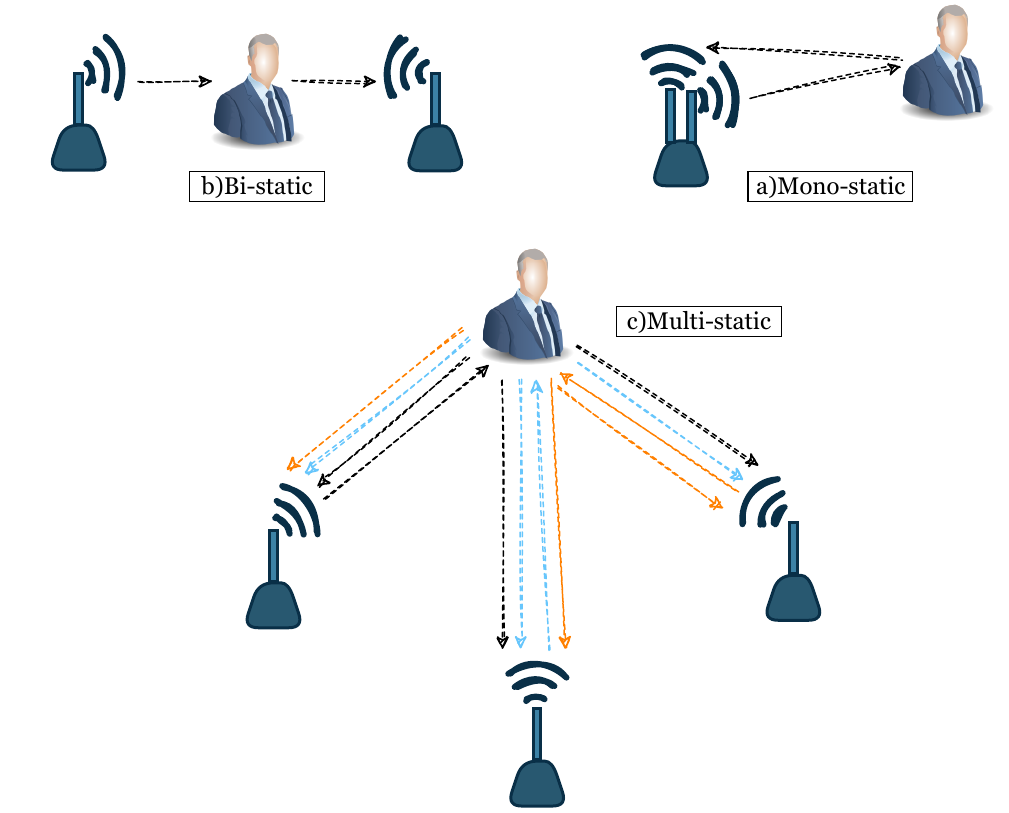}
\caption{
Different UWB radar system topologies. a) Mono-static; b) Bi-static, characterized by separate placements of the transmitter and receiver, providing varied angles to the target (in this case, with co-linear alignment); c) Multi-static, featuring multiple units that can be either co-linearly aligned on one side or strategically positioned to encircle the target for enhanced detection.}
\label{mono-bi-multi}

\end{figure}

\subsection{UWB radar Systems}

Radar systems can be classified into three main types based on the arrangement of their transmitters and receivers: mono-static, bi-static, and multi-static \cite{hanle1986survey,9449071}, as depicted in Fig~\ref{mono-bi-multi}.

\paragraph{Mono-static Radar Systems}
In mono-static radar systems, the transmitter and receiver are co-located. This setup is valued for its deployment simplicity and cost-effectiveness. The range of mono-static radars is dependent on factors such as transmitted power, receiver sensitivity, and transmit-to-receive isolation. These systems offer excellent Doppler sensitivity and are highly sensitive to weak signals. However, they are more sensitive to interference and jamming due to the single path for both transmission and reception. The granularity of target detection in mono-static systems is influenced by the resolution of the radar, which is a function of the bandwidth and the signal processing techniques used.

\paragraph{Bi-static Radar Systems}
Bi-static radar systems feature separate locations for the transmitter and receiver. This separation allows for exploiting spatial diversity, potentially offering extended detection ranges under certain geometrical conditions. Sensitivity and Doppler sensitivity in bi-static radars depend highly on the bi-static angle and can be comparable to mono-static radars. The physical separation in bi-static systems enhances target discrimination but may introduce deployment challenges due to the fixed distance between antennas. Interference in bi-static systems can be mitigated by spatial separation, though it can be more complex to manage due to the need for precise alignment and coordination.

\paragraph{Multi-static Radar Systems}
Multi-static radar systems involve multiple transmitters and receivers, providing extensive coverage, high target resolution, and excellent Doppler sensitivity. The complexity of multi-static systems is higher due to the need for tight synchronization among multiple units. They are typically more resistant to jamming and offer the most significant operational flexibility due to their distributed nature despite their higher cost and complexity. The granularity in multi-static systems is greatly enhanced due to the multiple perspectives available for target observation, allowing for finer resolution and more detailed target characterization. Interference management is also more effective in multi-static systems, as the multiple paths for transmission and reception can be leveraged to reduce the impact of jamming and ensure more reliable detection and tracking.

Each configuration—mono-static, bi-static, and multi-static—has specific benefits and limitations. Throughout the survey, we will discuss which of these options are most commonly used for relevant \ac{uwb} use cases such as device-free localization, vital sign monitoring, and activity recognition.

\subsection{Overview of UWB radar standards}
~\label{UWBStandards}

Driven by the increased industry interest in UWB, over the past years many standards have been proposed for the UWB physical layer, and to a lesser extent for UWB localization. An extensive overview of these standards can be found in \cite{UWBstandardsSurvey}. 
In contrast to standards focusing on UWB localization, only a limited number of UWB radar standards are currently available. 

\begin{figure}[htbp]
\centering
\includegraphics[width=0.45\textwidth]{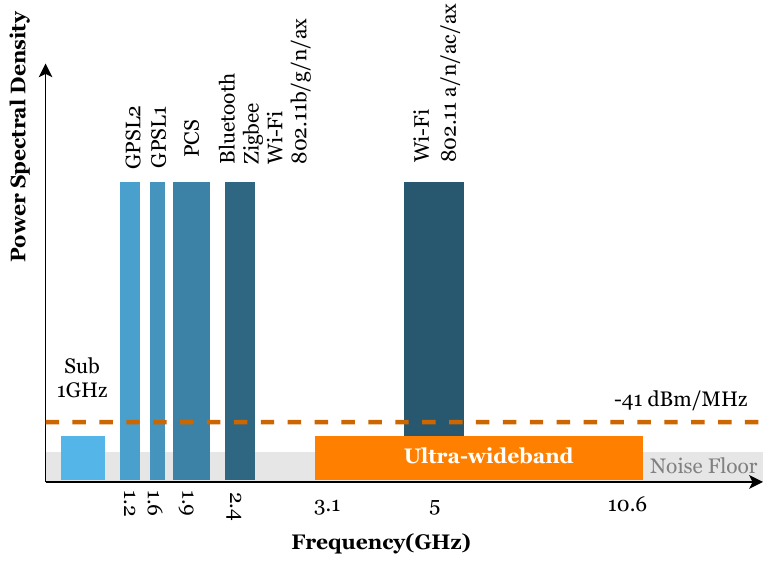}
\centering
\caption{Comparison of the power spectral density and bandwidth of UWB radio technology versus other technologies.}
\label{PED}
\end{figure}

At the physical layer, it is generally assumed that UWB radar systems will use the same standards as those already defined for UWB ranging and localization \cite{UWBstandardsSurvey}. Similar to UWB localization systems, these standards impose regulations for UWB radar, and the FCC has implemented several restrictions on the frequency and power of UWB devices \cite{UWBpropgationCHannel}. As a result, UWB devices must adhere to specific frequency channels and power limitations to prevent interference with other wireless technologies. This is illustrated in Fig.\ref{PED}, which shows the power restrictions that differentiate UWB from other technologies based on the bandwidth requirements.

When considering UWB radar-specific standards, the following ongoing standardization bodies are of interest. From the IEEE organization, the IEEE 1672-2006 standard provides definitions for promoting clarity and consistency in the use of UWB radar terminology. The definitions represent the consensus of a panel of radar experts \cite{1672-2006:IEEEStandard}. While limited standards are available describing the higher UWB radar layers, new and improved standards are likely to focus on a specific use case or a specific environment. For example, \ac{etsi} provides use case-specific standards for cases such as presence detection, through-air non-contact vital signs, location tracking, UWB parking lot sensors, gesture recognition, etc \cite{ETSI-EN-302-065-1, ETSI-EN-302-065-2, ETSI-EN-302-065-3}. 

A significant advancement to be mentioned is introducing the IEEE 802.15 WSN™ Task Group 4ab (TG4ab) 802.15.4 UWB Next Generation standard. This standard modifies the existing 802.15.4-2020 and its approved amendments, such as 802.15.4z-2020. It aims to improve the existing UWB physical layers and the MAC sublayer \cite{IEEE802.15WSN}. This amendment seeks to expand the utility of the standard for a broader range of UWB applications by improving aspects such as coding, modulation schemes, interference mitigation techniques, accuracy, precision, reliability, and interoperability for high-integrity ranging, among others.

Finally, at the service-level and integration level, Ripple is an open \ac{api} standard for general-purpose consumer radar that allows for interoperable software libraries to work across various radar hardware implementations, making it easier to build and integrate new radar applications \cite{Ripple-standard}. In addition, the standard enables developers to create specialized extensions to support their use cases. Google is one of several partners working on developing the standard alongside companies like Ford, Blumio, Infineon, NXP, and Texas Instruments. To concisely present these standards, a table is referenced (see Table~\ref{TBL:UWB:standards}), which summarizes the standards. It includes the scope and applications of the available standards.

\section{Applications Overview}\label{section:AppOver}

 In this section, we will provide a high-level overview of UWB radar use cases and compare them in terms of functionality. Afterward, the next sections will discuss in-depth the technical aspects of each use case.

We classify the applications into five main classes, namely vital sign monitoring, activity recognition, presence detection, distance estimation, device-free localization, and \ac{isac}. Each main category includes subcategories that have specific applications in diverse areas. In the following sections, we will go through all the main categories in detail.

In Fig.~\ref{Paper:structure:mindmap}, we present an outline of the various applications of UWB radar, which will be further dissected in Sections~\ref{section:presence} through \ref{section:jsac}. We will delve into each application category with subsections. These subsections will look into the UWB radar applications by reviewing specific scientific studies, discussing the techniques employed, and analyzing the current trends and observations in each area. Each application will be reviewed to give readers a comprehensive understanding of the state of UWB radar technology in these areas.

Simultaneously, analyzing the various UWB radar technology sector trends reveals distinct dependencies between radar systems and operational needs, unexpected in intended applications. For instance, presence detection generally employs a mono-static system configuration with minimal bandwidth and sampling rates. In contrast, device-free localization demands a more complex multi-static system, albeit with similar low sampling and bandwidth requirements. When delving into activity recognition, the complexity rises when both mono-static and multi-static setups are used, necessitating higher sampling rates and bandwidth. Vital sign monitoring parallels activity recognition regarding high sampling and bandwidth specifications, but it typically operates within a mono-static system framework. UWB technology is customized and adjusted in complexity and technical capability to meet the specific requirements of each application, which will be discussed in detail in the following sections.

\section{Presence Detection}\label{section:presence}

Identifying and examining changes in the movements or presence of objects or individuals within a particular area is often referred to as motion or presence detection. This is widely used across security, monitoring, robotics, and interactive gaming sectors. The application domains of presence detection span multiple uses, such as occupancy detection, people counting, and in-car occupancy detection in the current UWB radar studies.

Various technologies can be utilized for presence detection, ranging from computer vision to infrared, mmwave, and ultrasonic sensors. Amongst these technologies, the use of UWB radar for presence detection has several advantages. UWB radar offers good accuracy at a relatively low cost. When using sufficient bandwidth, UWB radar has a high degree of precision in detecting small changes in distance, motion, and velocity, making it ideal for presence detection. Another argument is that it utilizes very little power. UWB radar transmits and receives signals with low power, making it energy-efficient and suited for long-term monitoring and detection. In addition, UWB radar has the ability to pass through furniture, walls, and other obstructions, making it perfect for locating and following objects and people in cluttered surroundings. Since no identifying personal information is captured, UWB radar is also well-suited for applications where privacy is a concern. Finally, in contrast to e.g. video-based solutions, the installed UWB system can simultaneously be used for localization and/or communication. 

In the upcoming sections, we will examine the available papers related to various use cases in the field of presence detection and categorize the techniques employed for them. Finally, we will explore the trends and insights gathered from the current research in this domain.

\subsection{Overview of scientific publications focusing on presence detection}

The domain of presence detection research is broad. In the subsequent parts of this section, we will explore the scientific articles that focus on this area comprehensively. The following parts of this section are classified into three distinct categories: occupancy detection, people counting, and in-car seat occupancy detection. These papers, as presented in Table~\ref{table:Presence}, do not only note the publication year of each article but also offer a detailed context of the application, achieved accuracy percentages, and several technical specifications, including the experimental setup, employed bandwidth, central frequency, and the type of radio chip used.

\subsubsection{Occupancy detection}

The detection of occupancy or human presence using UWB radar has several practical applications beyond merely monitoring places. In crisis management circumstances such as earthquakes, this technology can save lives by allowing first responders to locate and rescue trapped individuals. Aside from disaster relief, the applications of UWB radar for human detection are diverse, from advanced home automation systems to security surveillance. Given its broad application spectrum, UWB radar is a vital study topic in human presence detection.

\cite{RobustPresence:IndoorUWbradar} introduces a robust detection method capable of operating effectively in diverse clutter environments. It examines various scenarios, such as detecting a standing individual in a typical clutter environment, an individual in prone position in heavily cluttered conditions, and detecting multiple individuals within a conventional clutter setting. On the other hand, the study \cite{AlgorithmHumanDetect} illustrates the utilization of a monostatic IR-UWB radar to detect human presence. The investigation focuses on recognizing individuals by analyzing their unique locomotion patterns. An innovative algorithm is proposed within the study, its primary function being to distinguish ambulatory humans from other moving objects, like automobiles, particularly within metropolitan settings. A monostatic UWB radar was installed 0.5 meters above California Blvd's curb level to gauge the methodology's efficacy. The device conducted 1778 scans when no one was around and 708 more scans when humans were present. The third paper, \cite{TinyMLpresence:UWBradar}, delineates a notable advancement in presence detection using UWB radar, introducing a method that harnesses the power of \ac{tinyml} for efficient resource usage. The proposed solution was tested in two critical scenarios: detecting occupants in a vehicle's rear seats and the presence of an individual in indoor settings. They achieved 98.5\% for indoor scenarios and 99.5\% accuracy for the vehicle's rear seat occupancy with highly efficient resource usage. 

The papers discussed focus on presence detection using UWB radar in mono-static settings, without considering the detection of humans without utilizing their vital signs. For further research, exploring the challenges and scenarios related to the detection of humans without relying on their vital signs in mono-static settings could be an interesting avenue to pursue.

\subsubsection{People Counting}
Applications of UWB radar-based people counting includes crowd management in public spaces, occupancy monitoring in smart buildings, and optimizing resource allocation in healthcare facilities. For example, \cite{WideAreaUWBRad:PeopleCount} delivers experiments involving two indoor environments: a lobby with an open layout and high ceiling and a hall with blocked walls and a low ceiling. Up to ten people were allowed to move freely within fan-shaped spaces, and an IR-UWB radar sensor was positioned near the apex of each space to capture the reflected signals. The accuracy of their measurements is based on two distances - five and ten meters - and an 80-degree field of view. When aiming for a relaxed accuracy of one person, their success rate is approximately 100\%, permitting counting errors of $\mp 1$. However, their success rate for strict accuracy falls between 82\% and 92\%. On the other hand, \cite{PassingPeopleCounting:UWBRadar} counts the number of passing people and utilizes two IR-UWB radar sensors with antennas with a narrow beam width horizontally. The radar's propagation direction is perpendicular to the direction humans are moving, creating two layers capable of simultaneously counting multiple individuals passing. This paper's conclusions for crowd management, urban planning, and smart city applications make this paper attractive at the intersection of advanced radar technology and urban planning.

\begin{table*}[htbp]
\caption{Overview of scientific papers utilizing UWB radar for Presence detection}
\label{table:Presence}
\fontsize{8pt}{8.2pt}\selectfont
\renewcommand{\arraystretch}{1}
\centering
\begin{tabular}{m{0.5cm}|m{0.5cm}  m{1.45cm} m{1.7cm} m{1.4cm} m{0.9cm}m{1.2cm} m{1.2cm}  m{3.4cm}} 
 \hline
 Paper & year  & Sub-type  & Accuracy & system & Devices & Bandwidth& Central F &  Chip\\ [0.5ex] 
 \hline\hline
 \cite{WideAreaUWBRad:PeopleCount} & 2020 & People counting & 81\% 99\% & mono-static & 1 & 1.5GHz & 7.29GHz & Novelda Xethru X4M03 \\

\hline          

\cite{UWBRadarSensor:PeopleCount} & 2017 & People counting & 36\% 70\% & mono-static & 1 &2.3GHz&6.8GHz& Novelda NVA-R661 \\

\hline

\cite{OccupancyFallShips:ReverberantEnv} & 2021 & People counting  Fall detection  & 88\% - 100\%  & multi-static & 8 & 0.9GHz & 4GHz & Decawave DW1000 \\

\hline   

\cite{PassingPeopleCounting:UWBRadar} & 2017 & People counting & 93\% & mono-static & 2 & -& - & Novelda  NVA-6201  \\

\hline

\cite{CNNcounting:UWB-radar} & 2017 & people counting & 99.9\% & mono-static &  1 & 2.3GHz & 6.8GHz & Novelda Xethru X2 \\
\hline

\cite{Bi-Motion-Model-Framework:UWBradar} & 2022 & people counting & 95.5\% - 98.8\% & mono-static &1& 1.4GHz & 7.29GHz & Novelda Xethru X4 \\
\hline

\cite{CNNPeopleCounting:UWBradar} & 2021 & people counting & 97\% & mono-static &1& 4.66GHz & 0.55GHz & -\\
\hline

\cite{CIR_based:Peoplecounting:UWBradar} & 2021 & people counting & 89\% - 100\% & multi-static & 2 & 0.5GHz & - & Decawave DWM1000\\

\hline

\cite{SPtempPhysiologicalVehmotion:UWBradar}& 2022 & People counting & 97.1\% & mono-static & 1 & 1.4GHz & 7.29GHz &  Novelda Xethru X4M03\\

\hline

\cite{JZhang2023}& 2023 & People counting & 86.66\% & mono-static &  1 &  1.4GHz & 7.29GHz & Novelda Xethru X4M02\\

\hline

\cite{GLee2023} & 2023 & People counting & 99\% & bi-static  & 2  & 0.5GHz  & 7.98GHz  &  Decawave DWM3000 \\

\hline

\cite{TinyMLpresence:UWBradar} & 2022 & Occupancy & 90.6\% & mono-static & 1 & 1.2GHz & 7.9GHz & ARIA LT103OEM\\

\hline

\cite{DNNVehiclePresence:UWBradar} & 2020  & Car seat Occupancy & 99\% & mono-static & 1 &6.3GHz& 3.85GHz & Novelda NVA6100\\

\hline

\cite{DetLocVwhiclwPresence:UWBradar} & 2020  & Car seat Occupancy & 91.9\% & mono-static & 1 & 6.3GHz& 3.85GHz & Novelda NVA6100\\

\hline

\cite{UWBCARGRAZ} & 2023 & Car seat Occupancy & -  & mono-static bi-static & 2 & 6.5GHz 7GHz & 0.5GHz  5GHz& ILMSENS m:explore\\

\hline

\cite{Trappedvictims:UWBradar} & 2009 & Trapped human detection & - & bi-static & 2 & 2GHz & 4.5GHz & custom hardware\\

\hline

\cite{RobustPresence:IndoorUWbradar} & 2020 & Occupancy & - & mono-static & 1 & 1.5GHz & 7.29GHz & Novelda Xethru X4M03\\

\hline

\cite{AlgorithmHumanDetect} & 2009 & Occupancy & False alarm: 1.58\% & mono-static & 1 & 3.2GHz & 4.7GHz & Timedomain  P210\\

\hline 

\cite{Ml:Vitalsign:Occupancy} & 2023  & Occupancy & 100\% & mono-static & 1 & - & - & NXP NCJ29D5\\

 \hline\hline
\end{tabular}
\end{table*}

Experiments were carried out in two rooms in the study conducted by \cite{CIR_based:Peoplecounting:UWBradar}. One room was $3\times4.5$ in size with furniture placed near the walls, while the other was $5\times5$ with furniture placed near the walls and in the middle, resulting in limited movement. In order to collect CIR vectors for up to five people in Room A and up to eight people in Room B, UWB modules were placed on desks in each room. These modules transmitted and received signals in a bi-static system, which sets this study apart. Volunteers were either moving randomly or standing still in the rooms while the data was collected. Their research revealed that UWB technology has an accuracy rate of 100\% for five people and 89\% for eight people. They also compared UWB with other technologies, such as Wi-Fi and \ac{lte}, which demonstrated higher accuracy and the ability to distinguish larger groups of people in spacious areas.

The three articles introduced novel strategies for people counting using UWB radar. Each paper considered different circumstances, including broad areas, people-passing counting, and multi-environment approaches, utilizing unique methods to count individuals per their respective scenarios. However, it is essential to address the hardware complexity caused by the high sampling frequency, and future research could focus on more complex scenarios with larger crowds in larger areas.

\subsubsection{In-car seat occupancy detection}

As of recently, UWB devices are also integrated into cars. The main driver of this trend is to allow fob-less car keys, instead unlocking cars with UWB devices from smartphones for improved convenience as well as improved safety. This trend is largely driven by the Digital Key standard from the \ac{ccc}, which utilizes UWB for hands-free, keyless access and location-aware features for cars. For example, Audi and BMW recently included UWB in their next-generation cars to allow unlocking doors with UWB-enabled smartphones. The automotive industry is already thinking of reusing this hardware for cabin sensing for (i) safety (monitoring the alertness of the driver, child left-behind detecting systems, etc.) and (ii) comfort and personalization (detecting which seats are occupied, gesture control for hands-free car interactions, etc..). 

Two recent papers have explored this topic using UWB radar. In the initial paper, \cite{DNNVehiclePresence:UWBradar}, a radar is installed inside a car at the rearview mirror to keep track of all seats. This results in 32 different seating scenarios for five seats, depending on whether they are occupied or not. The authors collected IR-UWB radar signals from these scenarios, varying the subjects, their number, location, and the type of vehicle. A \ac{dnn} was then used to train a classifier using these preprocessed radar signals, eliminating the need for a feature extraction stage. The proposed \ac{dnn}-based localization method improved performance over conventional \ac{ml} techniques, demonstrating its effectiveness in in-vehicle localization. In the second paper \cite{DetLocVwhiclwPresence:UWBradar}, a method was presented to monitor individuals inside a vehicle using the same dataset as the first paper. The paper explains an ensemble learning-based classification technique that uses feature extraction and selection to identify the signal's statistical characteristics as classification criteria. Additionally, it uses a neighborhood component analysis algorithm for feature reduction. They achieve high classification accuracy while keeping the computational complexity low. These studies demonstrate the potential of UWB radar technology for human presence and car seat occupancy detection in vehicles, which could have significant implications for improving safety and comfort in future automobiles.

\subsection{Presence Detection Techniques}

This section describes the two most commonly used techniques for UWB radar-based presence detection: background subtraction and anomaly detection.

\subsubsection{Background Subtraction}
UWB radar systems deployed in realistic environments are impacted by the presence of obstacles, walls, etc., even when no persons are present. Background reduction techniques aim to filter out the effects of these stationary obstacles, allowing to focus only on the person-specific CIR effects that arise when persons are present. In practical terms, they aim to characterize and subtract the time-invariant background CIR, with the aim of then detecting which obstacles are transient. Even without persons, the received CIR samples will vary over time due to imperfect clocks, antenna coupling, impedance mismatch response, and ambient static clutter \cite{M-sequence:Backremove:uwbradar}. As such, often a calibration phase is used to characterize the background using multiple samples in the absence of persons. For this purpose, there are numerous strategies, such as fundamental averaging techniques, adaptive variations, parametric or non-parametric multi-modal background models, and methods utilizing principal component analysis, filters, and predictors\cite{BackSubtractionTech:review}.

There are applications for background removal in UWB radar, as demonstrated in two papers. In \cite{UWBRadarSensor:PeopleCount}, the authors explore techniques such as adaptive thresholding and moving average filtering to eliminate background noise while preserving signals from moving objects. They also consider the use of \ac{pca} but find it does not enhance performance. Meanwhile, in \cite{WideAreaUWBRad:PeopleCount}, the focus is on thresholding as a straightforward yet effective method for background noise removal. The authors suggest the possibility of using adaptive thresholding with varying thresholds for different frames to improve people counting accuracy, particularly in noisy environments. These papers collectively highlight the significance of background removal techniques in achieving precise people counting using UWB radar sensors. Additionally, \cite{MovementThroughWall:UWBradar} employs the \ac{amf} to eliminate background interference. The \ac{amf}, recognized for its ability to mitigate noise and refine signals through the substitution of individual values with the median value of adjacent values, operates as a non-linear filter. Due to its efficacy and expeditious processing capabilities, it proves advantageous in scenarios demanding instantaneous processing. The research article applies the \ac{amf} to UWB radar imagery, amplifying the signal-to-noise ratio and streamlining human motion detection.

\subsubsection{Anomaly Detection}
The process of identifying unexpected incidents in datasets is known as anomaly detection. Anomaly identification is crucial in data analysis to find data points deviating from expected or standard patterns, which can e.g. represent the presence of new persons in a room.

The field of anomaly detection in UWB radar has been explored in three papers \cite{anomalPDF:UWBradar, Bigdata:Anomal:UWBradar, KPCA:PCA:Anomaly}. In \cite{anomalPDF:UWBradar}, the authors propose a \ac{kld} based anomaly detection method. This method effectively identifies anomalies by measuring the distance between observed data and a normal distribution using \ac{kld}. The paper demonstrates the application of this approach in through-wall human detection using UWB radar. The \ac{kld}-based method proves to be highly accurate and capable of detecting humans in various scenarios involving different wall materials.

Another study \cite{Bigdata:Anomal:UWBradar} introduces an anomaly detection method combining \ac{wpt} and \ac{spc}. The UWB radar system generates human body images through walls, which are then analyzed using \ac{wpt} to obtain wavelet coefficients. \ac{spc} is subsequently employed to monitor these coefficients for anomaly detection. The results reveal the effectiveness of the \ac{wpt} and \ac{spc}-based methods in accurately detecting humans behind walls, even in diverse scenarios with different wall materials. This approach benefits from the compact representation of signals through wavelet coefficients and the statistical process control techniques for identifying coefficient anomalies.

\cite{KPCA:PCA:Anomaly} focuses on a different approach, employing \ac{kpca} and \ac{pca} for anomaly detection in UWB radar-based through-wall human detection. \ac{kpca} maps the data into a higher dimensional space using a kernel function. At the same time, \ac{pca} reduces the dimensionality of the data by projecting it onto a lower-dimensional subspace. Both methods effectively identify anomalies in the generated human body images, demonstrating high accuracy in detecting humans across various scenarios. The utilization of \ac{kpca} and \ac{pca} enables enhanced data separation and visualization, leading to successful through-wall human detection using UWB radar.

\subsection{Trends and Observations}

This section provides an overview of the trends in the studies and observations for presence detection. It is notable that mono-static radar setups are commonly preferred in simple applications, such as presence detection, within scenarios where the ranges are smaller than 10 meters. However, studies highlight a shift towards multi-static configurations in response to the demand for higher resolution and precision in more challenging and larger environments. Currently, there is a lack of studies that critically analyze the cost versus benefits of deploying multi-static radar configurations.

The radar technology's evolution in bandwidth and central frequency selection provides insights but needs more depth in evaluating real-world applications. Bandwidths typically range from 500 MHz to 5 GHz to meet the demands of high-resolution tasks in challenging environments, although the necessity of such wide ranges for simpler scenarios is questionable. This tendency towards larger bandwidths may not always result in proportional benefits, raising concerns about efficiency and cost-effectiveness. On the other hand, while adopting lower central frequencies is better for their superior object penetration and extended range, suitable for detecting objects or individuals behind obstacles and over long distances, the papers do not address the potential resolution loss due to larger wavelengths. A refined assessment of these trade-offs would enrich the understanding of strategic technological choices in radar systems.

Most papers focus on obtaining high-accuracy predictions in laboratory conditions. As such, trade-offs such as accuracy versus complexity or computational demands are still open research questions. Similarly, a wide range of real-life technical limitations remain unexplored domains. Examples of unexplored real-life aspects include:
\begin{itemize}
    \item How to address signal interference management and what is the impact of interference on the accuracy of presence detection algorithms?
    \item How to optimally synchronize radios in multi-static setup, and what is the impact of synchronization errors on the accuracy?
    \item UWB radar systems usually select high sample rates to meet the Nyquist requirement, mainly when operating with broad bandwidths. However, studies investigating the trade-offs between sampling rate, accuracy and energy consumption are still lacking. 
    \item Similarly, the relation between the speed of moving targets and the required sampling rate is as of yet unexplored.
\end{itemize} 
Exploring the implications of these challenges for real-world application feasibility and performance could enrich the insights into the performance of such systems.

Finally, with \ac{ml} demonstrating superior performance compared to traditional algorithms, substantial research is now centered on identifying the most effective algorithms, often overlooking factors such as generalizability, trustworthiness, and the cost of data labeling.

\section{Device free localization}\label{section:DFL}

Conventional localization systems require the object or person to be tracked to be carrying some kind of tag which communicate with a number of fixed anchors. Example systems are \ac{gps} system commonly used in car navigation and UWB indoor localization. Some situations do not allow this. A possible alternative would be to use \ac{uwb} \ac{dfl}. Similar accuracies can be achieved to tag-based UWB localization in \ac{los} situations. The technologie allows for precise estimation of the distance and the angle of arrival of a signal. Eliminating the need for a tag could enable new application such as security via intelligent intrusion detection and individualized healthcare for elderly people. This section looks into the techniques and procedures used in \ac{dfl} utilizing \ac{uwb} radar, drawing on recent research trends and notable insights.

\begin{table*}[htbp]
\caption{Overview of scientific papers utilizing UWB radar for Device-Free Localization}
\label{table:DFL}
\fontsize{8pt}{8.2pt}\selectfont
\renewcommand{\arraystretch}{1}
\centering
\begin{tabular}{m{0.5cm}|m{0.5cm}  m{1.4cm} m{1.9cm} m{1.4cm} m{0.9cm}m{1.2cm} m{1.2cm}  m{3.4cm}} 
 \hline
 Paper & year  & Sub-type  & Accuracy & System & Devices & Bandwidth& Central F &  Chip\\ [0.5ex] 
 \hline\hline

\cite{PedesTrack:UWB} & 2022 & Tracking & MAE: 0.156m (50\%) 0.272m (90\%) & multi-static & 4 & 0.9GHz & 4GHz & Decawave DWM1000 \\

\hline

\cite{PassHumanTrackCOTS:UWBRadar} & 2022 & Tracking & RMSE: 26.12 – 29.24 cm  & multi-static & 4 & 0.9GHz & 4GHz &  Decawave DW1000 \\

\hline

\cite{DFL-CIR:UWB} & 2022 & Tracking & 0.64 m (50\%) 1.98 m (80\%)& multi-static & 3 & 0.5GHz & 4.5GHz &Decawave EVK1000  \\
\hline

\cite{MovementThroughWall:UWBradar} & 2012  & Tracking & - & mono-static  & 1 & 3GHz & 2GHz &SIR3000 radar system \\

\hline

\cite{MHTC:peopletracking:UWBradar} & 2010 & Tracking &-& mono-static &1& 2GHz & 4GHz & Humatics PulseOn P210\\
\hline

\cite{MultipersonTracking:PHDFilters:UWBradar} & 2014 & Tracking & AE: single person: $<$ 0.25m & mono-static &1& 2GHz & 4.3GHz & Humatics PulseOn P410\\
\hline

\cite{LocationTracking:movingtarget:UWBradar} & 2015 & Tracking & RMSE: 0.2260 - 0.2478 m & -static &2& 2.4GHz & 4.3GHz & Novelda Xethru X1 \\
\hline

\cite{One-Transmitter–Two-ReceiverUWBRadar} & 2022 & Tracking & MAE: 0.25 - 0.47m& -static &2&5.9 GHz&3.05GHz&Custom\\
\hline

\cite{PassiveLocalTrack:multi-static:UWBradar} & 2022 & Tracking & $<$ 0.60m (\%50), $<$ 1.20m(\%95) & multi-static &4& 0.9GHz & 4GHz & Decawave DWM1000\\
\hline

\cite{MA-RTI} & 2023 & Tracking & $<$ 1.00m (\%50), $<$ 1.80m(\%80) & multi-static & 4 & 500 MHz & 4.5 GHz & Decawave DW1000\\
\hline

\cite{gentner2023ranging} & 2023 & Tracking & MAE: 0.4m & multi-static & 9 & 500 MHz & 3.9 GHz & Decawave DW1000 \\
\hline

\cite{nomura2023device} & 2023 & Tracking & $<$ 1.27m & bi-static & 2 & 500 Mhz & 6.5/8 GHz & Decawave DWM3120\\
\hline

\cite{ACIR} & 2016 & Distance & -&  bi-static & 2  & 0.5GHz & 3.5GHz & Decawave DWM1000\\
\hline

\cite{MultObjLocalVital:UwbMimoRadar} & 2020 & Distance & rmse: 0.03 0.018& multi-static & 2 & 6.3GHz & 3.85GHz& Novelda XeThru X1 \\
\hline

\cite{DFpersonRange:UWBNet} & 2014 & Distance & RMSE: 0.12 - 1.8m & bi-static & 2 & 4GHz & 2.5GHz & Humatics PulseOn P400\\
\hline

\cite{Multi-StaticUWB:RadarNetwork} & 2020 & Distance & RMSE: 0.33 - 0.42m & multi-static & 3& 0.9GHz & 3.99GHz & Decawave DWM1000 \\
\hline

\cite{MAMPI-UWB:DFL:MamPhase} & 2020 & Distance &- & bi-static &2 & 0.499GHz & 4.49GHz & Decawave DWM1000 \\
\hline

\cite{ClutterSuppression:behindwall} & 2016 & Distance & $<$ 0.60m & mono-static &1 & 2.2GHz & 4.3GHz & Humatics PulseOn P410\\
\hline

\cite{ObstacleDetectionEntropy:UWBradar}& 2021 &  Car obstacles& F1: 68.68\% & mono-static & 1 & 1GHz & 4GHz &  UMAIN HST-D3\\
\hline 

\cite{LearningAOA:DFL} & 2022 & Angle & 3.5-7 degree error& bi-static & 2 & 0.5GHz & 3.49GHz & Decawave DWM1000\\
\hline

\cite{CELIDON:3D:loc} & 2020 & Angle & 20 degrees & multistatic & 4 & 0.499GHz & 6.48GHz & Decawave DWM1000\\

 \hline\hline
\end{tabular}
\end{table*}

As discussed in earlier sections, multiple reasons support using UWB radar for \ac{dfl} applications. The large bandwidth of UWB provides high-resolution distance measurements, which are critical for accurate positioning. The capacity of the UWB radar to penetrate through barriers and its resistance to multipath interference provides reliable performance in various demanding conditions.
\subsection{Overview of scientific publications focusing on Device free localization}
\textbf{Distance estimation}
The authors of \cite{DFpersonRange:UWBNet} present a strategy for identifying and estimating the distance of stationary individuals in indoor environments using UWB technology. The method used in the paper utilizes the inherent properties of the human presence and UWB signals, without needing a pre-existing training database, to enhance detection probabilities in indoor environments. In a different approach, \cite{MAMPI-UWB:DFL:MamPhase} proposes a \ac{dfl} system to determine a person's position within a target area. The system is designed to perform efficiently, with a comparable error probability to conventional systems, while only requiring two sensor nodes. Meanwhile, \cite{ClutterSuppression:behindwall} explores the use of UWB radar sensing for detecting static humans even when they are situated behind walls. This research showcases the potential for UWB technology in applications where visual detection is obstructed or impossible.

\textbf{Angle estimation}
\ac{aoa} estimation in UWB systems is an area that has gained significant interest due to its potential for precise localization. A \ac{ml}-based method for \ac{aoa} estimation was introduced in \cite{LearningAOA:DFL}, a classifier was used to identify multipath components in UWB systems and a \ac{pdoa} technique through a multilayer perceptron to reduce \ac{aoa} estimation errors. This methodology, validated using UWB array datasets, outperforms conventional approaches in \ac{aoa} estimation and \ac{dfl} performance. On the other hand, paper \cite{CELIDON:3D:loc} describes a 3D ad hoc localization system for assisting firefighters in low visibility situations by integrating UWB radar, two-way ranging range finding and \ac{pdoa} technique for \ac{aoa} estimation. The system, which can be conveniently incorporated into firefighter equipment, provides direction estimation with an accuracy of 20 degrees and position finding with an accuracy of 30 cm. It also highlights the potential of \ac{ml} for system reliability improvement in complex environments.

\textbf{Human Tracking}
Human tracking can have significant impact in various fields, including security and surveillance, sports analysis, and health monitoring. In this regard, several research studies have been conducted to improve the performance of UWB radar-based human tracking systems. For instance, \cite{PedesTrack:UWB} proposed a \ac{vats} mapping algorithm to relieve background interference and a particle filter algorithm to track the position likelihood changing. \cite{IdentificationTracking:UWBFusion} presented a real-time tracking system that leveraged the complementary of the uncertainty regions between UWB transceivers and cameras and demonstrated improvement in tracking accuracy through sensor fusion. \cite{MHTC:peopletracking:UWBradar} developed a multiple hypothesis tracking framework for UWB radar-based multiple human target tracking, addressing the complex observation clustering and data association problems using Bayesian inference. Finally, \cite{PassiveLocalTrack:multi-static:UWBradar} proposed using \ac{cir} for passive indoor localization of a moving target, which stands out against the background signal and does not require labor-intensive fingerprint updates. These studies show the potential of UWB radar-based human tracking and provide novel techniques to enhance the accuracy and robustness of such systems.

\subsection{Device free localization Techniques}
Device-free localization has mainly been tackled by breaking down the process into separate components. Three distinct components can be found, namely, background \& clutter removal, identification of target(s), locating and tracking of the targets. This section delves deeper into the methods utilized to execute these specific components. It is important to acknowledge that similarities will be present between this section and other sections that discuss applications.

\subsubsection{clutter subtraction}
Much like section \ref{section:presence}, \ac{dfl} also faces the challenge of extracting the multipath effects resulting from individuals from the noise and clutter that permeates the signal acquired from UWB devices.
The authors of \cite{One-Transmitter–Two-ReceiverUWBRadar} employ exponential averaging as a method to eliminate clutter from the signal. Unlike traditional averaging over all samples within a specific time window, exponential averaging assigns a higher weight to the most recent sample obtained. This approach offers several advantages, including quicker detection of environmental changes caused by people's movements compared to regular averaging.
\cite{LocationTracking:movingtarget:UWBradar} performed an analysis to determine the most effective clutter removal technique. Three different approaches were investigated: exponential averaging, \ac{svd}, and a specialized Kalman Filter designed for clutter removal. The results revealed that the Kalman Filter exhibited superior performance, with an average \ac{rmse} of 0.1029, compared to 0.2031 and 0.1342 for exponential averaging and \ac{svd}, respectively. It is important to note that noise poses a significant limitation for \ac{dfl}, primarily because humans absorb part of the radar signal, resulting in a weak reflection.

The studies conducted by \cite{Multi-StaticUWB:RadarNetwork}, \cite{ACIR}, and \cite{PassHumanTrackCOTS:UWBRadar} make use of the DW1000 chip developed by Qorvo. The DW1000 chip allows to get a \ac{cir} with a resolution of approximately 1ns per sample. An averaging and interpolation technique was used to effectively reduce noise while simultaneously enhancing the resolution. Specifically, it allows for a resolution up to 64 times greater than the normal resolution of 1 ns. The technique was initially proposed in \cite{ACIR}, where \ac{cir}s are grouped based on the timestamp of the received line-of-sight signal. These grouped bins are then averaged, and any missing bins are interpolated. Finally, the bins are recombined to obtain the final \ac{acir}.
\cite{PassiveLocalTrack:multi-static:UWBradar} presents an approach to address the clutter subtraction problem. They identify two types of clutter: noise-related clutter arising from measurement noise, and clutter caused by multipath reflections from objects, walls, etc., which tends to be static. The latter can be effectively removed using a background subtraction method such as exponential averaging, as seen in \cite{One-Transmitter–Two-ReceiverUWBRadar, Multi-StaticUWB:RadarNetwork}. However, the noise-induced clutter cannot be eliminated through this method alone. To tackle this issue, the authors of \cite{PassiveLocalTrack:multi-static:UWBradar} employ a bandpass filter and a technique known as \ac{rpca} after the background subtraction. In \ac{rpca}, the input matrix X (consisting of radar-scan data) is decomposed into the sum of two matrices, with one of them containing the necessary information for achieving the detection objective. This combination of techniques helps in effectively mitigating both types of clutter.

\subsubsection{target identification}
Once the clutter and noise, including background noise, have been removed from the signal, the subsequent step involves detecting the person or object within the signal. Despite background removal, it is still possible for peaks to emerge that surpass the signal emitted by the person being tracked. The authors of \cite{ObstacleDetectionEntropy:UWBradar} propose an solution to effectively detect a target despite the aforementioned problem by utilizing the Shannon's Entropy to detect regions of interest. The following algorithm was proposed: A sliding window of length \begin{math}W_{slide}\end{math} is chosen. The sliding window iterates over the signal and calculate the Shannon's Entropy. Afterwards the peaks are found and held against a threshold.
A basic target identification is chosen by the authors of \cite{Multi-StaticUWB:RadarNetwork}. similarly to the leading edge detection for first path recognition in a \ac{cir}, the paper investigates the possibility of using a leading edge detector on the residual signal after background removal and smoothing the signal. \cite{MAMPI-UWB:DFL:MamPhase} employed a fingerprinting method to localize an object or person. A detailed description of the feature vector composition is given and they used the following formula to calculate the position:

\begin{equation*}
    d_{\ell_1}(r_{p_0}, r_{p_r}) = \sum_{l=1}^{L_s} |r_{p_0,l}- r_{p_r,l}|
\end{equation*}

$r_{p_0}$ is the observed feature vector these are compared to  all reference feature vectors $r_{p_r}$. The Manhattan distance ($\ell_1$-norm) is calculated from all features and the one having the smallest distance is chosen as the position.
A Commonly used method is the \ac{cfar} detector. It is used in \cite{ObstacleDetectionEntropy:UWBradar,One-Transmitter–Two-ReceiverUWBRadar, PassiveLocalTrack:multi-static:UWBradar}. The Cell Averaging \ac{cfar} detector works as follows: \ac{cfar} employs a dynamic window technique that traverses through all samples in the squared signal. The algorithm is applied to each specific slow time instant (scan). A threshold is determined by estimating the noise floor level in the vicinity of the \ac{cut}. A target is identified within the \ac{cut} when its momentary power level surpasses a predetermined threshold. Another approach is taken by \cite{MHTC:peopletracking:UWBradar, LocationTracking:movingtarget:UWBradar}. The authors used the CLEAN algorithm to detect a person. The CLEAN algorithm takes a waveform shape template and a detection threshold as input. It initializes a residual waveform and iteratively detects signals by computing cross-correlation. The algorithm estimates time of arrival and amplitude, updating the residual waveform in each iteration until the amplitude falls below the specified threshold.
\subsubsection{target localization \& tracking}
Once a target is detected, a problem is obtained which is identical to the one encountered in more traditional (\ac{uwb}) localization systems, namely, how to localize and track a target. Different approaches exist to tackle this problem. \cite{LocationTracking:movingtarget:UWBradar,PassiveLocalTrack:multi-static:UWBradar} used an \ac{ekf}. The Kalman Filter is a recursive mathematical algorithm used for estimating the state of a dynamic system from a series of noisy measurements. The processes are modeled as linear processes with gaussian noise. The algorithm consists of two steps, a prediction step where a prediction is made based on the current state of the system. The second step (update step) consist of correcting this prediction by means of incorporating the measured data. While the traditional Kalman Filter assumes linear relationships between the system's state variables, the \ac{ekf} accommodates non-linearities by linearizing the system dynamics at each time step. This makes the \ac{ekf} applicable to a broader range of real-world problems where the underlying system dynamics may not follow a strictly linear model. The authors of \cite{LocationTracking:movingtarget:UWBradar} recognized that an \ac{ekf} is sometimes difficult to tune and implement, moreover, an \ac{ekf} is only reliable for systems where the first-order linearization is almost linear. The authors recognized this problem and therefore compared the \ac{ekf} to the \ac{ukf}. In an \ac{ukf} the state distribution is again represented by a Gaussian random variables. \ac{ukf} uses an \ac{ut}, in which a set of statistical points (sigma points) that propagate through nonlinear functions are used to parameterize the mean and covariance of the state distribution. These sigma points completely capture the true mean and covariance of the random variables. When these sigma points are propagated through the non-linear system, they accurately capture the posterior mean and covariance up to the 3rd order through the use of Taylor series expansion, applicable to any non-linearity \cite{UKF}. Particle Filter stands out as a widely recognized technique for both localization and tracking of a target. \cite{PassHumanTrackCOTS:UWBRadar, PedesTrack:UWB} utilized this algorithm in to track the target. A Particle Filter approximates the Bayesian posterior \ac{pdf} using a collection of randomly selected and weighted samples. Each state vector sample is termed a particle. Random samples (particles) derived from a distribution are progressed through the system equations to produce prior particles. These prior particles are then combined with measurement information to form the posterior distribution. Ensuring a sufficiently large number of particles ensures almost certain convergence to the true probability distribution function.\cite{RPMDoppler-based:UWBradar:TWR} investigated an adaptation to the \ac{gm-phd} filter named an EK-PHD. The EK-PHD is based on the idea of an Extended Kalman Filter. Opposite to the previous tracking algorithms, the \ac{phd} filter is a framework that deals with the estimation of the number and states of \textbf{multiple objects} in a dynamic environment. The PHD represents the probability density function of the number of objects and their states in the state space. Similarly to all other algorithms, the EK-PHD assumes that the target follows a linear Gaussian state and measurement model. 
\subsubsection{Exceptions}
Not all papers follows the regime state above. Some papers take an alternative route to directly determine the distance to or the position of a person. In contrast to the majority of papers, \cite{PassHumanTrackCOTS:UWBRadar} takes a distinct approach. Instead of employing a statistical model for clutter removal, they utilize a neural network. The network is trained using the background \ac{cir} and the dynamic \ac{cir} as inputs. Notably, the authors chose to use the \ac{acir} technique described previously, and they also included the standard deviation of both the background and dynamic \ac{cir}. This additional information is utilized by the network to effectively determine the \ac{tof} towards the object. Afterwards a Particle Filter is used for tracking. The paper by \cite{PedesTrack:UWB} proceeds by conducting a comparison among CNN-based, variance-based leading-edge detection, and \ac{vats} mapping. According to \cite{PedesTrack:UWB}, The variance proves to be a more resilient metric than the CIR for passive tracking of humans. The main idea is to map the variance differences between the background and dynamic scenarios from the delay domain to the spatial domain. Afterwards a particle filter algorithm is used to track the position likelihood and avoid ambiguity. The authors concluded that the \ac{vats} mapping had the best performance.

\subsection{Trends and Observations}

In this section, similar to section\ref{section:presence}, a brief summary is presented that highlights the current advancements and trends observed in the domain of \ac{uwb} \ac{dfl} systems.

Currently, the majority of papers in the field of \ac{uwb} device-free localization employ a multi-static radar setup. This choice is primarily driven by the fact that localization is typically not achievable with a mono-static configuration, unless the device has the capability to determine the angle of arrival of the signals and their multi-path reflections which is the case in papers \cite{LearningAOA:DFL, CELIDON:3D:loc}. It should be noted that most chips currently available do not support \ac{aoa} determination for incoming signals, as will be further discussed in section \ref{section:chips}.

Most papers utilize four distinct anchors for their UWB device-free localization experiments. There are two notable exceptions worth mentioning. The first one is presented in \cite{One-Transmitter–Two-ReceiverUWBRadar}, where the authors attempt to approximate a mono-static setup by placing three UWB devices in a straight line, spaced only 43 cm apart. The outer two nodes operate in \ac{rx} mode, while the center node operates in \ac{tx} mode. However, due to the poor \ac{dop}, the results obtained in this setup are not great. The second noteworthy paper is \cite{LearningAOA:DFL}, where a novel learning-based technique for \ac{aoa} estimation is employed to localize a person. As mentioned earlier, a mono-static or bi-static setup is sufficient for person localization. With the emergence of chips capable of performing \ac{aoa} estimation, it is expected that more papers will focus on mono-static device-free localization. This shift would significantly reduce costs and complexity for system integrators. 

Regarding bandwidth, a diverse spectrum of bandwidths is employed, spanning from 500MHz to 6.3 GHz. Despite the utilization of various bandwidths, drawing conclusions about the impact of larger or smaller bandwidths on the overall system performance proves challenging.

Currently only 3 papers \cite{MHTC:peopletracking:UWBradar, MultipersonTracking:PHDFilters:UWBradar, nomura2023device} localize and track more than 1 object/person. Their algorithms allow to track an arbitrary amount of objects/persons. As the field progresses the expectation is that there will be a shift from single object/person tracking towards multi-object/person tracking.

While the majority of studies concentrate on human localization, it is crucial to acknowledge that device-free localization should not be restricted to this; it can also prove beneficial in various other scenarios. One such instance is monitoring the movements of animals within confined habitats. This approach enables researchers to observe animal behavior without causing any disturbance to the subjects.

In conclusion, device-free localization using UWB predominantly emphasizes multi-static systems. Future efforts could explore the implementation of this technology in a mono-static setup. The optimal bandwidth remains inconclusive; however, it is noteworthy that a considerable number of papers utilize Qorvo chips. This prevalence is likely attributed to the success of Qorvo UWB chips in indoor localization systems with active tags, contrasting with the limited use in presence detection activities.

\section{Activity recognition}\label{section:activity}

\begin{table*}[htbp]
\caption{Overview of scientific papers utilizing UWB radar for Activity Recognition (part I)} 
\label{table:Act1}
\fontsize{8pt}{8.2pt}\selectfont
\renewcommand{\arraystretch}{1}
\centering
\begin{tabular}{m{0.5cm}|m{0.5cm}  m{1.6cm} m{1.2cm} m{1.4cm} m{0.9cm}m{1.2cm} m{1.2cm}  m{3.4cm}} 
 \hline
 Paper & year  & Sub-type  & Accuracy & System & Devices & Bandwidth& Central F &  Chip\\ [0.5ex] 
 \hline\hline

\cite{SleepPoseNet}& 2020 & 5 sleep postures  & 73.7\% & mono-static & 1 & 1.4,1.5 GHz & 7.29GHz & Novelda Xethru X4M03 \\

\hline
 \cite{SleepPostureDualradar:UWBradar} & 2022 & 4 sleep postures & 71.3\%-93.8\% & mono-static & 2 & 1.4GHz & 7.29GHz & Novelda Xethru X4M03\\

\hline
 
\cite{FallIncidentCompDPmContactL:UWBradar} & 2020  & fall detection & 80\%-100\% & mono-static & 1 & 1.5GHz & 7.29GHz & Novelda Xethru X4M03\\

\hline

\cite{CNN-LSTM-Falldetection:UWBradar} & 2020  & fall detection & 90\% & mono-static & 3 &  1.5GHz & 7.29GHz  & Novelda XeThru X4 M200\\

\hline

\cite{FalldetectionAdapCHselectCNN:UWBradar} & 2023 & fall detection & 95.7\% & mono-static & 1 & -& - & Novelda NVA-R631\\

\hline
\cite{LSTM-RNNhomeFallDetection:UWBradar} & 2018  & fall detection & 89.8\% & mono-static & 1 & 1.5GHz & 7.29GHz & Novelda Xethru X4M03\\
\hline

\cite{ResidualNeuralSLFallincident:UWBradar} & 2019 & fall detection & 93.7\% & mono-static & 1 & 1.5GHz & 7.29GHz & Novelda Xethru X4M03\\
\hline

\cite{ElderlyCareIndoorMont:UWBradar} & 2021  & 3 activities  elderly care& 99\% & mono-static & 1 & 1.7GHz & 3.95GHz & Humatics PulseOn 440\\

\hline

\cite{Qimeng2021} & 2023  & 5 activities  elderly care& 90.6\% & mono-static & 1 & - & - & Novelda NVA6100\\

\hline

\cite{RandomForestFilterDatabin:UWBradar} & 2022  & 15 activities (ADL) & 59.2\% & mono-static & 3 & 1.5GHz & 7.29GHz & Novelda XeThru X4 M200\\

\hline

\cite{Lafontaine2023} & 2023 & 14 activities (ADL) & 69.9\% & mono-static & 3 & - & - & Novelda XeThru X4 M200\\

\hline

\cite{EfficientNetLSTMActRec:UWBinput} & 2022  & 15 activities (ADL) & 73\% & mono-static & 3 & 1.5GHz & 7.29GHz & Novelda Xethru X4M200\\

\hline

\cite{ActivityDailyLifeDNN:UWBradar} & 2021 & 15 activities (ADL) & 90\% & mono-static & 3 & 1.5GHz & 7.29GHz & Novelda Xethru X4M200\\

\hline

\cite{Maitre2023}& 2023& 15 activities (ADL)& 64\% &mono-static & 3 & - & - &   Novelda Xethru X4M200\\

\hline

\cite{RadarDataFusion:ActMontMov} & 2021  &12 activities& 99.9\%  & mono-static & 2 &1.5GHz&7.29GHz& Novelda Xethru X4M03\\
\hline

\cite{HumanMotNoncontact:UWBRadar} & 2018  & 12 activities & 94.4\% 95.3\% & mono-static & 1 & 8.7GHz & - & Novleda NVA-6100  \\
\hline

\cite{CNN-RNNContHumanActRec:DistRadar} & 2022 & 9 activities & 90.8\% & mono-static & 5 & 2.2GHz & 4.3GHz & Humatics PulsON P410\\

\hline

\cite{FeatureRepAnalysisActRec:UWBradar} & 2020 & 10 activities & 92.6\% & mono-static & 1 & - &  - & developed TWR\\
\hline

\cite{Human-Human_InterRec:UWBradar} & 2020 & 6 activities & 89.48\% & mono-static & 1 & 1.7GHz & 3.95GHz & Humatics PulseOn 440 MRM\\

\hline

\cite{DevicefreeActRec:UWBradios} & 2019  & 4 activities & 95.6\% & bi-static & 2 & 0.5GHz & 4GHz & Decawave DW1000\\

\hline

\cite{ContactlessDailyActivityMonitoring:UWBradar} & 2020 & 5 activities & 97.7\% & mono-static & 2 & 1.5GHz & 7.29GHz & Novelda Xethru X4M300\\
\hline

\cite{HomeEmergencyDetectDL:UWBradar} & 2020  & 5 activities & 94\% & mono-static & 1 & 1.5GHz & 7.29GHz & Novelda Xethru X4\\
\hline

\cite{SynchMotionRecTVRD:UWBradar} & 2019  & 8 activities & 96.8\% & mono-static & 1 &  1.4GHz & 7.29GHz & Novelda Xethru X4M03 \\

\hline

\cite{HactClassonBreathingPattern:UWBradar} & 2020 & 4 activities & 85.25\% & mono-static & 1 & 1.5GHz & 7.29GHz & Novelda XeThru X4 M200 \\

\hline

\cite{WaveletPacketHActRecSVM:UWBradar} & 2014  & 8 activities & 97.6\% & bi-static & 2 & 2GHz & 4.2GHz & Humatics PulseOn 400 RCM\\
\hline

\cite{MultScResidualHACTrec:MicroDoppSig}& 2019  & 5 activities  Identification  &  98.2\%  95.87\% & mono-static & 1 & 1.8GHz & 4GHz & Humatics PulseOn 440 \\

\hline

\cite{uwbWiFiPassivOpportActSense}& 2021 & 5 activities & 95.53\% & bi-static & 2 & 4GHz & 0.5GHz & Decawave EVK1000 \\

\hline

\cite{XLi2023} & 2023 & 20 activities & 91.01\% & mono-static & 1 & 1.7GHz & 4.3GHz & - \\

 \hline
 
 \cite{HumanActRec:UWBcommunication}& 2015  & 8 activities & 99.2\% & bi-static & 2 & 2GHz & 4GHz &  Humatics Pulson 400 RCM\\
\hline

\cite{multiclassMOTIOn:IRuwbRADAR} & 2020 &  12 activities & 98\% & mono-static & 1 & 1.4GHz & 7.29GHz & Novelda  XeThru X4M300\\

\hline

\cite{2D-VMD-carrierfreeMotion:UWBradar} & 2019 &  10 activities & - & mono-static & 1 & - & - &  SIR-20 ground penetration radar \\

\hline

\cite{PersonIdentificationMotion:UWBradar} & 2019 & 6 activities Identification   &  70\%  95.21\%  & mono-static & 1 & - & - & - \\

\hline

\cite{ApplicationHumanActClass:UWBradar} & 2012 & 8 activities & 85\% & bi-static & 2 & 3.2GHz & 4.7GHz & Timedomain P220 \\
\hline

\cite{ConvGatedRecurrentNNActivityRec:UWB} & 2020 & 6 activities & 90\% 95\% & mono-static & 1 & - & - & - \\

\hline

\cite{MotionClassificationRangeinfo:CNN} & 2017 &  7 activities & 95.24\% & mono-static & 1 &6.3GHz&3.85GHz& Novelda NVA6100 \\

\hline

\cite{Brishtel2023}   & 2023 & Driver monitoring & 71.3\% & mono-static & 1 & 7.29GHz & 7.25GHz - 10.2GHz & Novelda Xethru X4M02 \\

\hline
\cite{HandGestureRec:UWBRadar}& 2020 &  8 gestures  & 95\% & mono-static & 1 & 2GHz & 8.75GHz & Novelda Xethru X4M06 \\
\hline          

\cite{HandGestureVehicular:UWBRadar} & 2017 &  6 gestures  & 88\% & mono-static & 1 & 2.32GHz & 6.8GHz & Novelda  NVA-6201 \\

\hline          

\cite{SignLanGesCUmDis:UWBRadar} & 2021 &  10, 15 SL and 10 gesture  & 82-85\%  & mono-static & 1 & 1.5GHz & 7.29GHz &  Novelda XeThru X4M300 \\

\hline\hline
\end{tabular}
\end{table*}

\begin{table*}[t!]
\caption{Overview of scientific papers utilizing UWB radar for Activity Recognition (part II)} 
\label{table:act2}
\fontsize{8pt}{8.2pt}\selectfont
\renewcommand{\arraystretch}{1}
\centering
\begin{tabular}{m{0.5cm}|m{0.5cm}  m{1.6cm} m{1.2cm} m{1.4cm} m{0.9cm}m{1.2cm} m{1.2cm}  m{3.4cm}} 
 \hline
 Paper & year  & Sub-type  & Accuracy & System & Devices & Bandwidth& Central F &  Chip\\ [0.5ex] 
 \hline\hline

\cite{FineHumanMotionDetect:UWBradar} & 2020  & Finger wiggling & - & mono-static & 1 &1.5GHz& 7.29 &  Novelda X4 \\
\hline
\cite{DeepLearningHandgesture:UWBradar} & 2020  & 14 gesture & 96\% &  mono-static  & 1  & 1.5GHz & 7.29GHz & Novelda Xethru X4M03 \\

\hline

\cite{BLi2023}  & 2023& 10 gestures & 61.67\% & mono-static & 1&  1.5GHz & 7.29GHz & Novelda Xethru X4M300 \\

\hline

\cite{AirCOntWritingGesture:UWBimpRadar} & 2020  & 10 numbers,  8 letters +2 numbers&100\%  93\% - 100\% &  mono-static  & 4 & 1.5GHz & 8.748GHz &  Novelda Xethru X4\\

\hline

\cite{SimThermalRadarDeepGesture:UWB} & 2020  & 14 gesture & 95.54\% &  mono-static  & 1 &1.5GHz & 7.29GHz &  Novelda Xethru X4M03
\\

\hline

\cite{FingureCountingGestureCar:UWBradar} & 2019 & 5 gesture & 95\% &  mono-static  & 1 & 1.5GHz & 7.29GHz &  Novelda Xethru X4 \\

\hline

\cite{Park2023} & 2023 & 5 gestures & multiple models & mono-static & 1 & 2.5GHz & 7.3GHz &  Novelda NVA-R661\\

\hline

\cite{MicroGestureSensFus:PressureUWB} & 2019 &  7 gesture  6 transitions  & 91.43\% &  mono-static  & 1 &1.5GHz & 7.29GHz & Novelda Xethru X4M300\\

\hline

\cite{DLHAndGestureThermalFusion:UWBradar} & 2021 & 14 gesture & 99\% &  mono-static  & 1 &1.5GHz & 7.29GHz &  Novelda Xethru X4M03 \\

\hline

\cite{PointingGestureMenueBoard:UWBradar} & 2019  & Menue board & 96.75\% & mono-static & 4 & 1.4GHz & 8.748GHz & Novelda Xethru X4\\

\hline

\cite{MidAirGestureWritingDigit:UWBradar} & 2019 & 10 numbers& 99.7\% & mono-static & 3 & 1.4GHz & 7.29GHz & Novelda Xethru X4\\

\hline

\cite{HumanGestureRecogApproach:UWB} & 2021 &  4 gesture + no activity & 99.48\% & multi-static & 4 & 0.5GHz & 3GHz & Decawave DWM1000\\

\hline

\cite{CNNincreaseAccuracyGesture:UWBradar} & 2022  & 5 gestures & 90\% & mono-static & 1 & 2.5GHz & 7.25GHz & Novelda NVA-R661\\
\hline

\cite{OfflineDataAugDNNgesture:UWBradar} & 2022  & 12 gestures & 98\% & mono-static & 3 & 2GHz & 8.745GHz & Novelda Xethru X4\\

\hline

\cite{HumanIDentificationUnsFeatL:UWBradar} & 2018 &  8 Human Identification & 80.34\% & mono-static & 1 & 3GHz & 4.5GHz &Novelda NVA-R640\\

\hline

\cite{SmartHomeNonWearIDsensor:UWB} & 2017 & 8 Human Identification & 88\% & mono-static & 1 & 3GHz & 4.5GHz & Novelda NVA-R640\\

\hline\hline
\end{tabular}
\end{table*}

Using multiple sensors, activity recognition automatically identifies and understands human actions and behaviors. Many uses for this technology exist, including fall detection, sleep monitoring, emergency response, elderly and child care, tracking daily activities, and hand gesture recognition. Activity recognition is significant because it can increase people's quality of life by assisting them when necessary, identifying potential health hazards, and encouraging independence. For instance, fall detection systems can notify emergency services or caretakers if an elderly or disabled person falls, enabling immediate medical intervention.

UWB radar is an excellent choice for activity recognition since it can rapidly gather massive amounts of data while providing outstanding resolution and accuracy in recognizing human activities. As a result, fine-grained movement data, such as minor changes in body posture, and gait analyses can be captured by UWB radar. This makes it suitable for detecting a wide range of activities such as walking, running, sitting, standing, and even more complicated movements such as sports or dance.

Moreover, UWB radar can work in real-time and can be used to monitor people continuously without having them wear any sensors or gadgets. This is especially beneficial for the elderly or crippled, who may be unable to wear sensors or remember to charge them. Additionally, UWB radar has a low power consumption and does not interfere with other wireless technologies, making it suitable for application in various contexts such as healthcare institutions, homes, and public areas.

\subsection{Overview of scientific publications focusing on activity recognition}

Table~\ref{table:Act1} and Table~\ref{table:act2} present a comprehensive compilation of current scientific papers that center around activity recognition. These tables provide extensive information, including the publication year, use case, typical accuracy rates, and technical aspects such as the setup, bandwidth usage, center frequency, and the specific radio chip employed. The subsequent subsections will provide a classified explanation of the papers listed in the tables.

\subsubsection{Sleep monitoring}

UWB radar has emerged as a promising technology for sleep monitoring and activity recognition due to its ability to capture physiological signals and movements accurately without requiring contact with the subject. In the quest for non-invasive and economically viable alternatives to traditional sleep monitoring techniques, the paper denoted by \cite{SleepPoseNet} provides an innovative method that harnesses data from both temporal and spectral domains, thereby augmenting UWB signal detection procedures. The UWB radar device was mounted onto the wall, positioned 0.8 meters above the bed in this research. The radar was directed at an estimated pitch angle of 45 degrees, oriented downwards towards the bed. The study was predicated on two distinct datasets from the observations of 38 volunteers. The first dataset comprised four \ac{spt}s classes. In comparison, the second dataset encompassed five categories, including the 4 \ac{spt}s mentioned earlier and additional background data. Their methodologies achieved an average accuracy of 73.7\%.

In contrast, \cite{SleepPostureDualradar:UWBradar} compares the performance of deep learning and traditional \ac{ml} approaches for feature extraction and classification in a dual radar system for sleep posture recognition. Two radars were utilized in the study involving 18 participants and four different postures to obtain top and side views of the bed. The authors conclude that the dual radar system outperforms single radar systems. Together, these papers highlight the potential of UWB radar in sleep monitoring and activity recognition and provide insights into the optimal approaches for feature extraction and classification using UWB radar data.

\subsubsection{Fall detection}
Several scientific publications, including \cite{FallIncidentCompDPmContactL:UWBradar, CNN-LSTM-Falldetection:UWBradar, FalldetectionAdapCHselectCNN:UWBradar, LSTM-RNNhomeFallDetection:UWBradar, ResidualNeuralSLFallincident:UWBradar}, have leveraged UWB radars for the detection of falls. A significant majority of these referenced works have reported accuracies exceeding 90\%. These research undertakings have primarily utilized a mono-static system featuring a bandwidth of 1.5GHz and a center frequency of 7.29GHz.

 In \cite{FallIncidentCompDPmContactL:UWBradar}, a fall detection method is presented employing compressed features of UWB radar signals obtained through deterministic row and column sense. By performing time-frequency analysis on the radar time series and projecting the resulting spectrogram onto a binary image representation, the proposed method enables constant monitoring of vulnerable older adults and timely detection of fall incidents. On the other hand, in the paper \cite{CNN-LSTM-Falldetection:UWBradar}, a technique for detecting falls in an apartment is introduced. This method uses three UWB radars and a neural network structure. The research included ten male participants placed in different spots in the apartment. The simulated falls were divided into two stages: first, the participants mimicked a fall lasting 5 to 10 seconds, then stayed still on the ground with minimal movement in the second stage. Ultimately, in the paper \cite{FalldetectionAdapCHselectCNN:UWBradar}, an adaptive channel selection algorithm is proposed to differentiate activity signals from background noise using UWB radar effectively. The study involved nine male volunteers who simulated three fall actions: standing to fall, bowing to fall, and squatting to fall. The proposed method achieved an accuracy of approximately 95.7\% in detecting and recognizing falls. These findings underscore the efficacy of UWB radar in fall detection and activity recognition and emphasize the potential of this technology for real-world applications.

\subsubsection{Elderly care}
UWB radar for elderly care has been explored in recent research with promising results. In particular, paper \cite{ElderlyCareIndoorMont:UWBradar} focuses on using UWB signals to recognize specific movements and postures of elderly individuals without requiring camera-type monitoring. The authors demonstrate that UWB signals can remotely detect small chest movements, such as coughing. Additionally, statistical data analysis can be used to recognize a person's posture in a steady situation. These findings suggest that UWB radar can improve elderly care by enabling non-intrusive monitoring of individuals' movements and health status.

\subsubsection{Human identification}
Human identification is crucial in various applications, including smart homes, security systems, and healthcare. Two papers have addressed the problem of human identification using UWB radar. 

The research in \cite{HumanIDentificationUnsFeatL:UWBradar} employed a UWB system composed of a transmitter and receiver mounted on top of a door frame to collect UWB data within an indoor setting. The study was conducted with eight participants who exhibited a range of body shapes and heights. The participants were asked to walk through the door in various manners. In this investigation, they developed a method for identifying individuals in multi-residential smart home scenarios with an accuracy of 80.34\%. On the other hand, \cite{SmartHomeNonWearIDsensor:UWB} discusses using the same setup to present two scenarios: distinguishing between two individuals and between more than two individuals. The system generates a unique UWB signature for each individual using the proposed \ac{roi} extraction approach based on their body figure and walking gait. These papers demonstrate the potential of UWB radar; however, the more significant number of people, faster-moving targets, and many other challenges can be considered to increase the robustness and reliability of these solutions.

\subsubsection{\ac{adl}}
\ac{adl}s consist of critical tasks routinely performed by individuals in their everyday lives, such as eating, drinking, sleeping, walking, cooking, and many other daily activities. The studies involved in \cite{RandomForestFilterDatabin:UWBradar, EfficientNetLSTMActRec:UWBinput, ActivityDailyLifeDNN:UWBradar} included ten healthy adults under the age of 40 performing 15 diverse activities representing \ac{adl}s. Collecting this data aims to learn about the slight differences and adaptability of systems involved in humans' everyday tasks. However, due to their complexity, these scenarios present significant challenges in classification. The studies mentioned above employed these datasets to classify \ac{adl}s, resulting in an accuracy evolution ranging from 59\% to 90\%. 

These three papers on \ac{adl} using UWB radar employ different methodologies. The first paper \cite{RandomForestFilterDatabin:UWBradar} compares the performance of three deep learning models (\ac{lstm}, CNN-LSTM, ResNet), with the CNN-LSTM model achieving the highest accuracy. The second paper \cite{EfficientNetLSTMActRec:UWBinput} processes radar data independently, applies filters, extracts features, reduces dimensionality, and tests the approach using the Random Forest algorithm and a leave-one-subject-out strategy. The third paper \cite{ActivityDailyLifeDNN:UWBradar} introduces a deep learning model combining \ac{lstm} and a customized EfficientNet model with transfer learning, data fusion, minimalist preprocessing, and training for activity and movement recognition using data from three UWB radars. It can be extremely challenging to generalize these solutions, making it an open area of research for future work.

\subsubsection{classification of activities}
The use of UWB radar for activity classification focuses on detecting and studying specific signal features related to diverse human movements. Such movements may encompass actions like bowing, falling, jumping, sitting, standing, turning, crawling, jogging, and walking. Individual actions are systematically identified, monitored, and classified, enabling detailed observation of human behavior in various scenarios, typically examining between 4 and 12 different types of movements.

The study \cite{HumanMotNoncontact:UWBRadar} investigates the classification of twelve typical human motions, divided into in-situ motions (occurring in place) and non-in-situ motions (covering a more comprehensive range with higher velocity). Experiments were conducted with thirteen volunteers of varied ages, weights, and heights performing each movement five times in an indoor laboratory. Using a processing scheme composed of four submodules, the motions were pre-screened, features were extracted based on their unique characteristics, and \ac{ml} classifiers were employed to yield the final classification results. The research showcased nearly 100\% accuracy in distinguishing between in-situ and non-in-situ motions, with Subspace \ac{knn} and Bagged Trees algorithms excelling in their respective classifications with accuracies of 94.4\% and 95.3\%. On the other hand,\cite{SynchMotionRecTVRD:UWBradar} presents a motion recognition architecture tested on 1657 motion examples across eight classes collected from four experimental subjects. The architecture employs two feature extraction methods for comparative analysis: a\ac{cae} and \ac{pca}. The \ac{cae} method achieves a higher accuracy rate of 96.80\% but requires more computational resources compared to the \ac{pca} method. A key advantage of the proposed approach is its ability to detect motion with minimal time delays while maintaining high-performance levels, compared to traditional \ac{tdi} methods.

Most of the papers in this area classify the motions and activities, and most of the contributions are on \ac{ml} solutions and feature extraction methods. In the coming years, we will require generalizable and more complex techniques for practical solutions.

\subsubsection{Hand gesture recognition}
Hand gestures are a form of non-verbal communication that conveys meaning and emotions through the movements and positions of the hands. They can either stand alone or complement spoken language. Different types of gestures are used to communicate emotions or ideas. Expressive gestures, such as thumbs up, thumbs down, and fist pumps, represent feelings. Descriptive gestures, such as air quotes, finger-pointing, and hand waving, illustrate or emphasize spoken words. The third group consists of regulatory gestures that guide the conversation or indicate when one should commence or stop speaking. Such gestures involve raising a hand, placing a finger on the lips, or nodding the head. The fourth group includes subconscious movements commonly used to alleviate stress or nervousness. Examples of these adaptors are actions like fidgeting, rubbing one's hands, and nail-biting.

The frequency range from 3 GHz to 8.75 GHz is considered ideal for hand gesture recognition using UWB radar, as per the publications in TABLE~\ref{table:act2}. The corresponding bandwidth range for this frequency is found to be generally between 1.5 GHz and 2.5 GHz. This frequency range is highly suitable for hand gesture recognition, as it offers high accuracy and precision in detecting slight hand movements.

There are many papers in the area of gesture recognition; however, most of them only focus on classifying hand gestures without considering specific applications. The first paper, \cite{DeepLearningHandgesture:UWBradar}, focuses on recognizing 14 typical hand gestures using the right hand. These gestures are captured under various radar orientations, speeds, and distances to introduce variability. The unique features of each gesture are extracted using a \ac{cnn}, and the highest accuracy is achieved by combining 2D-CNN with \ac{lstm}, resulting in an accuracy of 96.15\%. The second paper, cited as \cite{DLHAndGestureThermalFusion:UWBradar}, extends the approach by fusing UWB radar with thermal sensors for gesture recognition. Similar to the first study, 14 gestures are used but are single or double versions of 7 base gestures. Two parallel CNN-LSTM network paths are employed, and their outputs are combined using two different deep-learning methods, achieving an accuracy of 99.26\%. When comparing the two papers, sensor fusion in the second paper significantly enhances the classification accuracy.

On the other hand, some investigations solely focus on specific applications utilizing hand gestures. As an example, \cite{MidAirGestureWritingDigit:UWBradar} explores using three \ac{ir-uwb} radar sensors to recognize mid-air handwriting of numbers. The configuration involved two sensors placed 71.3 cm apart on a horizontal axis and a third sensor positioned midway between them at a height of 38.4 cm, and the algorithm's ability to identify different hand orientations is tested. It uses \ac{cnn} to process \ac{cir}'s chosen areas and shows that this method outperforms using raw data. Additionally, \cite{PointingGestureMenueBoard:UWBradar} presents a virtual grid menu board utilizing four sensors in a 2D plane, and the objective is to substitute public displays that require touch. This approach classifies various pointing gestures using a multiclass \ac{svm} and a \ac{hog} for feature extraction. It shows the system's scalability and achieves about 96\% accuracy.

To sum up, UWB radar shows a lot of potential for creating communication interfaces that are more intuitive and accessible, thanks to its high-frequency precision. Future research could build on these foundations by exploring how this technology can be integrated into real-world scenarios, improving gesture recognition and adapting it for broader applications.

\subsection{Activity Recognition Techniques}

In the upcoming section, we will delve into the field of techniques used for activity recognition. We've grouped the content into three categories: preprocessing techniques, feature extraction methods, and classification algorithms. These categories cover the initial data preparation, critical attribute identification, and activity categorization. Each core area will be discussed, providing a comprehensive overview of the methodologies used in activity recognition.

\subsubsection{Preprocessing Techniques}

Preprocessing techniques are vital for enhancing the quality of data and ensuring efficient computational processing. This section will explore two fundamental preprocessing strategies: \ac{roi} extraction, background clutter, and noise removal.

\textbf{Region of Interest Extraction:} Regarding UWB radar, the \ac{roi} concept is crucial in optimizing computational resources. The relevant data for recognizing activities or gestures is generally limited to specific segments within the \ac{cir} samples, as depicted in \cite{HandGestureRec:UWBRadar}. Many scholarly articles suggest using predefined \ac{roi}s to reduce computational overhead and speed up real-time processing. Focusing solely on these \ac{roi}s within the \ac{cir} can extract relevant information more efficiently, reducing computational limitations and achieving faster, near-real-time results.

The utility of selecting an \ac{roi} has two advantages. First, it helps to concentrate only on the most informative parts of the \ac{cir}, which often need to be clarified with irrelevant or redundant data. Second, by focusing on a smaller \ac{roi}, the system can perform fewer calculations, making it better suited for real-time or near-real-time applications where quick decision-making is critical. However, it adds additional computation at the start, which might be intensive based on the application, which should be evaluated within the context.

\textbf{Background Clutter and Noise Removal:} In gesture and activity recognition using UWB radar, it is crucial to address the issue of clutter and noise to improve the quality of the signals obtained. Numerous methods have been proposed in the literature to overcome these challenges. As described in \cite{Lazaro2014}, background subtraction is commonly utilized to eliminate stationary or slow-moving clutter from radar signals. Loop-back filters, discussed in \cite{HandGestureRec:UWBRadar}, serve a similar purpose by eliminating certain clutter types that may interfere with the desired signals. Additionally, Kalman filtering techniques, as mentioned in \cite{AirCOntWritingGesture:UWBimpRadar}, are crucial in reducing random noise and making the data more reliable for further analysis.

Various filtering techniques have been explored to remove noise. In a recent study \cite{Maitre2023}, the efficacy of high-pass, low-pass, and band-pass filters was tested in activity recognition. The study found that the Band-pass Chebyshev type I filter was the most effective in reducing noise. In addition, Convolutional Autoencoders have been identified as a powerful method for removing clutter and noise. This approach employs deep learning to learn features that effectively clean up signals, as demonstrated in work \cite{Lafontaine2023}.

\subsubsection{Feature Extraction Techniques}
Feature extraction is critical in determining classification outcomes in activity and gesture recognition applications utilizing UWB radar. As demonstrated in studies such as \cite{ADL:randomforest:UWBradar, HandGestureVehicular:UWBRadar}, statistical features of time series and frequency data are essential for capturing variations in time and frequency data. Another effective method of reducing dimensionality is \ac{pca} which reduces computational costs while preserving essential data characteristics, as demonstrated in \cite{MachineLgesstureREC:UWBradar}.

In addition to traditional methods, specialized techniques such as Cumulative Energy Distribution are used for specific purposes like sign language recognition \cite{SignLanGesCUmDis:UWBRadar}. Other methods, such as \ac{stft}, are suitable for dynamically evolving gestures as they provide both time and frequency information \cite{HactClassonBreathingPattern:UWBradar}. Advanced methodologies like \ac{2d-vmd} and \ac{scdae} offer adaptivity and deep learning to extract finer features \cite{2D-VMD-carrierfreeMotion:UWBradar, Lafontaine2023}. The choice of optimal feature extraction method depends on the application's specific requirements, taking into account factors like computational complexity and classification accuracy. As technology and methodologies advance, evaluating and integrating these feature extraction techniques is crucial to meet the demands of emerging applications.

\subsubsection{Classification Algorithms}

In recent years, \ac{ml} algorithms have shown significant promise in activity and gesture recognition using UWB radar sensors. Most cited works train machine-learning models using raw radar signals, time-series data, or feature-engineered inputs. Some papers also integrate information from multiple sensors, enhancing the model's generalization ability across diverse scenarios. \ac{cnn} are widely used, as evidenced by numerous studies \cite{Li2022, Han2020, PersonIdentificationMotion:UWBradar, Park2023, MidAirGestureWritingDigit:UWBradar}. In addition to standard \ac{cnn} models, some works explore enhancements such as \ac{se} blocks \cite{Li2022} and Capsule Networks \cite{Maitre2023}. Furthermore, researchers have leveraged established architectures like ResNet-18 coupled with \ac{lstm} \cite{Brishtel2023} and GoogLeNet \cite{HandGestureRec:UWBRadar} to improve recognition performance.

\ac{rnn} and \ac{lstm} models are influential for capturing temporal dynamics \cite{LSTM-RNNhomeFallDetection:UWBradar}. There is a growing trend towards hybrid models combining \ac{cnn} with recurrent architectures, aiming to capture spatial and temporal features. These include CNN-LSTM \cite{CNN-LSTM-Falldetection:UWBradar}, \ac{scgrnn} \cite{ConvGatedRecurrentNNActivityRec:UWB}, \ac{cnn} combined with various recurrent units like \ac{rnn}, \ac{lstm}, and GRU \cite{CNN-RNNContHumanActRec:DistRadar}, and \ac{convlstm} \cite{Ma2020}. The choice of algorithm selected for recognizing activities or gestures should be driven by specific requirements, resulting in more robust models with a balanced approach.

\subsection{Trends and Observations}
UWB radar technology has made notable strides in activity and gesture recognition in recent years, driven by the growing need for accurate monitoring and interaction systems. Research in this field encompasses a wide range of activities, from simple tasks like fall detection to more complex activities, covering up to 15 different scenarios (\ac{adl}). This highlights the extensive potential for various applications in the future, from elderly and childcare to enhancing life quality and security monitoring.

Most of the studies reviewed utilize Novelda devices in a mono-static setup, typically featuring a bandwidth of 1.5GHz and central frequencies between 4 and 7.29 GHz. This demonstrates a consistent trend that lacks parameter variations to assess their impact on the accuracy of UWB radar activity and gesture recognition. Additionally, many studies fail to report their sampling rates, which are crucial for these applications. This underscores the need for extensive research on these parameters to establish the foundation for genuinely optimal values and setups.

\ac{ml} is the primary solution employed in the majority of the studies, with \ac{cnn} and temporal methods like \ac{lstm} being the most commonly used techniques. However, several challenging points are often overlooked in these studies:

\begin{itemize}
\item Studies often do not consider the generalization capability of the models, whether from human to human or from environment to environment. This aspect is crucial for the applicability of these applications in real-life scenarios. Although transferring the learned model is one potential solution, it always requires data from new humans or environments, which is not feasible in all scenarios.
\item These models require extensive amounts of data to extract features, which is extremely challenging to provide. In real applications, we will face a large number of scenarios, making it difficult to gather sufficient data for each one in the near future.
\item If the hardware cannot provide the required time granularity, it raises the question of whether the models can handle the lack of detailed information and still classify activities accurately. 
\end{itemize}
Possible solutions to these challenges include utilizing unsupervised methods for training, which can reduce the dependency on labeled data and help models generalize better across different humans and environments. Additionally, implementing context-aware automatic transfer learning could be beneficial. This approach leverages side information to adapt the learned models to new scenarios without the need for extensive additional data collection, thereby enhancing the practicality and applicability of these systems in real-world situations.

In addition to the previously mentioned challenges, studies should consider improving and testing real-life insights. This includes evaluating the maximum distance of observation and how the system integrates with realistic objects and scenarios. It is crucial to account for unrestricted human movements to ensure the models are robust and effective in real-world applications. Additionally, a multi-static setup can significantly increase the range and covered area. Still, it presents challenges such as radar synchronization and the importance of the received signal from each node. By addressing these factors, researchers can develop more comprehensive and practical solutions for activity and gesture recognition using UWB radar technology.

\section{Vital sign monitoring }\label{section:vitalsign}

\begin{table*}[t!]
\caption{Overview of scientific papers utilizing UWB radar for Vital Sign monitoring} 
\label{table:VS1}
\fontsize{8pt}{8.2pt}\selectfont
\renewcommand{\arraystretch}{1}
\centering
\begin{tabular}{m{0.5cm}|m{0.5cm}  m{1.4cm} m{1cm} m{1.4cm} m{0.9cm}m{1.2cm} m{1.2cm}  m{3cm}} 
 \hline
 Paper & year  & Sub-type  & Accuracy & System & Devices & Bandwidth& Central F &  Chip\\ [0.5ex] 
 \hline\hline
 
 \cite{RadarDataFusion:ActMontMov} & 2021 & heart rate& 99.9\%  92.33\%  & mono-static & 2 & 1.5 GHz& 7.29 GHz& Novelda Xethru X4M03\\
\hline  

\cite{VitalSignDetect:noContactUWB} & 2018  & respiration heart rate&  95\% 90\% & mono-static & 1 & 2.2GHz & 4.3GHz & PulseOn 440 MRM\\

\hline   

\cite{HearRateMonitor:nonContactUWB} & 2018 & heart rate &  96.2\% & mono-static & 1 & 1.5GHz & 8.7GHz & Novelda Xethru X4M06 \\

\hline

\cite{HumanAnimalDistinguish:UWBRadar} & 2019  &respiration heart rate  & - & mono-static & 1 &1.4GHz&7.29GHz& Novelda Xethru X4M02  \\

\hline          

\cite{VitalSignCompare:UWBandFMCW} & 2020  & respiration heart rate & - & mono-static & 1 &1.5GHz& 8.7GHz& Novelda Xethru X4 \\

\hline          

\cite{RespHeartAutocorrelation:UWBRadar} & 2018  & respiration heart rate & - & mono-static & 1 & 2.2GHz & 4.3GHZ & Humatics PulseOn 410 \\

\hline          

\cite{VitalSignMontsta/nonsta:UWBRadar} & 2017 & respiration heart rate & 98\% & mono-static & 1 &2.3GHz & 6.8GHz & Novelda  NVA-6201 \\

\hline 

\cite{ECGmonitoringCNN:UWBRadar} & 2016  & ECG & 88.89\% & mono-static & 1 & 2.5GHz & 6.8GHz & Novelda NVA-R661 \\

\hline          

\cite{ShortRangeVitalSign:UWBRadar} & 2016  & respiration heart rate & - & mono-static & 1 & 2.3GHz & 6.8GHz & Novelda  NVA-6201  \\

\hline          

\cite{BreathHeartUWBRadar:FVPIEFandEEMD} & 2018  & respiration heart rate & - & mono-static & 1 & 2.2GHz & 4.3GHz & -\\

\hline 

\cite{AnalysisVitalS:IR-UWB-Rad} & 2010  & respiration heart rate & -  & mono-static & 1 & 7.5GHz & 5 GHz &Custom hardware \\

\hline 

\cite{VitalTh-Back:IR-UWB-CouAnt} & 2018 & respiration heart rate & 100\% & mono-static & 1 & 2GHz & 3.8GHz & Novelda Xethru X2 \\

\hline 

\cite{ vitalMobileCarCrash:UWBRadar} &  2017  & car crash prevention & 99\% & mono-static & 1 & 2.3GHz & 6.8GHz & Novelda  NVA-6201 \\

\hline 

\cite{MultObjLocalVital:UwbMimoRadar} & 2020 & respiration heart rate & rmse: 0.03 0.018& multi-static & 2 & 6.3GHz & 3.85GHz& Novelda XeThru X1 \\

\hline

\cite{HYu2023}  & 2023 & Heart rate respiration & Error $<$ 0.04Hz& mono-static& 1 &- & 7.29GHz & Novelda Xethru X4M03\\

\hline 

\cite{CorrelationRecVital:UWBradar} & 2010 & respiration heart rate & 2cm & multi-static & 1 & 7.5GHz & 7GHz & independent hardware \\

\hline 
\cite{DogCatVitalSign:UwbRadar} & 2020  & Animal health & 98.46\% & mono-static & 1 & 1.4GHz & 7.29GHz & Novelda Xethru X4 M02\\
\hline 
\cite{HMLDVitalSign:UWBRadar} & 2020  & respiration heart rate & 95.05\% 94.94\% & mono-static & 1 & 1.4GHz & 7.29GHz & Novelda Xethru X4 M200 \\
\hline

\cite{CmosRemoteMonitoring:UWBRadar} & 2016  & respiration & $>$ 95\% &mono-static  & 1 & 1.3GHz & 7.25GHz & independent hardware\\
\hline 
\cite{SHAPAvitalSign:UWBradar} & 2014  & respiration heart rate & RMSnE $<$ 0.268 & mono-static & 1 & - & 4.1GHz & independent hardware\\
\hline 
\cite{VitalHubmultiUser:VitalUWBradar} & 2021 & respiration heart rate & 98.5\% & mono-static  & 1 & 1.5GHz & 8.75GHz & XeThru x4m03 \\
\hline 
\cite{DenoisingThWallVitalSign:UWBRadar} & 2018  & respiration heart rate & - & mono-static  &1 &0.4GHZ& - & independent hardware\\
\hline 
\cite{OptCentFreqVitalSign:monoUWBradar} & 2020  & respiration heart beat & - &  mono-static   & 1 & 2-5GHz & 3GHz &  independent hardware\\
\hline 
\cite{BreathingRateInVehicle:UWBradar} & 2018  & Respiration & 4bpm error & mono-static & 1 &  - &  - & Novelda Xethru X2M200 \\
\hline 
\cite{HumanPetDescrimination:EnsureVital:UWBradar} & 2021 & human, pet distinguishing & 90.38\% & mono-static & 1 & 1.4GHz & 7.29GHz & Novelda Xethru X4 M02\\
\hline 
\cite{PeriodEstimationVitalSign:UWB} & 2010  & respiration heart rate  & - & mono-static  & 1 & 2.2GHz & 4.2GHz & TD PulsON 210\\
\hline 
\cite{CircularPolarVitalSign:UWBradar} & 2013  & respiration heart rate & 100\% \textpm 2bpm & mono-static & 1 & 7.5GHz & - & independent hardware  \\

\hline

\cite{Khan2022} & 2022  & Heart rate & 97.4\% & mono-static & 1 & 1.5GHz & 7.29GHz & Novelda Xethru X4\\

\hline

\cite{Ml:Vitalsign:Occupancy} & 2023  & respiration heart rate & 100\% & mono-static & 1 & 6.4GHz & - & NXP NCJ29D5\\

\hline

 \cite{Multi-Breath:respirationmonit} & 2019  & respiration & 93-97\% & mono-static & 1 & 2.5GHz 2.9GHz & 7.3GHz 8.8GHz& Novelda Xethru\\
\hline

 \cite{ApproachDetVitalSign:UWBradar} & 2021 & Heart rate respiration & 85\%-99\% & mono-static & 1 & - & - & BIOPAC MP150\\
\hline

\cite{MlaugmentedLearningVitalSign:UWBradar}& 2018 & Heart rate respiration rate  & 98.03\% & mono-static & 1 & 2.2GHz & 4.3GHz &  Humatics PulsON P410\\

\hline

\cite{SleepVitalSignMonitoring:UWBradar}& 2022 & Heart rate respiration rate & 99.28\%  96.29\%  & mono-static & 1 & 1.5GHz & 7.29GHz & Novelda XeThru X4 M200 \\

\hline

\cite{ChildrenRespirationVital:UWBradar}& 2021 & Respiration & - & mono-static & 1 &  1.5GHz & 7.29GHz & Novelda Xethru X4M300 \\

\hline

\cite{Liang2023} & 2023 & Heart rate respiration & multiple methods & mono-static & 1 & 1GHz & 0.5GHz & Custom hardware\\

\hline\hline
\end{tabular}
\end{table*}

Vital sign monitoring is the procedure of measuring and monitoring various physiological parameters of a person or an animal, such as \ac{hr}, including heart rhythm and strength of the pulse, \ac{rr}, and blood pressure, to assess their overall health and well-being \cite{PhyDiagnisis:History}. These factors are essential in healthcare because they enable professionals to identify and diagnose potential health issues quickly and precisely. The primary vital sign monitoring devices include \ac{ecg} machines, pulse oximeters, and blood pressure monitors \cite{ECG:Oximeter:vitalMont}. Nevertheless, all these devices require a direct connection to the body, making them hard to use in particular scenarios.

The capability of UWB radar technology to function across a wide frequency range and offer enhanced time scale resolution is a significant advantage in monitoring vital signs. UWB radar systems operate in the frequency range of 3.1 to 10.6 GHz, providing high accuracy and resolution in measuring vital signs. Additionally, the broad frequency range helps reduce the impact of multi-path propagation and interference, which can be significant issues in enclosed and urban environments.

One of the crucial aspects of vital sign monitoring using UWB radar is measuring the distance covered by the radar signal within the body. This distance, known as $d(t)$, is not constant but varies due to the movements caused by respiration and heartbeat. The equation that captures this variation is:

\begin{equation}
d(t) = d_0 + m(t) = d_0 + m_b \sin(2\pi f_b t) + m_h \sin(2\pi f_h t)
\label{eq:vital_signs}
\end{equation}

In Equation \ref{eq:vital_signs}, $d_0$ represents the nominal distance, $m_b$ and $m_h$ are amplitudes of the sinusoidal components corresponding to breathing and heartbeat, and $f_b$ and $f_h$ are their respective frequencies. The variations in this distance provide essential data for monitoring vital signs \cite{AnalysisVitalS:IR-UWB-Rad}.

The appropriate distance between a person and a UWB radar system will vary depending on the characteristics of the radar system and the environment in which the heart rate and respiration are being measured. However, the closer the radar is to the subject, the more precise the measurement is \cite{CorrelationRecVital:UWBradar}. For example, for heart rate monitoring, the radar system needs to detect small changes in the person's body movement caused by the heart's beating. Typically, the radar must be positioned just a few meters away from the subject to enable this application to function \cite{ComparisonIR-UWBandFMCWradar}. On the other hand, since the radar must be able to pick up on even the slightest variations in the person's chest movement brought on by breathing, respiration rate tracking often necessitates placing the radar closer to the subject. The ideal distance for this would be around one to two meters.

In \cite{PreclinicalEvalVitalSign:UWBRadar}, the authors describe developing and evaluating a real-time vital sign monitoring system using \ac{ir-uwb} radar technology. The paper's authors assessed the radar's accuracy in measuring \ac{hr} and \ac{rr} and investigated the factors that can affect the accuracy of radar-based measurements. These factors include age, sex, height, weight, and \ac{bmi}. They concluded that the radar could be accurate in a clinical setting; however, the accuracy of the measurements may be reduced in certain situations, such as sitting position, low respiratory and heart rate. In \cite{VitalSignDetect:noContactUWB}, the authors extract time-frequency information from heart and respiratory signals using the Hilbert transform. Additionally, they compare the outcomes of the \ac{uwb} radar sensor using the \ac{ecg}. The report also describes the capability of this sensor technology to detect a human target through walls and detect the respiratory based on the UWB radar sensor reliably.

\subsection{Overview of scientific publications focusing on vital sign monitoring}

Table~\ref{table:VS1} offers research articles focused on vital sign monitoring. These tables highlight the year of publication, application scenarios, accuracy levels, and key technical details like configuration, bandwidth, central frequency, and the particular radio chip used. The following section will provide a classified analysis of the studies featured in these tables.

\subsubsection{Emergency and Natural Disaster }

During emergencies such as natural catastrophes, building collapses, and accidents, emergency rescue vital sign monitoring employing UWB radar can rapidly and precisely identify and monitor people's vital signs. Without physical contact, the radar can sense a person's presence and monitor their vital signs at a distance. This can be useful for quickly locating and providing medical assistance to individuals in need.

\cite{Zhu2021} carries out a total of 175 scans to test human respiratory signal detection through different types of obstacles, such as foam boards, glass, wooden doors, and brick walls. It employs various analytical techniques like \ac{hht}, \ac{fft}, and correlation analysis to extract features from the normalized echo signals. These features include micro-Doppler characteristics, spectrum features, and correlation metrics. A classification model based on \ac{svm} is then used to identify the respiratory signals. The results indicate that the chosen features effectively capture human respiratory signals. The technique can locate the human body with high accuracy and low detection error, even when different obstacles are present. On the other hand, \cite{ApproachDetVitalSign:UWBradar} employs an edge-cloud computing approach to monitor vital signs. This architecture allows the system to handle high-priority tasks on the edge. The system is designed to offer fast and accurate services in an emergency. The results include an \ac{mae} of 0.09 bpm for respiration and 1.43 bpm for heart rate.

While both papers utilize different methodologies and target different applications, they share a common focus on vital sign monitoring. The first is tailored more towards emergency rescue scenarios involving obstacles. At the same time, the second is suited for telehealth services and broader disaster management, particularly for monitoring mildly ill patients.

\subsubsection{Child and Elderly Care}
UWB radar technology has been deployed in various contexts to monitor the vital signs of both older adults and children, serving as a non-contact method of tracking health metrics and detecting emergencies like falls. The study \cite{Ziganshin2010} focuses on monitoring infants to prevent \ac{sids}. They developed a baby monitor placed 50-70 cm above the cot and conducted sleep tests on a 6-month-old girl. The device monitors respiratory and heart rates continuously, and the recorded data is sent to a PC for further analysis. The study found that respiration was clear 86\% of the time and masked by motion 14\% of the time, while heart rate was clear 83\% and masked by motion 17\% of the time.

Additionally, \cite{Lingyun2015} also utilizes UWB radar but focuses on detecting obstructive sleep disorders like sleep apnea in babies. They introduce a respiration-enhanced algorithm that can detect breathing signals even during slight body motions. Unlike conventional monitors, the system provides a timely alarm if any sleep abnormalities are detected.

On the other hand, \cite{Qimeng2021} extends the application of UWB radar technology to elderly care. Their system uses a framework that combines edge computing for critical task processing and cloud infrastructure for additional data storage and processing. Data was collected from 30 participants with an average age of 31.5 years and an average weight of 55.3 kg to evaluate the accuracy of heartbeat and respiration signals. Participants were recorded in both sitting and lying positions for three minutes each. The system's accuracy in vital sign monitoring was validated through Pearson correlation coefficients for heart rate and respiratory rate, which were 0.98 and 0.96.

Thus, these papers present different but equally crucial applications of UWB radar technology for vital sign monitoring in vulnerable populations like infants and older people.

\subsubsection{Vehicular Monitoring}

In the first paper, \cite{BreathingRateInVehicle:UWBradar}, to ensure that regular driving motions do not influence the breathing rate estimation, the authors examine UWB radar placement in 16 different vehicle positions, ultimately recommending the rearview mirror as the optimal location for accurate results. All experiments were conducted in powered-on vehicles to include the effects of vehicle vibrations. The paper evaluates two signal processing methods and reports average errors of 0.86, 1.07, 1.02, and 1.30.

The second paper, \cite{Khan2022}, focuses on heart rate monitoring, and the human back reflects the radar signals. Data were collected from four male and one female subjects, keeping the distance between the human and radar sensor within the 0.5 to 1-meter range to simulate real-world vehicular conditions. A deep learning classifier detects heart rate patterns, achieving an average error of 2.32\% and a classification accuracy of 97.4\%. However, the study acknowledges certain limitations, such as the fact that the experiments were performed in stationary vehicles, and the effect of seat attenuation has yet to be considered for sensor placement.

\subsubsection{Animal health monitoring}
One of the essential functions of UWB radar technology is its ability to monitor the health of animals by tracking vital signs, all without the need for physical contact with the animals, while also distinguishing between humans and animals.

The first paper, \cite{DogCatVitalSign:UwbRadar}, introduces a non-contact method using \ac{uwb} radar to monitor the vital signs of pets such as dogs and cats. The paper aims to improve animal welfare by removing the need for invasive procedures like anesthesia, hair removal, and surgical implants. The radar accurately detects both the respiratory and heart rates of anesthetized animals. Regarding the data collection, the radar was placed within one meter of the dog. Gas anesthesia was used to minimize its effects on the animals. The paper also claims a high accuracy rate of over 95\% for measuring these animals' respiration.

Furthermore, \cite{Ma2019} studies other animals and focuses on the vital sign monitoring of a New Zealand White Rabbit under water and food deprivation conditions. This paper uses various measurement techniques, including UWB radar, bandage sensors, and contact measurements. One of the critical contributions is identifying variations in the respiratory waveform that can serve as a reference for different life states in non-contact scenarios, like earthquake rescue operations. For the experiments, the rabbit was tied to an experiment bench at a height of 0.9 m, and the UWB radar was positioned 0.6 m away from it.

Despite the previous studies, \cite{HumanAnimalDistinguish:UWBRadar} proposes a scheme using IR-UWB radar to distinguish between humans and common pets such as dogs, cats, and rabbits. The paper uses a 20-second radar echo signal for detection. It keeps the distance at 1 m for all human or animal subjects. A new metric called the \ac{rher} has been introduced to classify the target into its respective category. This classification process is divided into four steps: signal preprocessing, initial identification, vital sign acquisition, and final classification using \ac{rher}.

\subsection{Vital Sign Monitoring Techniques}

This section discusses various analytical and computational methods used to extract, filter, and interpret vital sign data from UWB radar signals. We explore multiple techniques, from traditional signal processing methods to advanced \ac{ml} algorithms, to broadly review the currently available solutions. The following subsections provide a detailed review of each category, illustrating their strengths and applications in UWB radar-based vital sign monitoring.

\subsubsection{Data Filtering}
One of the techniques used for filtering raw data in vital sign monitoring with UWB radar is Kalman filtering. This method is known for making optimal predictions, especially when dealing with incomplete and noisy data. It uses a mathematical framework to estimate the system's state over time, making it a helpful tool for removing noise and clutter \cite{VitalIntegTrack:KNNkalman:UWBradar , VitalSignMontsta/nonsta:UWBRadar}. Curvelet Transform is a powerful tool, especially effective for multi-scale analysis, particularly non-stationary data where signal characteristics may change over time \cite{JLi2014}. Background Subtraction is explicitly used to remove unwanted signals, including environmental noise and irrelevant information, thus isolating the vital sign signals for more precise analysis \cite{VitalSignDetect:noContactUWB}. \ac{wtd} operates in the frequency domain, focusing on eliminating noise by setting a threshold for waveforms likely to be noise components \cite{Ml:Vitalsign:Occupancy}. Lastly, Adaptive Filtering techniques, such as \ac{eemd} and \ac{eemd} combined with \ac{cwt}, adapt to time-varying and non-stationary data, thus offering greater flexibility in filtering signals that have complex or changing characteristics \cite{BreathHeartUWBRadar:FVPIEFandEEMD, DenoisingThWallVitalSign:UWBRadar, ShortRangeVitalSign:UWBRadar}.

\subsubsection{Signal Separation and Feature Extraction}
In vital sign monitoring using UWB radar, the methods for signal separation and feature extraction are unique and different from other applications. This is because the vital signs information is mostly present in the frequency domain of the signals. 

Lower time series analysis is utilized, such as Maximum Likelihood period estimators, to identify the most likely periodicity in signals. Additionally, \ac{vmd} methods based on \ac{tdca} are used for robust separation of overlapping signal components \cite{PeriodEstimationVitalSign:UWB, Pan2023}. Frequency Analysis techniques, such as the Chirp Z-transform and Fourier Transform, are commonly used to break down signals into their frequency components. This process provides a spectral view, making it easier to identify vital signs \cite{AnalysisVitalS:IR-UWB-Rad, DesignIssues:UWBradar, Khan2014}. Wavelet Transform methods, including \ac{emd} and  \ac{cwt}, offer multi-resolution analysis capabilities that allow for the capture of features at various scales and resolutions. This, in turn, facilitates detailed signal interpretation \cite{He2020, ShortRangeVitalSign:UWBRadar}. Time-frequency analysis approaches like Pulse-Doppler time-frequency analysis and \ac{vmd} combined with Hilbert Transform are uniquely suited to analyze non-stationary signals, capturing how frequency components change over time \cite{CmosRemoteMonitoring:UWBRadar, VitalSignDetect:noContactUWB}.

\subsubsection{\ac{ml} and Data Modeling}
Although traditional signal processing methods have been widely used in the field of vital sign monitoring with UWB radar, \ac{ml} techniques are gaining traction. \ac{stft} inputs combined with \ac{svm} have shown potential in analyzing and classifying time-dependent spectral data \cite{MlaugmentedLearningVitalSign:UWBradar}. \ac{dnn}s have further improved performance by incorporating subject-specific physical information for more accurate and nuanced monitoring \cite{Yuan2023}. Encoder-Decoder architectures specialize in robustly recovering respiratory waveforms even when the subject is in motion, thereby solving one of the biggest challenges in remote vital sign monitoring \cite{Zheng2021}. \ac{cnn}-based models are versatile, processing both time series and 2D signal representations to isolate vital signs in different monitoring scenarios \cite{Khan2022, Choi2023, Liu2022}. It should be noted that \ac{ml} solutions are used much less frequently in this context than other applications like activity and gesture recognition. Still, their utilization is rising as they offer advanced capabilities and adaptability.

\subsection{Trends and Observations}
UWB radar technology shows great promise in non-contact vital sign monitoring, particularly when sensor-based or video-based alternatives do not provide sufficient comfort or privacy. The possibility of wirelessly detecting slight chest movements makes UWB vital sign monitoring ideal for monitoring respiratory and heart rates without attached devices, which is beneficial for patients with burn injuries, elderly people, kids, or in sterile environments like operating rooms.

Most research utilizes a mono-static configuration with relatively high (1.4 GHz or higher) bandwidth, necessitating precise radar placement to measure vital signs accurately. Future studies could explore a multi-static configuration, allowing for switching between nodes to achieve accurate measurements while providing flexibility in human movement. On the other hand, exploring different center frequencies and bandwidths can improve UWB radar detection capabilities. Higher bandwidth offers better resolution for distinguishing fine movements, while a lower center frequency enhances penetration through obstacles, beneficial for through-clothing or through-wall monitoring.

Similar to the other use cases, several real-life aspects are unexplored. 
\begin{itemize}
    \item For many use cases, it is important to distinguish between the vital signs of multiple persons, for example between patient and doctor or multiple persons sharing a bed. To this end, recently advanced signal processing algorithms and \ac{ml} techniques have been developed to isolate individual signals in crowded environments \cite{Yuan2023}. However, scenarios like monitoring individuals with unrestricted movements and multiple people at the same distance from the radar setup remain challenging.
    \item Range limitations remain a concern, typically effective within a few meters, requiring strategic device placement. 
    \item Monitoring individuals in motion or from moving vehicles presents additional challenges, necessitating further research into algorithms to automatically remove vibrations and background noise from the signal of interest.
    \item Many commercial use cases assume the radar will be built into existing furniture such as beds, steering wheels, light switches, etc. The impact of these integration aspects on the accuracy of existing solutions is as of yet unexplored.
\end{itemize}

In conclusion, while UWB radar technology for vital sign monitoring is progressing rapidly, addressing challenges related to complex environments, range constraints, and signal differentiation is crucial. As the technology evolves, it will likely find broader applications and greater adoption in various healthcare settings.

\section{Integrated Sensing and Communication (\ac{isac})}\label{section:jsac}
UWB technology is widely known for its ability to provide high-speed communication, precise localization, as well as UWB radar. However, these three domains are currently being researched independently. \ac{isac} (sometimes also referred to as ``Joint Sensing and Communication or JSAC'') is the trend toward combining both sensing and communication functionalities \cite{wang2022integrated}. 

Two main approaches are possible:
\begin{itemize}
    \item \textbf{Using transmitted signals for multiple purposes simultaneously.} Wireless signals used for communication can also be reinterpreted to provide distance estimations or a radar-like view of the environment using the multipath components of the signal. While reusing the same radio signals is optimal from a radio spectrum usage point of view, it may require compromises in pulse shape, radio settings, and other parameters, which must be co-optimized for all functionalities simultaneously, potentially resulting in degraded performance. 
    \item \textbf{Sequential use of radio platforms.} Alternatively, a single radio platform can be employed sequentially for localization, communication, and radar functions. This method allows optimization of the radio settings (like pulse shape and transmission settings) for each specific function, but it is less efficient in terms of energy consumption and spectrum usage.
\end{itemize}

The possibility to exchange data reliably and sense the environment accurately simultaneously results in reduced energy costs, less need for separate hardware components (thereby reducing costs and weights) and allows to more optimally use the scarce spectrum. As such, the demand for realizing multiple functions within a single system is especially relevant for autonomous robot platforms, self-driving cars, and immersive \ac{ar}/\ac{vr} in Industry 4.0 scenarios, which can potentially replace communication units and multiple sensors by a single radio platform. As such, this concept involves more than just a combination of existing functions, hardware, and software elements. Instead, it encompasses a comprehensive transformation involving simultaneous redesign of systems, hardware, and algorithms. 



\begin{table*}[ht]
\caption{Overview of scientific papers utilizing UWB radar for \ac{isac}}
\label{table:jsac}

\renewcommand{\arraystretch}{1.1}
\centering
\begin{tabular}{m{0.5cm}|m{0.5cm}m{2.3cm}m{5cm}m{1.6cm}} 
 \hline
 Paper & Year  & Focused Area  & Considered Features & Theoretical/ Experimental\\ [0.5ex] 
\hline\hline
 
\cite{Zhang2023FundamentalLO}  &  2023 & Physical Layer & Decoupling Sensing and Communication + Energy Allocation &   Theoretical \\

\hline

\cite{10437359}  &  2023 & Physical Layer &  Integration Channel Sensing, Doppler measurement, and Communication  &   Theoretical \\

\hline

\cite{10333549}  &  2023 & Physical Layer & Modulation impact for sensing in \ac{isac} &  Theoretical  \\

\hline

 \cite{9652951}  &  2021 & Physical Layer & Integration of communication and localization &  Theoretical  \\

\hline

\cite{electronics12020330}  &  2023 & Antenna Design & UWB Sensing + \ac{mimo} Communication & Experimental   \\

\hline

 \cite{UWBChip:Survay} &  2022 & Hardware Design & Survey of Integrated sensing and communication devices & -   \\

\hline\hline
\end{tabular}
\end{table*}

 The field of \ac{isac} utilizing \Rev{\ac{uwb}} technology is still in its early stages of development. However, it is attracting increasing scholarly attention and research. In this section, we will focus on the fundamental aspects of chip-scale innovations and antenna redesigns necessary to enhance the performance and integration of UWB systems. We will explore various methodologies that enable the merging of communication and sensing capabilities, providing examples of integrated systems. Finally, we will discuss trends and challenges in this field. 
\subsection{Overview of scientific publications focusing on \ac{isac}}

In this section, we will explore the scientific articles that focus on this recent research domain. Table~\ref{table:jsac}provides an overview of relevant works, including  the publication year of each article, the specific area of focus, the multi-functionality features considered, and whether the work is theoretical or experimental. 


Starting with scientific work investigating the impact of the physical layer, the first study, \cite{Zhang2023FundamentalLO}, explores the fundamental trade-offs between sensing and communication in UWB systems. The paper demonstrates that resource sharing between these functions influences system performance and design decisions by utilizing the \ac{fim} and the \ac{crb}. The analysis reveals that while inherent trade-offs exist, optimized system design can control performance issues. The second paper, \cite{10437359}, extends this discussion by examining the simultaneous estimation of delay and Doppler shifts within UWB \ac{isac} systems. It highlights the critical role of waveform design and modulation schemes, such as \ac{ppm} and \ac{bpsk}, in optimizing system performance. The \ac{efim} is introduced to understand the complex interdependencies between transmitted symbols and channel parameters, emphasizing the challenges in achieving accurate parameter estimation without compromising the system's communication capabilities. Adding to these studies, \cite{10333549} delves further into the intricate balance between communication data and sensing capabilities within IR-UWB \ac{isac} systems. Their research utilizes the \ac{fim} and the \ac{crb} to critically evaluate how unknown transmitted data affects channel sensing parameters under different modulation schemes, specifically \ac{ppm} and \ac{bpsk}. This work presents a pilot-based strategy to decouple data from sensing parameters, improving the estimation accuracy of key channel characteristics such as delay and Doppler shifts, thus providing a theoretical foundation for selecting appropriate modulation schemes in varying \ac{isac} scenarios. Finally, \cite{9652951} explores the integration of communication and localization functionalities using Impulse Radio \ac{ir-uwb} signals, notable for their high time resolution and broad bandwidth. The proposed asynchronous IR-UWB system, employing a non-coherent system model, addresses challenges in complex environments by implementing a sequential detection-based data demodulation strategy and innovative soft information-based solutions for reliable pulse detection and \ac{toa} estimation. These system enhancements promise to simultanuously improve the accuracy and reliability of communication and localization tasks in environments where traditional synchronous methods are ineffective or impractical. In conclusion, these studies collectively underscore the need for optimized UWB physical layer configurations for facilitating the integration of sensing and communication and they delineate the technical challenges and the necessary considerations for designing efficient \ac{isac} physical layer configurations.

In addition, innovations in antenna designs and chip-scale RF transceivers are similarly vital for achieving efficient, compact, and energy-effective solutions that meet the growing demands for simultaneous communication and radar sensing capabilities. A notable advancement is presented in \cite{electronics12020330} for designing a \ac{cr} integrated antenna system featuring a single sensing antenna coupled with a 24-element communication array across the UWB spectrum from 2 to 12 GHz. This system enables effective spectrum utilization by allowing dynamic access and allocation based on real-time spectrum environment analysis. Each segment of the communication array operates within specific sub-bands, thereby minimizing interference and enhancing signal clarity. This design presents a strategic unification of multiple communication functions within a unit, enabling advancements in CR technology and spectrum efficiency. Simultaneously, the survey on integrated wideband chip-scale RF transceivers \cite{UWBChip:Survay} delves into the application of \ac{fmcw}, \ac{cw}, and \ac{uwb} in radar sensing and communications. The survey discusses the integration of radar sensors. These semiconductor advancements allow the integration of complex RF functionalities into compact, low-power devices and the implications of AI-enhanced signal processing techniques that enable precise localization and multimodal sensing. However, to advance the UWB \ac{isac}, a specific review focused on analyzing the critical next steps for hardware and antenna studies is necessary. Such a review would open new routes for research and accelerate studies for applying \ac{uwb} \ac{isac}. These studies highlight the critical role of integrated antenna systems and chip-scale transceivers in enabling the next generation of communication and sensing solutions.

\begin{table*}[htbp]
\caption{Overview of available open-source UWB radar datasets, specifying the application domains they can be used for and the dataset characteristics (number of devices, bandwidth, center frequency, and radio chip used for data capturing).}
\label{table:dataset}
\fontsize{8pt}{8.2pt}\selectfont
\renewcommand{\arraystretch}{1.1}
\centering
\begin{tabular}{m{0.5cm}|m{0.5cm}  m{2.5cm} m{2cm} m{1.4cm} m{0.9cm}m{1.2cm} m{1.2cm}  m{2.8cm} m{0.6cm}} 
 \hline
 Paper & year  & Application  & Sub-type & System & Devices & Bandwidth& Central F &  Chip & Open-source\\ [0.5ex] 
 \hline\hline
 
 \cite{DatasetHM:UWBrad} &  2021 & Motion detection & Through wall motion detection &   mono-static    &   1 & - & 0.5GHz &  independently designed hardware & \Checkmark  \\
          
\hline

\cite{DatasetUWB:gesture} & 2021 & Activity recognition & 12 hand gestures & mono-static & 3 & 2GHz & 8.745GHz & Novelda XeThru X4 & \Checkmark  \\
          
\hline

\cite{uwb:DENSEPEOPLE:COUNT} & 2018 & Motion detection & People counting 
& mono-static & 1 & 2.3GHz & 6.8GHz & Novelda NVA-R661 & \Checkmark\\

\hline  

\cite{SleepPoseNet} & 2020 & Activity recognition & Sleep pose recognition & mono-static & 1 & 1.4GHz, 1.5GHz & 7.29GHz & Novelda Xethru X4M03 & \Checkmark \\

\hline

\cite{Humandetect:ML:AcGen} & 2020 & Motion detection & NLOS presence detection & mono-static & 1 & 2.5GHz & 6.8GHz & Novelda  NVA-R661 & \Checkmark\\

\hline 

\cite{OPERAnet:activityrecg}& \multirow{2}{*}{2022} & \multirow{2}{*}{Activity recognition} & \multirow{2}{2.5cm}{Presence, counting, and multiple activities} &  \multirow{2}{*}{multi-static} &  \multirow{2}{*}{4} & 1.3GHz & 4GHz & Decawave EVK1000 & \Checkmark\\[9pt]

&  &  & & &  & 0.5GHz & 4.5GHz & Decawave MDEK1001 & \Checkmark\\

\hline  

\cite{DLUWBradar:AirWirtung}& 2022 & Activity recognition & Air-writing recognition & mono-static & 1 & 1.5GHz & 7.29GHz & Novelda Xethru X4M03 & \Checkmark\\

\hline

\cite{UWBCARGRAZ:dataset} & 2023 & Presence detection \& Activity recognition&Car occupancy and 3 activities & mono-static bi-static & 2 & 0.5GHz  5GHz & 6.5GHz 7GHz & ILMSENS m:explore & \Checkmark \\

\hline

\cite{Brishtel2023}   & 2023 &  Activity recognition & Driver monitoring & mono-static & 1 & 7.29GHz & 7.25GHz - 10.2GHz & Novelda Xethru X4M02 & \Checkmark \\

\hline

\cite{PassHumanTrackCOTS:UWBRadar} & 2022 & Device Free localization & tracking  & multi-static & 4 & 0.9GHz & 4GHz &  Decawave DW1000 & \Checkmark \\

\hline

\cite{Multi-StaticUWB:RadarNetwork} & 2020 & Device Free localization & Distance & multi-static & 3& 0.9GHz & 3.99GHz & Decawave DWM1000 & \Checkmark \\
\hline\hline
\end{tabular}
\end{table*}

\subsection{Trends and Observations}

Although the first UWB-specific scientific publications are being published, the field of \ac{isac} for UWB is very new and only a limited number of studies have been performed. Several \ac{isac} specific challenges remain as of yet unaddressed in scientific literature.

\begin{itemize}
    \item \textbf{Optimized antenna and radio architectures:}     One of the significant challenges lies in designing systems that can effectively handle the interference and signal integrity issues stemming from the dual functionality. Innovations in antenna design, waveform generation, and signal processing for self-interference cancellation are critical to overcoming these hurdles. At analog front-end level, self-interference cancellation requires judicious selection and thorough optimization of an appropriate antenna (array) topology to obtain sufficient isolation over a wide bandwidth. In addition, careful design of the RF routing network is essential in order to maintain the isolation level offered by the antennas. Decoupling structures and metamaterials, such as electromagnetic band-gap (EBG) or defected ground structure (DGS) \cite{DRA_decoupling1,DRA_decoupling2}, as well as arrays with parasitic elements or a dielectric superstrate \cite{Decoupling_parasitic, DRA_4}, have been developed but must be optimized for wideband SIC in a compact footprint. 
    \item \textbf{Adaptive \ac{isac} algorithms:} Most of current \ac{isac} research focuses on new physical layer and antenna designs. Scientific work that explores adaptive algorithms that can dynamically optimize system performance based on real-time environmental and operational conditions are still lacking. To this end, the trade-offs for reusing signals for multiple functionalities have to be described in more detail, allowing to make informed optimization choices.
    \item \textbf{Experimental validation:} There is a significant disparity between the theoretical models and real-world validation. Especially in regards to new physical layers, no experimental studies have been performed. While the initial models and simulations demonstrate promising results, practical feasibility and reliability must be established through field tests. Hence, investigation is required to bridge this gap and validate the integrated approach.

    \item \textbf{Potential Applications:} The application spectrum for \ac{isac} integrated with UWB is vast. In high-stakes environments such as autonomous vehicular systems, disaster management, and smart cities, the ability to perform high-resolution radar imaging while maintaining robust communication channels could deliver outstanding safety and efficiency improvements. There are as of yet no scientific papers providing a clear overview of the requirements of new \ac{isac} use cases in terms of required form factors, performance requirements, constraints, commercial viability, etc. As such, it is very difficult to understand the exact needs of \ac{isac} solutions for future applications.


\end{itemize}

As the field progresses, the intersection of UWB radar and \ac{isac} is poised to open new avenues for multidisciplinary research and practical innovations that could redefine the landscape of wireless technology.


\section{UWB radar datasets}\label{section:dataset}

UWB radar has been used in various applications, and the availability of datasets is crucial to advancing research in this field. As depicted in Table~\ref{table:dataset}, nine available datasets use UWB radar for different applications. \cite{DatasetHM:UWBrad} presents a dataset of three test subjects measured in different environments and motion statuses using an IR-UWB through-wall radar system. The dataset includes scenarios of people standing, walking, and empty space in free space, as well as through-wall radar placement. \cite{DatasetUWB:gesture} introduces UWB-Gestures, a public dataset of dynamic hand gestures acquired with UWB impulse radars. The dataset contains 9,600 samples gathered from eight different human volunteers for twelve different dynamic hand gestures, eliminating the need for UWB radar hardware to train and test the algorithm.

In \cite{uwb:DENSEPEOPLE:COUNT}, they employed a radar system mounted at a height of 1.8 meters, capable of detecting objects within a 5-meter radius and a 90-degree angle. The objective was to assess how well this radar could measure the density of people in a restricted space. They conducted three different scenarios to test this. In the first two scenarios, groups of up to 20 people were observed walking in an area. Depending on the scenario, the density varied between 3 and 4 persons per square meter, and the available space expanded with the number of individuals to maintain this density. The third scenario presented a different challenge: they had up to 15 individuals standing in a line, each about 10 centimeters apart, to see if the radar could effectively detect stationary objects nearby. Across these tests, they used 44 individuals to gather a wide range of data. The radar recorded data for five seconds in each test, capturing 200 signals. This resulted in 3,360 radar samples for each of the first two scenarios and 2,560 samples for the third, providing a comprehensive dataset for analyzing the radar's performance in different crowded environments. \cite{SleepPoseNet} a dataset of \ac{spt}s utilized a UWB radar system, positioned 0.8 meters above a bed and angled at 45 degrees downwards, to gather two sets of data. The first dataset involved participants performing six motions, including various sleeping positions such as supine, prone, and lateral. It focused on four sleeping postural transitions: supine to side, supine to prone, side to supine, and prone to supine. The second dataset was conducted in two parts; the first was in a static environment, and the second was with a moving fan to create a more realistic setting. In this dataset, participants performed the same postural transitions as in the first dataset. They engaged in activities like limb movements and smartphone use, categorized as background activities.

Detection of human beings in \ac{nlos} conditions using UWB radar is explored in \cite{Humandetect:ML:AcGen}, where an extensive measurement campaign is performed in realistic environments, considering different body orientations, obstacle materials, and radar-obstacle distances. Two main scenarios are examined based on the radar position: placed on top of a mobile cart and handheld at different heights. \cite{OPERAnet:activityrecg} presents a dataset consisting of RF data extracted from a WiFi \ac{nic}, \ac{pwr}, UWB signals acquired via commercial off-the-shelf hardware, and vision/Infra-red based data acquired from Kinect sensors. The dataset includes approximately 8 hours of annotated measurements collected from 6 participants performing six daily activities, which can be exploited to advance WiFi and vision-based activity recognition \ac{har} and potentially used for indoor human tracking.

The authors of \cite{DLUWBradar:AirWirtung} investigate air-written numbers from 0 to 9, using a uni-stroke writing technique, which means each number is written in one continuous motion. The samples vary in writing angles, speeds, and positions, enriching the dataset. The air-writing was done in a small area, about 20 cm by 20 cm. The radar used to detect this writing could cover a range from 0.4 m to 1.55 m, known as range bins. They expanded this range by an extra 95 cm to improve the monitoring of the Doppler spectrum and capture a broader range of writing movements. This enhancement makes the dataset more versatile for analyzing different numeral writing styles.

\section{Available radar chips}\label{section:chips}
This section provides an overview of UWB radar chips and devices from leading manufacturers. Table~\ref{table:chips} included in this study presents comprehensive information about the UWB radar chips. The table includes both generic, IEEE-standard-compatible \ac{uwb} radio chips and radar-specific devices. Generic radio chips such as the Qorvo or NXP radio chips are suitable for a wide range of applications such as communication, localization and sensing, offering versatility and precision for different use cases. They are most typically equipped with wide-field-of-view antennas by default and support the minimum required bandwidth of 500MHz (or in some cases up to 1GHz). On the other hand, UWB radio chips from manufacturers such as Novelda, Pulson, Time Domain, and Umain are designed mainly for radar purposes. Devices from this last category are characterized by having a larger bandwidth (typically in the order of 2GHz) and are most often equipped with directional antennas offering a narrow field of view. 

\begin{table*}[htbp]
\caption{Overview of commercial UWB radar chips and devices, as well as their characteristics (center frequencies, bandwidth, antenna type, ...)}
\label{table:chips}

\renewcommand{\arraystretch}{1.2}
\centering
\begin{tabular}{|m{1.8cm}| m{1.2cm} | m{0.5cm}| m{2.6cm} |m{2cm} |m{2.3cm} |m{1.5cm}|m{2.5cm}|} 
 \hline
 Manufacturer & chip  & year &Devices  & Central F & Bandwidth& Mono-static /multi-static & Antenna \\ [0.5ex] 
\hline\hline
ARIA Sensing& LT103 && ARIA LT103OEM & 7.9GHz & 1.2GHZ&mono-static & Integrated patch antennas\\
\hline
\multirow{3}{*}{Novelda}& X1 (NVA610x) &2011& NVA-R631, NVA-R641 & 
3.85GHz, 5.2GHz & 6.3GHz, 8.7Ghz&mono-static & Connectorized (Vivaldi and sinuous antennas provided)\\
\cline{2-8}
& X2 (NVA620X) &2013& X2M200, NVA-6201,  NVA-R661 & 5.3GHz, 5.4GHz, 5.7GHz, 6.1GHz, 6.4GHz, 6.8GHz, 7.3GHz, 7.7GHz, 7.8GHz, 8.2GHz, 8.8GHz & 1.75Ghz, 1.8GHz, 1.85GHz, 2.05GHz, 2.15GHz, 2.30GHz, 2.35GHz, 2.50GHz, 2.65GHz, 3.10GHz &mono-static & Connectorized (Vivaldi and sinuous antennas provided)\\
\cline{2-8}
& X4 &2017& X4F103, X4M02, X4M03, X4M05, X4M06, X4M07, X4M200, X4M202, X4M300, X4SIP02, SLMX4 & 
7.29GHz, 8.748GHz & 1.4GHz, 1.5GHz &mono-static & Integrated differential patch antennas\\
\hline
\multirow{3}{*}{\shortstack[l]{Pulson/\\Humatic/ TDSR}} & P400 &2009&  & 4GHz &2GHz &mono-static/ bi-static & Connectorized (BroadSpec wideband monopole antenna provided)\\
\cline{2-8}
 & P410 &2012&  & 4.3GHz &2.2GHz &mono-static/ bi-static & Connectorized (BroadSpec wideband monopole antenna provided)\\
\cline{2-8}
 & P440 &2015& PulsON P440, Humatic P440, TDSR P440 & 3.95GHz &1.7GHz & mono-static/ bi-static/ multi-static & Connectorized (BroadSpec wideband monopole antenna provided)\\
\hline
\multirow{2}{*}{Time Domain} & P210 && TD P210 & 4.7GHz & 3.2GHz & bi-static/ multi-static & Connectorized (BroadSpec wideband monopole antenna provided)\\
\cline{2-8}
 & P220 && TD P220 & 4.3GHz & 2.3GHz & mono-static/bi-static & Connectorized (BroadSpec wideband monopole antenna provided)\\
\hline
Umain & HST-C1R && HST-S1M-CT, HST-D3 & 4.27GHz, 3.95GHz & 1GHz, 1.7GHz & mono-static & Connectorized (dipole antenna provided)\\
\hline
\multirow{6}{*}{Qorvo} & \multirow{3}{*}{DW1000} & \multirow{3}{*}{2013} & DWM1000, DWM1004C& 3.5GHz, 4GHz, 4.5GHz, 6.5GHz &500MHz, 1GHz &bi-static/ multi-static & Integrated ceramic chip antenna\\
\cline{4-8}
 & & & DWM1001C, MDEK1001,  DWM1001-DEV & 3.5GHz, 4GHz, 4.5GHz, 6.5GHz &500MHz, 1GHz &bi-static/ multi-static & Integrated monopole antenna\\
\cline{4-8}
 & & & EVK1000 & 3.5GHz, 4GHz, 4.5GHz, 6.5GHz &500MHz, 1GHz &bi-static/ multi-static & Connectorized (Wideband monopolar spline antenna provided)\\
\cline{2-8}
 & \multirow{3}{*}{DW3000} & \multirow{3}{*}{2020} & DWM3000, DWM3000EVB & 6.5GHz, 8GHz & 500MHz & bi-static/ multi-static & Integrated ceramic chip antenna\\
\cline{4-8}
 & & & DWM3001C, DWM3001CDK, QM33120WDK1 & 6.5GHz, 8GHz & 500MHz & bi-static/ multi-static & Integrated monopole antenna\\
\cline{4-8}
 & & & QM33120WDK1 & 6.5GHz, 8GHz & 500MHz & bi-static/ multi-static & Connectorized\\
 \hline
 \multirow{5}{*}{NXP} & SR040 & 2020 & & 6.5GHz, 8GHz &500MHz &bi-static/ multi-static &  \\
 & SR150 & 2020 & & 6.5GHz, 8GHz &500MHz &bi-static/ multi-static & \\
 & SR100 & 2019 & & 6.5GHz, 8GHz &500MHz &bi-static/ multi-static & \\
 \cline{4-8}
  & NCJ29D5  &  2019 & LID2434-R4-SMA &  6.5GHz, 8GHz & 500MHz & bi-static/ multi-static & Connectorized\\
   & NCJ29D6   & 2023 & & 6.5GHz, 8GHz & 500MHz & mono-static/ bi-static/ multi-static & \\
\hline\hline
\end{tabular}
\end{table*}

Since most scientific research papers use only a single device type for their research, it is currently not yet clear to what extent the reported accuracies from the scientific literature are impacted by specific hardware aspects, which minimum bandwidth is required for specific use cases and if all proposed scientific innovations from scientific papers generalize well towards other platforms.

\subsection{Trends and Observations}
Initially, UWB systems adopted very wide channels in excess of 500 MHz, or a fractional bandwidth over 20 $\%$, until 16 standardized channels were allocated by the 2007 IEEE 802.15.4a standard \cite{IEEE_UWBSTD_2007}. This limited the channel bandwidth to standardized values of approximately 500 MHz, 1.1 GHz, or 1.35 GHz, depending on the considered channel. Nonetheless, as shown in Table~\ref {table:chips}, several non-standardized commercial solutions were developed over time, with bandwidths up to 8.7 GHz (Novelda X1) to achieve very accurate ranging (see Table~\ref{table:DFL}). In general, UWB systems are shifting towards higher operating frequencies for several reasons. First of all, operation in UWB channel 9 (7.99 GHz) ensures the highest level of regulatory clearance \cite{Samsung_UWBchannel9, JSC_UWBchannel59}, which drives many standards towards this higher-frequency channel. Furthermore, the frequency spectrum exploited by the newly introduced WiFi 6E systems overlaps with UWB channels 5, 6, and 7, which discourages the use of these lower-frequency bands \cite{monkaewe2024coexistence} and motivates research towards WiFI 6E/UWB coexistence and interference mitigation techniques. Finally, in 2023, additional spectral resources (6.425-7.125 GHz) were allocated to 5G systems in the frequency range 1 (FR1), further pressurizing these UWB channels \cite{monkaewe2024coexistence}. This trend is reflected by the most important standardization institutes \cite{UWBstandardsSurvey}, often favoring channel 9, which satisfies global regulations and is free from potential WiFi interference, in combination with a lower-frequency channel.

Several trends are also observed with respect to the used antenna hardware. First of all, it is observed that all commercial UWB radar modules listed in Table~\ref{table:chips} are either two-antenna modules (Novelda, Aria, Humatics, Umain) in a monostatic radar architecture, with one \ac{tx} and one \ac{rx} antenna located at opposite sides of the radar IC to limit self-interference, or single-antenna modules in a bistatic or multi-static radar setup with multiple synchronized boards (Qorvo, Humatics). Despite the stringent design requirements in the frequency, time, and spatial domains, a variety of robust and high-performance antenna topologies have been developed over time, as shown in Table~\ref{table:chips}. Nevertheless, several antennas, in particular the monopole and dipole antennas, exhibit omidirectional radiation, which is typically not desirable in UWB radar systems since it leads to significantly more clutter, a higher sensitivity to electromagnetic coupling with their integration platform, and angle estimation ambiguity in linear and planar two-dimensional antenna arrays \cite{GJ_33, GJ_73}. In addition, several of the more directive designs are bulky and challenging to integrate with UWB transceivers, including the popular Vivaldi antennas and sinuous antennas. As such, these antenna types are often included with connectorized modules. UWB patch antennas constitute the most widely used solution for integrated UWB radar modules, such as the Novelda XeThru X4-based modules \cite{datasheetNoveldaX4M200_21, datasheetNoveldaX4M300_22, datasheetNoveldaX4M06_23} and Aria Sensing module \cite{datasheetAriaLT103_24}, driven by their main advantages: ease of integration in standard PCB manufacturing technology and compact integration within radar IC modules. To achieve sufficient bandwidth in a compact footprint, either thick PCB substrates \cite{Huang23}, or multilayer PCB substrates are leveraged to implement stacked patch configurations \cite{Zeng22,Shakib15}. Antenna efficiency is limited in both cases by the dielectric losses in the antenna substrate, urging the need for high-frequency PCB laminates. In current-generation commercial UWB radar modules \cite{datasheetNoveldaX4M200_21, datasheetNoveldaX4M300_22, datasheetNoveldaX4M06_23, datasheetAriaLT103_24}, the number of patch antenna elements is fixed and limited as a result of side-by-side active electronics integration, which hinders upscaling to larger UWB arrays and leads to large system footprints. 

While several solutions have been proposed in the literature to alleviate the aforementioned issues, these have not been implemented by commercial UWB radar modules so far. First of all, to alleviate the issues with side-by-side active electronics integration, a similar approach as in \cite{Vanveerdeghem14,Baelen2021} could be pursued, in which all active electronics are integrated behind the antenna system, although this results in a more complex and expensive PCB stackup. Furthermore, recently, novel UWB antennas based on \ac{afsiw} technology have been proposed \cite{GJ_Sensors} that are simultaneously compact, wideband, and highly efficient. Compactness is achieved by adopting a quarter-mode cavity-backed in combination with a capacitively coupled feed. High efficiency is obtained by leveraging an air substrate, almost entirely eliminating dielectric loss. Its performance is thoroughly optimized in the time, frequency, and spatial domains to ensure a stable radiation pattern with minimal pulse distortion, phase fluctuations, and range bias. The adopted cavity-backed topology results in a high antenna/platform isolation, maintaining high performance when deployed on a variety of platforms, such as the metallic hull of a ship \cite{GJ_Sensors}. In addition, performance is also maintained in 1D and 2D array configurations, resulting in a scalable building block for planar multi-antenna UWB systems. The cost-effective antenna array is implemented using standard low-cost PCB manufacturing technology and allows scalable integration of active electronics behind its ground plane, making it a perfect candidate for future scalable multi-antenna UWB systems. \ac{dra}s constitute another interesting candidate. DRAs consist of a dielectric resonator mounted on a metal ground plane. The lack of metal in the resonating structure minimizes conductor losses and enables high-efficiency operation at high (\ac{mmwave}) frequencies. By applying multi-resonator techniques, e.g. by adding a resonating slot to realize an aperture coupled-fed DRA, they can offer a wide bandwidth and hemispherical radiation pattern in a compact footprint. Furthermore, they facilitate the integration of active electronics, such as UWB radar ICs, on their ground plane backside, paving the way for compact and scalable radar arrays \cite{DRA_1}. However, mutual coupling between adjacent elements is typically large and heavily influenced by the inter-element distance, although isolation enhancement techniques have been studied \cite{DRA_4}. Moreover, none of the DRA arrays in the literature have been optimized nor analyzed in the time domain. In fact, only a few DRAs have been analyzed for their time-domain performance \cite{DRA_2, DRA_3}, and in those cases, the effects of array integration or the integration environment were not considered.
Finally, a conformal UWB array consisting of 52 titanium 3D-printed Vivaldi elements was proposed in \cite{THREED_PRINTING}. By leveraging a doubly curved surface, extreme scan angles of up to $120^{\circ}$ can be realized, which cannot be achieved by means of a planar array. Unfortunately, the integration of active electronics, such as UWB transceiver ICs, remains problematic with such structures.

\section{Future Research Challenges and Open Problems}\label{section:Future}

Finally, this section will give an overview of the technological, standardization, and legislative future research challenges in the domain of UWB radar research. In no particular order, some of these challenges are the following.

\textbf{Lack of UWB radar standards} \\Most upcoming UWB radar standards (see Section~\ref{UWBStandards}) focus on PHY layer standardization or (in the case of Ripple) backend APIs. No current standards describe the setup and configuration of UWB radar systems or standardized methodologies to combine, process, and interpret UWB (CIR) radar signals. This lack of ongoing standardization efforts makes it difficult to achieve compatibility among existing UWB radar solutions, hindering widespread industry adoption. A standardized framework for the interoperability and certification of commercial UWB radar devices would promote the utilization of UWB radar technologies. 

\textbf{Lack of UWB datasets and competitions} \\Similarly, the availability of open-access UWB radar datasets is scarce (see Section~\ref{section:dataset}). Access to such datasets is crucial for researchers, as they will facilitate the development and evaluation of new methodologies:
\begin{itemize}
    \item New advanced methodologies can be evaluated on the same general datasets for fair comparison.
    \item Datasets obtained from realistic use cases and environments (in-car, residential housing, etc.) can be used to evaluate to what extent proposed techniques perform well and generalize well in realistic environments.
    \item Datasets that utilize more radios and radio configurations would give new insights regarding PHY layer configuration trade-offs (bandwidth and center frequency versus accuracy) for different use cases. 
\end{itemize}
Furthermore, the research community would benefit from organized competitions that challenge participants to advance state-of-the-art UWB radar methods. These community-driven initiatives would greatly expedite the progress of UWB radar technologies.

\textbf{Robustness of UWB radar in realistic conditions} \\With the increasing deployment of UWB radar solutions in a wide range of industrial applications and locations, robustness becomes another concern. UWB systems might be deployed in cars, industrial processes, integrated beds for elderly care, etc. Currently, there is a lack of insights into the performance of UWB radar under various environmental influences such as temperature, vibrations caused by industrial machinery, humidity, clothing, wall materials, different body poses and orientations, etc. To improve the robustness and reliability of UWB radar systems, researchers should explore the adaptability of UWB radar in typical and extreme environments. For example, the work of \cite{Zheng2021} shows that breathing rate detection is possible even for runners on a treadmill, where the breathing rate can be distinguished from other body movement patterns. However, most scientific papers focus on setups in controlled environments with very few non-controlled parameters. To improve robustness, the techniques outlined in this overview paper will have to be extended with processing techniques for external noise suppression, body orientation estimation, repression of non-relevant signal frequencies, etc.

\textbf{Coexistence and interference mitigation} \\Most UWB radar systems operate at a higher center frequency than traditional communication technologies such as BLE, WiFi, and Zigbee and at a lower center frequency than mmwave radar or FMCW radar. However, several (upcoming) technologies share the spectrum or have overlapping frequency bands with UWB \cite{Adeogun2019}. For instance, the Wi-Fi 6E standard operates in the 6 GHz band, similar to UWB. Wi-Fi 6E has been shown to negatively impact the performance of UWB signals \cite{Brunner2022}, although this impact can be mitigated by the optimal selection of the PHY layer parameters. Similarly, the upper mid-band frequency range (FR), spanning from 7 GHz to 24 GHz (also known as FR3), has emerged as a focal point in 6G communications, to the point where FR3 has been named the ``golden band'' by Nokia. If these bands are generally accepted for 6G communications, this will, on the one hand, likely result in legislation that is well suited for worldwide UWB deployments but, at the same time, potentially result in millions of interfering devices. Finally, deploying a large number of UWB radar systems can lead to concurrent UWB signals, which will compromise the quality of the \ac{cir} and, therefore, detection accuracy. Recent work has shown that the use of scrambled time sequences (STS) in recent UWB standards strongly increases the robustness of UWB in the presence of interference \cite{TiemannIEEE}. However, there is a strong need to investigate further the impact of interference on UWB radar as well as to design novel, robust interference mitigation strategies that can dynamically adapt the channel and PHY settings and transmit power depending on the wireless conditions.

\textbf{Ease of deployment and calibration} \\Standards for automatic discovery and configuration of UWB ranging and localization are currently being proposed by the FiRa standardization organization. However, such facilities are still lacking for UWB radar and activity recognition. In this survey, we found that applications often use noise removal and background subtraction techniques. Both techniques require an understanding of the environment and a calibration phase. For mobile devices, this calibration phase is not always feasible, and the device should adapt itself to the environment by changing the physical parameters of the UWB signals (power, frequency, etc.) and learning the environment during deployment. There is, hence, a need for solutions that dynamically discover the required functionality and capabilities of nearby UWB radar devices and identify the current user requirements (e.g., required radar range, type of activities that should be recognized, maximum power consumption, etc.) and use these to automatically select the best settings for each use case.

\textbf{Multi-target tracking and monitoring} \\The state-of-the-art algorithms discussed in this survey mostly focus on accurate prediction of the activity and/or vital signs of one subject. Current solutions cannot differentiate between multiple people moving in the same area, especially under occlusion conditions. A further research direction is the expansion of the current algorithms to distinguish multiple targets and track their position simultaneously with the same \ac{uwb} radar installation. One popular approach is the use of multi-static setups spread out over a larger area that can differentiate between multiple targets \cite{J.S.Park2012}. However, several improvements are possible. These works focus mainly on multi-target tracking, but similar work focusing on multi-person activity recognition or multi-person vital sign monitoring is still lacking. In addition, achieving similar goals using a single anchor node would strongly reduce deployment costs. A promising approach is to use a single multi-antenna UWB radar system, whereby AoA and AoD transmissions allow a reduction of occlusions due to other moving people and objects present in the environment. Different targets can be discovered and monitored simultaneously by exploiting spatial information such as distances and angles towards the radar.

\textbf{Self-supervised \ac{ml} models}\\ \ac{ml} models are already popular for various UWB radar use cases, such as gesture recognition, activity recognition, etc. However, current work focuses mostly on traditional supervised neural network architectures from traditional domains such as vision processing. To train these supervised models, sufficient and representative data should be available. There are many opportunities to apply more advanced neural network architectures, especially for tasks where generalization is important. Generalization is important when there is a need to recognize unseen persons' activities, deploy the UWB radar in previously unseen environments, or learn new activities after deployment. Such use cases can benefit from novel self-supervised architectures like foundation models, deep reinforcement learning, etc. This will require innovations that do not require pre-labeled data but instead rely on ``hints'' from the environment to correctly classify unseen behavior. Such hints might be derived from integrating UWB radar models with large language models (allowing users to describe their activities using natural language), from recognizing likely sequences of activities (e.g., identifying typical sequences of household activities), from other deployments in similar environments (using federated learning), etc. Similarly, the use of unsupervised (clustering) techniques could allow retraining of the neural network when new conditions arise. Finally, current \ac{ml} approaches for UWB radar do not include domain knowledge in their training phase. Novel physics-informed neural networks (PINNs) architectures could incorporate domain knowledge related to propagation characteristics, ray-tracing, etc.

\Rev{\textbf{\ac{ml} models in constrained devices} \\ 
Integrating large \ac{ml} models into constrained devices used in \ac{uwb} radar systems is challenging due to these devices' computational and memory limitations. However, directly executing the neural networks on the devices has several advantages: it reduces the network load (since radar systems typically generate large amounts of raw \ac{cir} data), it reduces the overall latency, and it improves privacy-awareness. To this end, one approach towards embedded machine learning is to use model compression techniques such as pruning, quantization, and knowledge distillation. Pruning removes redundant neurons or connections, while quantization reduces the precision of model weights, and knowledge distillation transfers knowledge from a large model (teacher) to a smaller model (student). Another promising solution is using efficient model architectures, such as MobileNets \cite{howard2017mobilenetsefficientconvolutionalneural}, SqueezeNets \cite{iandola2016squeezenetalexnetlevelaccuracy50x}, early-exit neural networks \cite{LATTANZI2023106035} and TinyML \cite{TinyMLpresence:UWBradar} models, which are designed to reduce the computational load while preserving a balance between accuracy and speed. This makes them suitable for real-time applications like activity recognition. Finally, federated learning is a solution for scenarios where data privacy is crucial. By training the model across multiple devices without sharing raw data, federated learning distributes the learning process and reduces the computational load on individual devices. These strategies address current challenges and offer promising directions for future research in integrating ML models into compact devices.}

\textbf{Privacy, security, and bias concerns} \\ As UWB radar technologies become more widespread, concerns about privacy and security can arise. Although UWB radar inherently offers more privacy for end users than vision-based tracking systems, unwanted activity and location monitoring can still occur. With the advent of advanced AI methods, it is difficult to predict if UWB radar technology will advance to the point where individuals can be recognized based on their behavior, which would again significantly impact privacy. Additionally, radar systems often use complex machine-learning algorithms for target recognition. If the training data used to develop these algorithms is biased, the system may exhibit discriminatory behavior, leading to the misidentification or profiling of certain individuals or groups. Novel techniques from the field of explainable AI could be used to identify such biases in neural networks that have been trained for activity recognition and presence detection. Such explainable AI techniques could assist with improving privacy and identifying bias that can be solved through the design of next-generation neural networks. Addressing these challenges requires a multidisciplinary approach involving technologists, policymakers, ethicists, and other stakeholders to strike a balance between security needs and the protection of individual rights and privacy and to promote responsible use of radar technology.

\textbf{Adaptive PHY layer settings} \\ All discussed scientific publications use pre-set PHY configurations. There is a lack of research investigating the impact of PHY layer settings on overall performance. Although PHY layer configurations such as transmission power, bandwidth, modulation, etc., strongly influence energy consumption, it is currently unclear which PHY configurations are required to realize specific use-case requirements such as accuracy, range, etc. Depending on the environment and its dimensions, maintaining the desired radar accuracy may require adapting different transmitted power levels within the regional regulatory limits. The ideal power level and PHY layer settings should be dynamically and intelligently selected, depending on factors such as the specific use case, environmental conditions, the device's power supply (e.g., whether it is battery-powered or not), and other relevant variables. Dynamically adapting PHY layer settings has previously been shown to be beneficial for both communication and localization but has not yet been investigated for UWB radar \cite{Coppens2023}. Moreover, it might be interesting to investigate if the use of multiple PHY settings (different center frequencies, modulations, etc.) could even improve the accuracy of UWB radar applications by offering increased frequency and time diversity, allowing the design of novel algorithms with better capabilities to accurately filter out noise.

\textbf{System-level design of UWB multi-antenna systems} \\ State-of-the-art UWB antenna design frameworks currently focus on full-wave/circuit co-design techniques to analyze and optimize single-TX single-RX UWB radar systems in the time, frequency and spatial domains to achieve peak performance by mitigating orientation-specific pulse distortion and reducing self-interference. With the surging interest in UWB antenna arrays to further improve spatial resolution and enhance interference rejection, there is a strong need for system-level design frameworks that  incorporate mutual coupling and radiation pattern deformation effects in closely-spaced multi-antenna configurations to accelerate the development of such advanced multi-antenna UWB radar nodes. Moreover, with UWB radar systems entering the stage of mass production and deployment in a wide variety of challenging environments (IoT, Healthcare, Industry 4.0, …), special attention should be devoted to include integration platform and radome effects from the early design stages to guarantee stable installed performance.  Physics-based machine learning techniques show great potential to tackle these challenges, but there is still a long road ahead. 

\textbf{Polarization reconfiguration and advanced UWB arrays} \\ To address the increasing demand for finer resolution sensing in the smallest possible footprint, there is a pressing need for novel UWB antennas and arrays. Current high-performance UWB antenna designs are typically bulky and impede compact integration with UWB transceivers, whereas miniaturized designs are easily detuned by nearby objects and suffer from low isolation hindering large-scale array deployment. The efficient roll-out of UWB radar systems critically depends on compact, mass-manufacturable antenna designs that maintain high performance in challenging operating conditions, and facilitate scaling to large UWB arrays. Furthermore, the potential of polarization reconfiguration to further improve reliability and/or to implement pose estimation on top of distance and angle estimation must be explored. Even more challenging is the path towards shared-aperture full-duplex operation. By leveraging the same antenna (system) for transmit and receive, the footprint can be significantly reduced with respect to current UWB radars leveraging distinct transmit and receive systems. In addition, novel sparse array configurations should be investigated to achieve high resolution and low mutual coupling, while reducing the number of antenna elements compared to fully populated arrays. Finally, novel fabrication technologies should be developed to realize cost-effective 3D (conformal) UWB antenna arrays to achieve maximum flexibility and enable precise 3D spatial awareness. 

\textbf{Towards multi-function UWB systems} \\ Last,  but certainly not least, an important emerging research area focusses on simultaneously providing multiple functions (such as radar, localization, and sensing) within a single system. These innovations are becoming increasingly important with the rapid rise of autonomous robot platforms in Industry 4.0 scenarios, self-driving cars, and immersive AR/VR, all of which need to reliably exchange data and accurately sense the environment at the same time. Conceiving novel methods to simultaneously provide multiple functions in the same frequency spectrum is crucial to deal efficiently with scarce and costly hardware resources and frequency spectrum. Yet, unleashing the full potential of such multi-function UWB nodes requires innovations in and across multiple research domains. First, algorithmic and hardware co-optimization is essential to translate physical-layer innovations to unprecedented system-scale performance. Finally, the development of \Rev{\ac{uwb}} self-interference cancellation methods for next-generation UWB systems, equipped with tens or even hundreds of UWB antennas, will likewise prove to be a challenging multi-disciplinary task.

\section{Conclusion \& Lessons Learned}\label{section:Conclusion-lessons-learned}

\begin{table*}[t]
\caption{Overview of UWB radar application domains, as well as their typical system setup, sampling rate requirements, bandwidth requirements, their typical post-processing techniques, and example use cases. Detailed descriptions are given in Sections \rom{5} - \rom{6}}
\label{table:applications:usecase:example}
\fontsize{8pt}{8.2pt}\selectfont
\centering
\renewcommand{\arraystretch}{1.5}
\begin{tabular}{|m{1.8cm}||m{1.5cm}|m{1.1cm}|m{1.5cm}|m{4cm}|m{3.3cm}|m{1.5cm}|}
\hline
\textbf{Applications} & \textbf{Typical system} & \textbf{Sampling rate} & \textbf{Bandwidth requirement} & \textbf{Most Common Applied Techniques} & \textbf{Cases and Examples} & \textbf{Common used device} \\ \hline\hline

Presence Detection (Section~\ref{section:presence}) & mono-static & low & low & Background Subtraction, Anomaly Detection & Occupancy detection, People counting, and In-car seat occupancy detection  & Novelda Xethru X4M03 \\ 
\hline

Device-free localization (Section~\ref{section:DFL}) & multi-static & low & low & Clutter subtraction, Target detection, Target localization, Target tracking & Distance estimation, Angle estimation, Human tracking & Decawave DW1000 \\ 
\hline

Activity recognition (Section~\ref{section:activity}) & mono-static / multi-static & high & high & Preprocessing techniques (clutter and noise removal - region of interest extraction), Feature extraction techniques, Classification Algorithms (Pattern recognition and neural networks) & Sleep monitoring, Fall detection, Elderly care, Human identification, ADL, Hand gesture recognition & Novelda Xethru Series \\ 
\hline

Vital sign monitoring (Section~\ref{section:vitalsign}) & mono-static & high & high & Data filtering, Signal separation and feature extraction, \ac{ml} and data modeling & Emergency and natural disaster, Child and elderly care, Vehicular monitoring, Animal health monitoring & Novelda Xethru Series \\ 
\hline \hline

\end{tabular}
\end{table*}

This survey presents a comprehensive overview of UWB radar, delving into its principles and various applications. Unlike previous survey papers, our focus encompasses the entire system, covering hardware, software, datasets, UWB standards, commercial implementations, and recent standardization efforts. We have determined five application domains for UWB radar: presence detection, device-free localization, activity recognition, vital sign monitoring, and \ac{isac}. For each of these domains, we have summarized the relevant information in TABLE~\ref{table:applications:usecase:example}, along with the critical lessons learned for each application, as detailed below:

\subsubsection{Presence Detection}
From our study of presence detection using UWB radar, we learned that reliable detection systems must adapt to varying environmental conditions, such as physical obstructions and electromagnetic interference, to maintain high detection accuracy. The low sampling rate and low bandwidth requirements are crucial to reduce power consumption and operational costs, making the technology viable for widespread use. However, the low sampling rate imposes limitations on data resolution, necessitating detection algorithms, including \ac{ml} and anomaly detection techniques, to differentiate between various movements and objects effectively. To tackle the issue of insufficient information in the samples, transitioning to a multi-static setup can significantly enhance detection capabilities, albeit with synchronization challenges that must be addressed. Multi-static radar systems can provide more comprehensive spatial coverage and distinguish between multiple targets in cluttered environments by using data from various angles. This setup can also help accurately distinguish between multiple moving objects in real-time, a task where current methods struggle.

\subsubsection{Device-free Localization}
From our study of device-free localization using UWB radar, we learned that localization systems must utilize a multi-static setup with low sampling rates and low bandwidth requirements to achieve accurate localization in various environments. Techniques such as clutter subtraction, target detection, target localization, and target tracking are crucial for enhancing the precision of these systems. However, current methods face challenges in complex scenarios, such as accurately tracking multiple targets simultaneously in environments with significant clutter and interference. For example, the use of \ac{aoa} estimation techniques in UWB radar has not been studied extensively, which limits the ability to locate targets precisely using UWB radar. Additionally, enhancing the robustness of clutter and background subtraction techniques—such as adaptive filtering, online environment learning, and context-aware filtering—and developing more advanced algorithms for real-time multi-target tracking will be essential. These enhancements will help effectively manage complex environments using \ac{ml} methods, where more significant features impact the accuracy and complexity of the models.

\subsubsection{Activity Recognition}
From our study of activity recognition using UWB radar, we learned that effective activity recognition systems often require mono-static and multi-static setups, considering the environment, with high sampling rates and bandwidth requirements. These systems rely on preprocessing techniques such as clutter and noise removal, region of interest isolation, and advanced feature extraction. Classification algorithms, including pattern recognition and neural networks, are crucial for accurately identifying activities. However, current methods face challenges in complex scenarios, such as distinguishing between similar activities and recognizing activities in dense environments. Integrating deep learning techniques could significantly improve the system's ability to adapt to new activities, but this area still needs to be explored. Approaches like continuous and incremental learning allow the AI to update and refine its models, ensuring robust performance across complicated scenarios like \ac{adl} allowing retraining for unseen environments and persons. For instance, in activity recognition, these learning techniques must adapt to the dynamic changes in signal reflections caused by different movements, such as environments like cars. Moreover, AI solutions for UWB radar require extensive data from various environments based on application and a diverse range of human subjects to test their generalizability and applicability. It is essential to ensure that these methods are robust and can handle environmental changes, such as variations in indoor and outdoor settings, as accuracy on specific data alone is insufficient. 

\subsubsection{Vital Sign Monitoring}
From our study of vital sign monitoring using UWB radar, we learned that effective monitoring systems typically employ a mono-static setup situated in front of the person, with high sampling rates and bandwidth requirements. These systems utilize data filtering, signal separation, and \ac{ml} techniques to monitor vital signs accurately. Devices from the Novelda Xethru Series are commonly used in these applications due to their high resolution and reliability. However, current methods face challenges in accurately measuring vital signs in real-time and ensuring privacy in sensitive applications. For example, the different orientations and locations of the humans in front of the radar and the monitoring of multiple humans simultaneously present significant challenges. Investigating these aspects further and exploring synchronized multi-static setups could enhance the system's applicability and effectiveness in diverse scenarios. Privacy remains a crucial concern, as UWB radar systems must ensure that personal health data is securely processed and stored without compromising individual privacy.

\subsubsection{\ac{isac}}
In our overview of \ac{isac} using UWB radar, we discussed that this emerging field, characterized by recent papers, is extremely new and requires extensive further studies. Effective \ac{isac} systems must integrate advancements in the physical layer with innovations in chip design, antenna design and algorithm design. Current research focuses mainly on novel physical layer design, but the lack of experimental studies highlights the need for more comprehensive datasets. Studies analyzing the trade-offs in antenna designs, radio configuration settings and algorithm choices for simultaneous radar and communication are still lacking. In addition, the lack of standardized use case requirements or benchmarks makes it difficult to understand which inherent performance trade-offs should be targeted in new publications. 

Our overview also addresses the evolution of UWB radar hardware. Whereas in the past, mostly custom-built hardware platforms were used for UWB radar research, a significant number of commercially available radio chips have recently become available on the market, indicating a growing maturity of the domain. Our overview provided an examination of available radio hardware platforms as well as their characteristics in terms of supported center frequencies and available bandwidth.

Finally, the survey highlights various promising areas for future research. Topics that include rich opportunities for exploration include, among others, better standardization, more open datasets, improved robustness in non-ideal conditions, challenges related to coexisting radio transmission, auto-calibration solutions, the extension of existing work towards multiple targets, improved bias and privacy, dynamic adaptation of PHY layer settings, and innovations that allow joint sensing and communication.

The contributions of this survey are diverse, providing a clear-eyed view of UWB radar's fundamentals, a categorical analysis of its diverse applications, and a technical overview that includes development datasets and hardware selection. As such, this overview has laid the groundwork for researchers to build upon, offering a comprehensive resource that charts a clear course for those navigating the complexities of UWB radar research.

\section*{Acknowledgment}
This research has been funded by Fonds Wetenschappelijk Onderzoek Vlaanderen (FWO) through grant number G018522N for the FWO PESSO project.

\bibliographystyle{unsrt}
\bibliography{references}

\end{document}